\documentclass[%
 reprint,
 superscriptaddress,
 numerical,
 showpacs,
 amsmath,amssymb,
 aps,
 prx,
 longbibliography,
 floatfix
]{revtex4-2}

\usepackage{graphicx}
\usepackage{bm}

\usepackage[dvipsnames]{xcolor}
\usepackage{hyperref}
\hypersetup{
  colorlinks   = true,
  urlcolor     = Blue,
  linkcolor    = BrickRed,
  citecolor   = PineGreen
}

\usepackage[bb=dsserif]{mathalpha}

\usepackage{braket}
\usepackage{pifont}

\let\Re\relax

\DeclareMathOperator{\tr}{tr}

\DeclareMathOperator{\Re}{Re}

\newcommand{\logX}{X_\text{L}}

\newcommand{\logZ}{Z_\text{L}}
\newcommand{\xsf}{\mathsf{x}_j}
\newcommand{\xsfb}{\overline{\mathsf{x}}_j}
\newcommand{\qseta}{\eta_j^{(q,s)}}
\newcommand{\qsetab}{\overline{\eta}_j^{(\bar{q},s)}}

\newcommand{\thr}{\mathrm{th}}
\newcommand{\calM}{\hat{\mathcal{M}}_{q\bar{q},s}}
\newcommand{\MWPM}{\text{(MWPM)}}
\newcommand{\coh}{\text{(coh)}}
\newcommand{\inc}{\text{(inc)}}

\newcommand{\calMdiag}{\hat{\mathcal{M}}_{qq,s}}
\newcommand{\calH}{\hat{\mathcal{H}}_{q,s}}
\newcommand{\calZ}{\mathcal{Z}_{q\bar{q},s}}

\def\tcm{T.C.M. Group, Cavendish Laboratory, University of Cambridge, J.J. Thomson Avenue, Cambridge, CB3 0HE, UK}
\def\DAMTP{DAMTP, University of Cambridge, Wilberforce Road, Cambridge, CB3 0WA, UK}

\begin{document}

\title{The surface code beyond Pauli channels: Logical noise coherence, information-theoretic measures, and
errorfield-double phenomenology}

\author{Jan Behrends}
\affiliation{\tcm}

\author{Benjamin B\'eri}
\affiliation{\tcm}
\affiliation{\DAMTP}

\begin{abstract}
We consider the surface code under errors featuring both coherent and incoherent components and study the coherence of the corresponding logical noise channel and how this impacts information-theoretic measures of code performance, namely coherent information and quantum relative entropy. Using numerical simulations and developing a phenomenological field theory, focusing on the most general single-qubit X-error channel, we show that, for any nonzero incoherent noise component, the coherence of the logical noise is exponentially suppressed with the code distance. We also find that the information-theoretic measures require this suppression to detect optimal thresholds for Pauli recovery; for this they thus require increasingly large distances for increasing error coherence and ultimately break down for fully coherent errors. To obtain our results, we develop a statistical mechanics mapping and a corresponding matrix-product-state algorithm for approximate syndrome sampling. These methods enable the large-scale simulation of these non-Pauli errors, including their maximum-likelihood thresholds, away from the limits captured by previous approaches. 
\end{abstract}

\maketitle

\section{Introduction}

Quantum error correction (QEC) aims to protect quantum information by significantly suppressing noise~\cite{Shor:1995fj,Calderbank:1996ja,Steane:1996kg,Preskill:2018gt}.
Reducing noise is believed to be necessary to achieve quantum advantage~\cite{Dalzell2021,StilckFranca:2021cb,Hangleiter:2023cq}, i.e., for quantum computers that cannot be simulated classically.

A major source of noise is the coupling between a quantum system and its environment~\cite{Zurek:2003fm}.
The resulting decoherence of quantum information can be described by stochastic noise that acts probabilistically~\cite{Nielsen:2010ga}.
However, decoherence is not the only factor that deteriorates quantum information~\cite{Chamberland:2017dm,Bravyi:2018ea,Gottesman2019}:
Generally, noise acts as a completely positive trace-preserving~\cite{Nielsen:2010ga} (CPTP) quantum channel, which can combine stochastic (incoherent) noise and unitary (coherent) rotations of qubits.
Coherent noise channels, which may arise, e.g., due to spurious gate rotations, can add up constructively, which makes them potentially challenging for QEC~\cite{Bravyi:2018ea,Gottesman2019,Gutierrez:2016dm,Huang:2019gj,Greenbaum:2018ce,Iverson:2020fe}.
They can be a relevant, or even the primary, source of noise~\cite{Bluvstein:2024ht,gross2024characterizing,Kurilovich_corr,nigmatullin2025experimental,Bluvstein_arch,Cong_PhysRevX.12.021049},  depending on the physical platform, and even in cases where they can be directly suppressed, this comes with gate and time overhead~\cite{Wallman:2016be,Cong_PhysRevX.12.021049,Zhang_PhysRevApplied.17.034074,Hashim:2021gr}.  

A simple example combining incoherent and coherent contributions is the error channel $\mathcal{E} = \bigotimes_j \mathcal{E}_j$ with 
\begin{equation}
 \mathcal{E}_j [\rho] = (1-p_j) \rho + p_j X_j \rho X_j + i \gamma_j \sqrt{(1-p_j)p_j} [X_j ,\rho] 
 \label{eq:error_channel}
\end{equation}
on qubit $j$; $\mathcal{E}_j$ is CPTP~\cite{Nielsen:2010ga} when $\gamma_j^2 \le 1$.
Eq.~\eqref{eq:error_channel} is the most general single-qubit channel one can build using $X_j$. 
When $\gamma_j=0$, it describes incoherent bit flip errors, and when $\gamma_j = \pm 1$, we get coherent errors, namely $\mathcal{E}_j^\coh [\rho] = U_j \rho U_j^\dagger$ with $U_j = e^{\pm i\vartheta_j X_j}$ and $\vartheta_j= \arcsin \sqrt{ p_j }$.
The error channel~\eqref{eq:error_channel} thus interpolates between incoherent and coherent limits, which have qualitatively distinct QEC behavior~\cite{Dennis:2002ds,Bombin:2012km,Wootton:2012cb,Chubb:2021cn,Bravyi:2018ea,Gutierrez:2016dm,Huang:2019gj,Gottesman2019,Greenbaum:2018ce,Iverson:2020fe,Venn:2020ge,Venn:2023fp,Behrends:2024bs,behrends2024statistical,Chen2024,Lee2024,Bao2024}.

Studying the impact of such a channel on quantum codes is important both conceptually and practically. Practically, it allows one to assess whether an error threshold exists; if so, then below theshold the logical error rate decays exponentially with code distance~\cite{Dennis:2002ds}.
Conceptually, it gives key insights for the rapidly emerging field of error-corrupted quantum many-body states and mixed-state phases of matter~\cite{Fan:2024ku,Lee:2023fe,Chen:2024jh, Sang2023,Myerson:2025bx,Su:2024je,Li2024,Chen2024,Bao2024,Lavasani2024,Lyons2024,Sohal:2025ec,Lu2024,Sang2024, Ellison:2025kp,Zhang2024,Lee2024,Oshima2024,Lessa_PRXQuantum.6.010344,RGK_PhysRevX.14.041031}, and the relation of frequently used information-theoretic measures to QEC. For example, topological order in mixed states has recently been characterized using quantum relative entropy~\cite{Umegaki:1962ee}, coherent information~\cite{Schumacher:1996gt,Lloyd:1997if}, or topological entanglement negativity~\cite{Zyczkowski:1998ks,Lee:2013ia}, but the focus has mostly been on fully incoherent channels.
For these, the maximum-likelihood threshold coincides with a topological phase transition in the mixed state~\cite{Bao2023,Lee:2023fe,huang2024coh} and the coherent information can detect this already for small systems~\cite{Colmenarez:2024iz,huang2024coh,Colmenarez2024er}.
For coherent errors, however, measures like the coherent information are much more subtle; while they detect general unitary recoverability (which can be guaranteed for coherent errors) their use for Pauli-string recoverability, relevant for QEC, is unknown.

For fully coherent errors~\cite{Bravyi:2018ea,Gottesman2019,Greenbaum:2018ce,Iverson:2020fe,Venn:2020ge,Venn:2023fp,Behrends:2024bs,behrends2024statistical,Chen2024,Lee2024,Bao2024} in surface codes~\cite{Kitaev:1997kq,Bravyi1998quantum,Freedman1998projective}, previous works found a surprisingly high error threshold and an unusual above-threshold phase with power-law decaying logical error rate~\cite{Venn:2020ge,Venn:2023fp,Behrends:2024bs,behrends2024statistical}. 
The robustness of these features against a small but nonzero incoherent component is however also unknown, as is the nature of the error-corrupted many-body state, not only when coherent and incoherent error components act in tandem, as in channel~\eqref{eq:error_channel}, but even in the fully coherent case post syndrome measurements.

In this work, we study the surface code~\cite{Kitaev:1997kq,Bravyi1998quantum,Freedman1998projective} under error channel \eqref{eq:error_channel}.
We focus on the surface code due to its experimental relevance~\cite{Acharya:2023fl,Bluvstein:2024ht,Acharya2024,eickbusch2024dynamic}, and known maximum-likelihood thresholds for incoherent~\cite{Dennis:2002ds,Merz:2002gj} and coherent~\cite{Venn:2023fp} $X$ errors.
Our main interest is conceptual: given the ensemble $\rho_0'$ of error-corrupted post-syndrome-measurement states, how does the residual coherence $\gamma_\text{L}$ in the logical channel impact the characterization in terms quantum relative entropy $S_\text{rel}$ and coherent information $I_\text{C}$, and the relation of these to QEC?

To assess this, we must establish maximum likelihood thresholds---i.e., establish the QEC phase diagram. 
To this end, and to compute $S_\text{rel}$, $I_\text{C}$ and $\gamma_\text{L}$,  we develop a statistical mechanics mapping for the decoding problem. This expresses the probabilities of equivalence classes of errors as partition functions of classical interacting random-bond Ising models (RBIM).
To evaluate a partition function numerically, we use the transfer matrix~\cite{Schultz:1964fv}, i.e., a (1+1)D many-body quantum circuit.
We show, generalizing our findings from the coherent case~\cite{Venn:2023fp,Behrends:2024bs,behrends2024statistical}, that below threshold and above it away from the coherent limit the entanglement entropy of a 1D state evolved by the quantum circuit exhibits an area law, and we can thus efficiently simulate the evolution numerically using matrix product states (MPS)~\cite{Hauschild:2018bp,Cirac:2021gx}. We leverage this to develop a circuit-based sampling algorithm to sample error syndromes  according to their approximate probabilities (with the error being controlled by that of the MPS approximation); this enables the simulation of the non-Pauli channel~\eqref{eq:error_channel} away from the fully coherent and almost fully incoherent regimes accessible to previous methods~\cite{Ma:2023ia,Bravyi:2018ea,behrends2024statistical,Bao2024}. 

\begin{figure}
\includegraphics[scale=1]{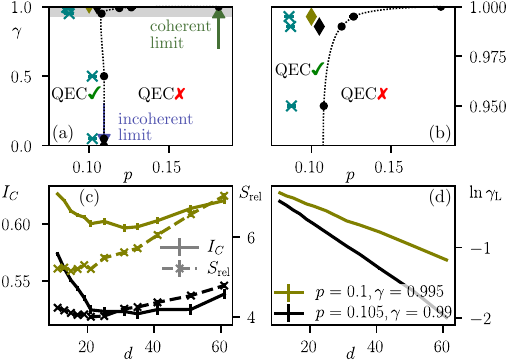}
\caption{(a) Approximate QEC phase diagram where bottom and top boundaries correspond to the incoherent ($\gamma=0$) and coherent ($\gamma=1$) limits.
The dotted line denotes the maximum-likelihood threshold separating the error-correcting phase (QEC\ding{51}) from a non-correcting phase (QEC\ding{55}), where black dots represent numerically calculated values including error bars.
Teal crosses show the minimum weight perfect matching threshold.
Green and blue arrows show the incoherent~\cite{Dennis:2002ds} and coherent~\cite{Venn:2023fp} maximum-likelihood thresholds, respectively.
(b) Part of the phase diagram, indicated in (a) as a gray rectangle.
Olive and black diamonds denote the parameters used in panels (c) and (d).
(c) Coherent information $I_\text{C}$ (solid lines) and quantum relative entropy $S_\mathrm{rel}$ (dashed lines and crosses) as a function of code distance $d$ for $p=0.1$, $\gamma= 0.995$ (olive markers) and $p=0.105$, $\gamma = 0.99$ (black markers).
(d) Logical noise coherence $\gamma_\text{L}$ versus $d$ for the same $p$ and $\gamma$ as in (c).
(For coherent errors, $\gamma_\text{L}=1$.)
Data are averaged over 1000 to 10000 syndromes and the error bars showing the standard error of the mean are imperceptible.
}
\label{fig:overview}
\end{figure}

We show the QEC phase diagram in Fig.~\ref{fig:overview}(a) and (b), for $\gamma_j=\gamma,\, p_j=p,\, \forall j$.
The maximum-likelihood threshold separates an error-correcting phase from a phase where QEC fails.
The threshold is largely independent of the coherent contribution $\gamma$  until it is close to the coherent limit $\gamma = 1$ [Fig.~\ref{fig:overview}(b)].
The above-threshold phase that we find away from the coherent limit is similar to the above-threshold phase of incoherent errors, i.e., it is characterized by a logical error rate that increases with system size up to a limiting value~\cite{Fowler:2012fi} ($1/2$ for $X$ errors), and by area-law entangled 1D states from the long-time evolution by the (1+1)D quantum circuit~\cite{Bravyi:2014ja,Behrends:2024bs}.
The above-threshold phase for coherent errors, on the other hand, is characterized by a power-law decreasing logical error rate~\cite{Venn:2023fp,behrends2024statistical} and a logarithmically increasing entanglement entropy~\cite{Behrends:2024bs,behrends2024statistical}.
From our numerical data we conclude that this above-threshold behavior is however a special property of the coherent limit, unstable to a small incoherent noise component.

For the information-theoretic measures, we find new results already in the fully coherent limit: For the surface code graph class that our system exemplifies~\cite{Venn:2020ge}  $S_\text{rel}$ is simply the Shannon entropy of syndrome probabilities, while $I_\text{C}$ is maximal, hence constant and thus cannot detect the QEC, i.e., Pauli-correction, threshold.
Close to, but not at, the coherent limit, we show numerically that both $S_\text{rel}$ and $I_\text{C}$  suffer from finite-size effects and thus require large systems to detect the threshold.
In particular, as we show in Fig.~\ref{fig:overview}(c) for two pairs of
subthreshold $p$ and $\gamma$, with $\gamma$ close to unity, both $S_\text{rel}$ and $I_\text{C}$ initially decrease with code distance $d$ before increasing again for sufficiently large systems.
Since the expected subthreshold behavior is an increase with $d$, both measures need large systems to signal the subthreshold regime.
To interpret these findings, we consider the logical error channel and numerically show that $\gamma_\text{L}$ is exponentially suppressed with code distance for any $|\gamma|<1$, as illustrated in Fig.~\ref{fig:overview}(d).
This suggest than $S_\text{rel}$ and $I_\text{C}$ become performant only once  $\gamma_\text{L}$ is sufficiently suppressed.

To provide a theory for the qualitative features of the system, we develop a phenomenological description using the errorfield double approach~\cite{Bao2023} based on surface code anyons $e$ and $m$. Our theory provides a new perspective already in the fully coherent case, and establishes a picture for how the instability of the above-threshold critical fully coherent regime sets in.
We first show how, upon a spacetime rotation~\cite{Bao2023}, the $X$-error nature of channel~\eqref{eq:error_channel} allows one to relate the problem to a non-Hermitian 1D Ising chain at the boundary of the 2D surface code. [As we shall discuss, this theory is closely related to our Ising statistical mechanics model, and its corresponding quantum circuit.] This invokes a holographic ``symmetry topological field theory'' (SymTFT)~\cite{WenWei15,aasen2016topological, JiWen2020categorical,lichtman2020bulk, ChatterjeeWen23,FreedMooreTeleman22,TH23,Bhardwaj_PtII, Lootens_PRXQuantum.4.020357,Lootens_PRXQuantum.5.010338,Bhardwaj_PhysRevLett.133.161601,Fechisin_PhysRevX.15.011058,Barbar_PhysRevLett.134.151603,Sun_PRXQuantum.6.020333,Bottini_PhysRevLett.134.191602,Seifnashri_PhysRevLett.133.116601,Okada_PhysRevLett.133.191602,SymTFT_TC} correspondence. 
We then take a phenomenological formulation of this boundary theory as a non-unitary compact boson field theory of $\mathbb{Z}_2$ topological order boundary dynamics. In both the 1D Ising and field theory approaches, the error-correcting phase corresponds to a gapped edge condensing $e$ anyons.  Using the field theory, we also find a conformal field theory (CFT) state emerging from coherent errors above threshold, which is however unstable to perturbations by incoherent noise. The resulting gapped phase corresponds to the area-law entanglement observed in our numerics, and the errorfield double also predicts, consistently with Fig.~\ref{fig:overview}(d), that $\gamma_\text{L}$ decays exponentially with code distance in the presence of even a small incoherent noise component.

This work is organized as follows: After reviewing, in Sec.~\ref{sec:error_correction}, relevant features of the surface code, in Sec.~\ref{sec:information} we show  how information-theoretic measures can be expressed for channel~\eqref{eq:error_channel}.
We introduce our statistical-mechanics mapping in Sec.~\ref{sec:statmech} and the corresponding (1+1)D quantum circuit in Sec.~\ref{sec:quantum_circuit}.
In particular, we discuss symmetry-breaking and entanglement properties of this circuit's long-time state in Sec.~\ref{sec:entanglement} and introduce our sampling algorithm in Sec.~\ref{sec:error_strings}.
We compute the error thresholds, giving the phase diagram [Fig.~\ref{fig:overview}(a) and (b)], and information-theoretic quantities [Fig.~\ref{fig:overview}(c) and (d)] in Sec.~\ref{sec:numerics}, and develop our phenomenological theory in Sec.~\ref{sec:FT}. We conclude in Sec.~\ref{sec:conclusion}.

\section{Error correction in the surface code}
\label{sec:error_correction}

\begin{figure*}
\includegraphics[scale=1]{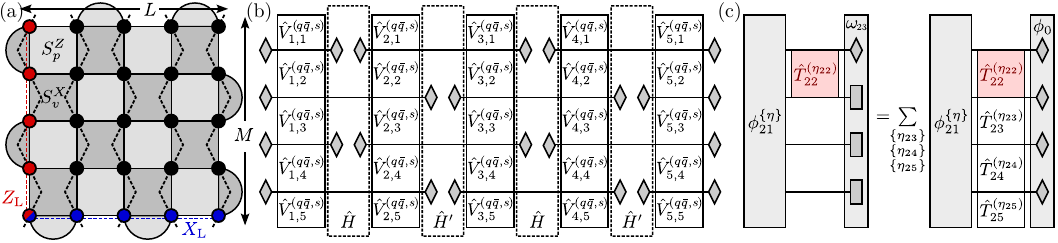}
\caption{(a) Surface code ($L=M=5$) with physical qubits on the vertices (black discs), and alternating $S_v^X$ and $S_p^Z$ stabilizers on the faces of the lattice.
The logical $\logX$ and $\logZ$ are denoted by blue and red dashed lines, respectively.
We map the surface code to a complex RBIM with two Ising spins $\sigma_v$ and $\bar{\sigma}_v$ on each $S_v^X$ site.
The dashed black lines connect $S_v^X$ sites, constituting vertical slices of the RBIM.
(b) Quantum circuit with many-body gates $\hat{V}_{l,m}^{(q\bar{q},s)}$ on physical qubits, where diamonds denote state projection or initialization to $2\ket{++}$ (hence adjacent diamonds are projective $x$-basis measurements with outcome $+1$).
The $\hat{V}_{l,m}^{(q\bar{q},s)}$ alternate with layers $\hat{H}$ and $\hat{H}'$, denoted by dashed rectangles.
(c) To sample an error string's weight on site $j$ via $\eta_j = \pm 1$ [shown here: $j=22$ with the gate $\hat{T}_{22}^{(\eta_{22})}$ in red; $\hat{T}^{(\eta_j)}_j$ is a shorthand notation for $\hat{V}_{l,m}^{(qq,s)}$, see the discussion of Eq.~\eqref{eq:Tnotation}], we evaluate its conditional probability $P_j \propto \braket{ \omega_{j+1} | \hat{T}_j^{(\eta_j)} | \phi_{j-1}^{\{\eta\}} }$  [Eq.~\eqref{eq:conditional}] using that $\ket{\omega_{j+1}}$ is a product state of $2\ket{++}$ states (denoted by diamonds) and $\ket{00} + \ket{11}$ Bell states (denoted by rectangles). The state $\ket{\omega_{j+1}}$ is the superposition of all $\{\eta\}$ configurations of the state $T_{j+1}^{(\eta_{j+1})} \dots T_N^{(\eta_N)} \ket{\phi_0}$, shown on the right, cf.\ Eq.~\eqref{eq:omega_sampling}.
}
\label{fig:layout}
\end{figure*}

We investigate the surface code under generic single-qubit $X$ errors using the geometry shown in Fig.~\ref{fig:layout}(a).
The code consists of mutually commuting stabilizers $S^X_v = \prod_{j \in v} X_j$ and $S_p^Z = \prod_{j \in p} Z_j$, where $X_j$ and $Z_j$ are physical qubit operators on the vertices of a square lattice.
The stabilizers form a checkerboard pattern on the faces of the lattice~\cite{Bombin:2007ed,Horsman:2012ga}.
Bulk stabilizers act on four adjacent qubits, and boundary stabilizers on two qubits.
The logical subspace, taken as the $+1$ eigenspace of all stabilizers, consists of one logical qubit:
The logical $\logX = \prod_{j \in \zeta}X_j$, connecting left and right boundaries, and $\logZ = \prod_{j\in \zeta'} Z_j$, connecting top and bottom boundaries, commute with all stabilizers but mutually anticommute, $\{\logX ,\logZ \}=0$.
Choosing the operators as in Fig.~\ref{fig:layout}(a), we denote $\logX$'s path length by $L$ and $\logZ$'s path length by $M$.
The total number of qubits is $N = LM$.

The QEC recovery procedure consists of two steps: First, all stabilizers are measured, which projects the error-corrupted state $\mathcal{E}[\rho]$ onto the syndrome $s$.
Second, a Pauli string $C_s$ is chosen according to $s$ (typically by a decoder), which returns the state back to the logical subspace.
(For $X$ errors, $C_s$ is equivalent, up to multiplication by some of the $S_v^{X}$, to one of $\mathcal{X}_s$ and $\logX\mathcal{X}_s$, where $\mathcal{X}_s$ is an arbitrary reference Pauli-$X$ string consistent with $s$.)
The error channel and subsequent recovery operation act on an initial logical state $\rho$ as
\begin{equation}
D_s [\rho] = \Pi_0 C_s \mathcal{E}[\rho] C_s \Pi_0 ,
\label{eq:effective_channel}
\end{equation}
with $\Pi_0$ the projector onto the logical subspace, and where we have used that the projector onto the syndrome $s$ is $\Pi_s = C_s \Pi_0 C_s$.
Due to the projection onto the logical subspace, $D_s$ must be of the form
\begin{equation}
 D_s [\rho] = \mathcal{Z}_{00,s} \rho + \mathcal{Z}_{11,s} \logX \rho \logX + \mathcal{Z}_{01,s} \rho \logX + \mathcal{Z}_{10,s} \logX \rho
 \label{eq:Dpartition}
\end{equation}
with real coefficients $\mathcal{Z}_{qq,s}$ and generally complex $\mathcal{Z}_{01,s}=\mathcal{Z}_{10,s}^*$.
As the $\mathcal{Z}_{qq',s}$ merely collate the contributions of the $X$-error strings consistent with $s$, they are independent of $\rho$. 
The syndrome probability $P(s|\rho) = \tr[ D_s[\rho]]$, however, generally does depend on $\rho$, 
\begin{equation}
P(s|\rho) = \mathcal{Z}_{00,s}+\mathcal{Z}_{11,s}+2\mathrm{Re} \mathcal{Z}_{01,s}\tr(\logX \rho).
\end{equation}

For lattices with only even-weight stabilizers and odd-weight logical operators, as the layout we consider [Fig.~\ref{fig:layout}(a)], we  have $\mathrm{Re} \mathcal{Z}_{01,s}=0$, as we show in Appendix~\ref{sec:phase_relation}.
In this case, $P(s|\rho) = \mathcal{Z}_{00,s} + \mathcal{Z}_{11,s} = P(s)$ is independent of $\rho$. Hence, $D_s [\rho]/P(s)$ implements channel~\eqref{eq:error_channel} in logical space with parameters $p^{(s)}_\text{L}=\mathcal{Z}_{11,s}/P(s)$ and $|\gamma^{(s)}_\text{L}|=|\mathcal{Z}_{10,s}|/\sqrt{\mathcal{Z}_{00,s}\mathcal{Z}_{11,s}}$.
We define for convenience the syndrome-averaged $\gamma_\text{L} = \langle |\gamma^{(s)}_\text{L}| \rangle_s$.

For lattices with $\mathrm{Re} \mathcal{Z}_{01,s}\neq 0$, we define $P(s)$ as the marginal $P(s) = \int d \rho P(s|\rho)P(\rho) = \langle P(s|\rho)\rangle_\Omega$ with uniform distribution $P(\rho)=1/(4\pi)$ on the Bloch sphere $\Omega$; this again yields $P(s) = \mathcal{Z}_{00,s} + \mathcal{Z}_{11,s}$. However, now $D_s [\rho]/P(s)$ is not normalized since  $P(s)\neq \tr[ D_s[\rho]]$ in general.

The syndrome probability $P(s)$ is the sum of probabilities $P_{0,s} = \mathcal{Z}_{00,s}$ and $P_{1,s} = \mathcal{Z}_{11,s}$, from $C_s$ and $C_s \logX$ (and equivalent string) contributions, respectively.
To assess the feasibility of QEC, we consider the infidelity
\begin{align}
 r_{q,s}
 &= 1- F(\rho_0,\rho_{0}^{\prime (q,s)})
 = 1-\frac{\braket{0_\text{L}| \logX^q D_s[\rho_0]\logX^q | 0_\text{L}}}{P(s)}  \nonumber \\
 &= \frac{P_{1-q,s}}{P(s)}
\end{align}
with $\rho_0 = \ket{0_\text{L}}\bra{0_\text{L}}$ and the post-error and post-correction state $\rho_0^{\prime(q,s)} = \logX^q D_s[\rho_0] \logX^q/P(s)$, where $q=0,1$ denote the two distinct correction operations $C_s$ and $C_s \logX$.
(For lattices with $\mathrm{Re} \mathcal{Z}_{01,s}\neq 0$, $(2/3) r_{q,s}$ is the Bloch-sphere average of the infidelity between a generic logical state $\ket{\psi_\text{L}}$ and $\logX^q D_s[ \ket{\psi_\text{L}} \bra{\psi_\text{L}}] \logX^q$~\cite{Venn:2020ge,Venn:2023fp}.)
The logical error rate
\begin{equation}
 P_\text{L} = \langle \min_q r_{q,s} \rangle_s = \sum_s \min_q P_{q,s}
 \label{eq:logical_error}
\end{equation}
equals the syndrome-averaged minimum infidelity, where $\langle \dots \rangle_s = \sum_s P(s) [\dots]$.
In the error-correcting phase, one of the $P_{q,s}$ dominates on average such that $P_\text{L}$ decays to zero exponentially with the code distance.

\section{Information-theoretic measures}
\label{sec:information}

For incoherent Pauli errors, the QEC maximum-likelihood threshold coincides with a phase transition of the error-corrupted mixed state and can be characterized with information-theoretical diagnostics~\cite{Fan:2024ku}, such as the quantum relative entropy or the coherent information.
We now study what these information-theoretic concepts can reveal if one goes beyond Pauli channels. To this end, we consider the statistical ensemble of post-error and post-measurement states~\cite{Eckstein:2024ev}. 

Syndrome measurements project the error-corrupted state $\mathcal{E}[\rho_0]$ onto $s$ via $\Pi_s = C_s \Pi_0 C_s$, giving block-diagonal mixed state
\begin{align}
 \rho_0'
 &= \sum_s \Pi_s \mathcal{E} [\rho_0] \Pi_s = \sum_s C_s D_s [\rho_0] C_s \label{eq:block_diagonal_mixed}\\
 &= \sum_{sq\bar{q}}\mathcal{Z}_{q\bar{q},s} C_s \logX^q \rho_0 \logX^{\bar{q}} C_s,
\end{align}
with the effective channel $D_s[\rho]$ from Eq.~\eqref{eq:Dpartition}.
Using the orthonormal states $\ket{\psi_{qs}} = \logX^q C_s \ket{0_\text{L}}$, which form a basis in the $S_v^{X}=1$ subspace (i.e., that accessible by $X$-errors), we obtain the block-diagonal mixed state
\begin{align}
\rho_0' = \sum_s ( | \psi_{0,s} \rangle , |\psi_{1,s} \rangle ) \underbrace{ \begin{pmatrix} \mathcal{Z}_{00,s} & \mathcal{Z}_{01,s} \\ \mathcal{Z}_{01,s}^* & \mathcal{Z}_{11,s} \end{pmatrix}}_{=\mathcal{Z}_s} \begin{pmatrix} \langle \psi_{0,s} | \\ \langle \psi_{1,s} | \end{pmatrix}
\label{eq:ensembleZ}
\end{align}
that arises after syndrome measurements, upon discarding the measurement record. The blocks $\mathcal{Z}_s$ are proportional to the density matrices for a given syndrome outcome $s$. The coherence in $\mathcal{Z}_s$ is also measured by $\gamma_\text{L}^{(s)}$. 

We shall compare $\rho_0'$ with $\rho_0'' = \logX \rho_0' \logX$.
To this end, we diagonalize $\mathcal{Z}_s$ to obtain a different basis $\ket{\psi_{n,s}'}$ (with $n=0,1$), such that $\rho_0' = \sum_{n,s} \lambda_{n,s} \ket{\psi_{n,s}'} \bra{ \psi_{n,s}'}$ with eigenvalues
\begin{equation}
 \lambda_{n,s} = \frac{1}{2} \left(P(s) -(-1)^n \sqrt{(\mathcal{Z}_{00,s}-\mathcal{Z}_{11,s})^2 + 4 |\mathcal{Z}_{01,s}|^2}\right) .
\end{equation}
In the incoherent limit, the $\lambda_{n,s}$ equal the probabilities $P_{q,s}$.
In the coherent limit, the matrix elements factorize as $\mathcal{Z}_{00,s}=|\tilde{\mathcal{Z}}^\coh_{0,s}|^2$, $\mathcal{Z}_{11,s}=|\tilde{\mathcal{Z}}^\coh_{1,s}|^2$, and $\mathcal{Z}_{01,s}=\tilde{\mathcal{Z}}^\coh_{0,s} [\tilde{\mathcal{Z}}^\coh_{1,s}]^*$, where $\tilde{\mathcal{Z}}_{q,s}^\coh$ is a complex single-spin-species RBIM partition function that arises for coherent errors~\cite{Venn:2023fp}.
Hence, $\lambda_{0,s} = 0$ and $\lambda_{1,s} = P(s)$.

We now give analytical expressions for the quantum relative entropy and coherent information based on the block-diagonal mixed state, Eq.~\eqref{eq:block_diagonal_mixed}.

\subsection{Quantum relative entropy}

The quantum relative entropy between the error-corrupted state $\rho_0'$ and the state $\rho_0''=\logX \rho_0'\logX$ takes the form
\begin{align}
 S_\mathrm{rel} (\rho_0' \parallel \rho_0'')
 &= -\tr [ \rho_0' \left( \ln  \rho_0' - \ln \rho_0'' \right)] \label{eq:quantum_entropy} \\
 &= -\sum_{n,s} (1-\kappa_s) \lambda_{n,s} \ln \frac{\lambda_{n,s}}{\lambda_{1-n,s}}
\end{align}
where the second equality holds for full-rank density matrices, i.e., away from the coherent limit.
We introduced $\kappa_s = 4(\mathrm{Re} \mathcal{Z}_{01,s})^2/[(\mathcal{Z}_{00,s}-\mathcal{Z}_{11,s})^2+4 |\mathcal{Z}_{01,s}|^2]$ to simplify the squared overlap $|\braket{\psi_{n,s}'|\logX|\psi_{n',s}'}|^2 = \kappa_s \delta_{nn'} + (1-\kappa_s)\delta_{n,1-n'}$ that is needed to evaluate $\tr [ \rho_0' \ln \rho_0'' ]$.
Since in our setup $\mathrm{Re} \mathcal{Z}_{01,s}=0$, we have $\kappa_s = 0$.

For incoherent errors, since $\lambda_{0,s}=\min_q P_{q,s}$ and $\lambda_{1,s}=\max_q P_{q,s}$, the quantum relative entropy equals the Kullback–Leibler divergence~\footnote{In the statistical-mechanics language, this equals the typical disorder correlator~\cite{Kadanoff:1971fm,Read:2000hu,Merz:2002ia}, i.e., the typical energy cost of flipping bonds along the logical $\logX$.} between the probability distributions $P_{q,s}$ and $P_{1-q,s}$,
\begin{align}
 S_\mathrm{rel}^\inc (\rho_0' \parallel \rho_0'')
 &= -\sum_{q,s} P_{q,s} \ln \frac{P_{1-q,s}}{P_{q,s}},
\end{align}
which signals the error-correcting threshold~\cite{Venn:2023fp}---below threshold, $\exp[S_\mathrm{rel}^\inc]$ decreases exponentially to zero with code distance~\cite{Merz:2002ia}.

For coherent errors, since $\lambda_{0,s} = 0$ and $\lambda_{1,s} = P(s)$, the density matrices in Eq.~\eqref{eq:quantum_entropy} will only contain sums over the eigenstates $\ket{\psi_{1,s}'}$ with weight $P(s)$, which generally gives $S_\mathrm{rel}^\coh = -\sum_{s}(1-\kappa_s) P (s) \ln P (s)$.
In geometries with only even-weight stabilizers and odd-weight logical Pauli operators, coherent errors followed by error correction results in logical subspace rotations, $D_s^\coh [\rho] = e^{i \logX \vartheta_s} \rho e^{-i \logX \vartheta_s}$~\cite{Bravyi:2018ea,Venn:2020ge,Darmawan2024,Cheng2024}, which implies
\begin{equation}
 \left. S_\mathrm{rel}^\coh \right|_\mathrm{even} = -\sum_{s} P (s) \ln P (s) ,
\end{equation}
equaling the entropy of the probability distribution, but does not correspond to any QEC threshold indicator.

\subsection{Coherent information}

The coherent information is a measure for the recoverability of quantum information~\cite{Schumacher:1996gt,Gullans:2020eg}.
To define it, we consider a system consisting of two parts:
The system $Q$, initially in a state in the logical subspace, and an ancilla $R$. 
We consider the following procedure to define the coherent information:
First prepare $Q$ and $R$ in a Bell-pair superposition (using for $Q$ states in the logical subspace), then act on the system by $\mathcal{S}\circ \mathcal{E}$, where $\mathcal{S}(\cdot)=\sum_s \Pi_s (\cdot) \Pi_s$. This is the same process on $Q$ as that leading to $\rho_0'$, but now gives the density matrix $\rho_{RQ'}$ because of the Bell initial state instead of $\rho_0$.
The coherent information is
\begin{equation}
 I_\text{C} = S [\rho_{Q'}] - S [\rho_{RQ'}] ,
 \label{eq:coherent_info_def}
\end{equation}
where $S[\rho] =\tr[ \rho \ln \rho]$ is the von Neumann entropy of the density matrix $\rho$, and $\rho_{Q'} = \tr_{R} [\rho_{RQ'}]$. 

In terms of the entries of the block-diagonal density matrix~\eqref{eq:ensembleZ}, the coherent information can be expressed as
\begin{align}
 I_\text{C} =& - \sum_{s,\pm} \left(\frac{P(s)}{2} \pm \Re[\mathcal{Z}_{01,s}] \right)\ln \left(\frac{P(s)}{2} \pm \Re[\mathcal{Z}_{01,s}] \right) \nonumber \\
 & + \sum_{n,s} \lambda_{n,s} \ln \lambda_{n,s} ,
\end{align}
which for even-weight stabilizers and odd-weight logicals, by  $\Re[\mathcal{Z}_{01,s}] =0$, simplifies to
\begin{align}
 \left. I_\text{C} \right|_\mathrm{even} =  \sum_{n,s} \lambda_{n,s} \ln \frac{2 \lambda_{n,s}}{P(s)} .
 \label{eq:coherent_info}
\end{align}

From Eq.~\eqref{eq:coherent_info}, we can take the coherent limit, which always gives
\begin{equation}
 \left. I_\text{C}^\coh \right|_\mathrm{even} = \ln 2,
\end{equation}
which indicates perfect recoverability.
To understand this, note that the effective error channel $D_s^\coh [\rho] = e^{i \logX \vartheta_s} \rho e^{-i \logX \vartheta_s}$ acts unitarily on $\rho$~\cite{Bravyi:2018ea}. It is---in principle---always possible to correct such rotations.
(For even-weight stabilizers and odd-weight logical Pauli operators,  this also holds 
for generic single-qubit coherent errors~\cite{Darmawan2024,Cheng2024}.)
However, this does not necessarily imply that the channel is Pauli-string correctable.
Considering correctability via Pauli strings yields a threshold at a critical rotation angle~\cite{Venn:2023fp}, which is not captured by the coherent information.

\section{Statistical mechanics mapping}
\label{sec:statmech}

We now map the coefficients $\mathcal{Z}_{q\bar{q},s}$ to a statistical mechanics model.
To this end, we rewrite each local error channel [Eq.~\eqref{eq:error_channel}] as a sum
\begin{equation}
\mathcal{E}_j [\rho] = \sum_{\substack{ \xsf=\pm1 \\ \xsfb = \pm 1}} e^{J_j^{(0)} + J_j^{(1)} \xsf + J_j^{(2)} \xsfb + J_j^{(3)} \xsf \xsfb} X_j^{\frac{1-\xsf}{2}} \rho X_j^{\frac{1-\xsfb}{2}}
\end{equation}
with couplings (using $\gamma_j>0$ for simplicity)
\begin{align}
 J_j^{(0)} =& \frac{1}{2} \ln \left[ \gamma_j p_j (1-p_j)\right],
 & J_j^{(1)} = -\frac{1}{4} \ln \frac{p_j}{1-p_j} - \frac{i \pi }{4}, \nonumber \\
 J_j^{(3)} =& -\frac{1}{2} \ln \gamma_j
  & J_j^{(2)} =-\frac{1}{4} \ln \frac{p_j}{1-p_j} + \frac{i\pi}{4} .
\end{align}
This enables us to write the total error as a sum over error string configurations
\begin{align}
 \mathcal{E}[\rho]
=& \sum_{\{\xsf, \xsfb \}} e^{\sum_j (J_j^{(0)} + J_j^{(1)} \xsf + J_j^{(2)} \xsfb + J_j^{(3)} \xsf \xsfb)} \nonumber \\
 & \times \mathcal{P}(\{\xsf\}) \rho \mathcal{P}(\{\xsfb\})
 \label{eq:error_string_configurations}
\end{align}
with the Pauli string $\mathcal{P}(\{\xsf\}) = \prod_j X_j^{(\xsf-1)/2}$.
Using the same notation, the error strings $\logX^{q} C_s = \mathcal{P}(\{\qseta \})$ and $\logX^{\bar{q}} C_s = \mathcal{P}( \{ \qsetab \})$ with the configuration of signs $\{\qseta \}$ and $\{\qsetab\}$.
The configurations of $\{\qseta \}$ and $\{\qsetab\}$ can only differ on sites with support of $\logX$.

This implies that the coefficients in $D_s[\rho_0]$ [Eq.~\eqref{eq:Dpartition}] can be expressed as
\begin{align}
 \mathcal{Z}_{q\bar{q},s} &= \braket{0| \logX^q D_s [\rho_0] \logX^{\bar{q}} | 0 } \\
= &\sum_{\{ \xsf , \xsfb \} } e^{\sum_j (J_j^{(0)} + J_j^{(1)} \xsf + J_j^{(2)} \xsfb +  J_j^{(3)} \xsf \xsfb)} \\
\times & \braket{0| \mathcal{P} (\{ \qseta \}) \mathcal{P} (\{ \xsf \}) |0}\braket{0| \mathcal{P} (\{ \xsfb \} ) \mathcal{P} (\{\qsetab \})| 0} \nonumber ,
\end{align}
where the matrix overlap is nonzero only when both Pauli strings $\mathcal{P}(\{\eta_j\}) \mathcal{P} (\{ \xsf \})$ and $\mathcal{P}(\{\bar{\eta}_j\}) \mathcal{P}(\{ \xsfb \})$ form closed loops of $X_j$ operators, i.e., when $\mathcal{P}(\{\eta_j\}) \mathcal{P}(\{ \xsf \}) = \prod_{v} (S_v^X)^{n_v}$ and $\mathcal{P}(\{\bar{\eta}_j\}) \mathcal{P}(\{ \xsfb \}) = \prod_{v} (S_v^X)^{\bar{n}_v}$, respectively.
Importantly, the configurations $\{n_v\}$ and $\{ \bar{n}_v \}$ do not need to be the same.
We describe the implications for $\xsf$ ($\xsfb$ is analog):
Each sign $\xsf$ equals $\eta_j$ unless exactly one neighboring stabilizer $S_v^X$ is part of the closed-loop configuration.
This restricts the sum over configurations of $\xsf$ to fewer terms, namely to configurations $\{n_v\}$ of closed loops.
Introducing classical Ising spins $\sigma_v = (-1)^{n_v}$, we write $\mathsf{x}_{vv'} = \eta_{vv}^{(q,s)} \sigma_{v} \sigma_{v'}$, where $v$ label the positions of the $S_v^{X}$ stabilizers and $vv'$ the bond neighboring $v$ and $v'$.
Writing the sum over configurations of closed loops as a sum over Ising spins, the coefficients have the form of a partition function
\begin{align}
 \mathcal{Z}_{q\bar{q},s} =  \sum_{ \{ \sigma_v,\bar{\sigma}_v \}} \exp [H_{s,q\bar{q}}(\{\sigma_v,\bar{\sigma}_v \} )]
 \label{eq:partition_function}
\end{align}
with
\begin{align}
H_{s,q\bar{q}} =& \sum_{\langle v,v'\rangle} \left( J_{vv'}^{(0)} + J^{(1)}_{vv'} \eta_{vv'}^{(q,s)} \sigma_v \sigma_{v'} + J^{(2)}_{vv'} \bar{\eta}_{vv'}^{(\bar{q},s)} \bar{\sigma}_v \bar{\sigma}_{v'} \right.\nonumber \\
&\left. + J^{(3)}_{vv'} \eta_{vv'}^{(q,s)} \bar{\eta}_{vv'}^{(\bar{q},s)} \sigma_v \sigma_{v'} \bar{\sigma}_v \bar{\sigma}_{v'} \right),
\label{eq:hamiltonian}
\end{align}
where we labeled all couplings $J^{(\mu)}$ ($\mu=\{0,1,2,3\}$) by their neighboring $v$, and where $\langle v,v'\rangle$ denotes the sum over nearest neighbors.
$\mathcal{Z}_{q\bar{q},s}$ is the partition function of a random-bond Ising model with two interacting spin species.

We recover previous results for both coherent and incoherent errors:
For coherent errors ($\gamma_j = 1$), the interaction term $J_{vv'}^{(3)} = 0$, which implies that the Ising spins $\sigma_v$ and $\bar{\sigma}_v$ are decoupled.
The partition function thus factorizes into independent sums for each spin species whose couplings $J_{vv'}^{(1)} = (J_{vv'}^{(2)})^*$ are related via complex conjugation, i.e., $\mathcal{Z}_{q\bar{q},s}^\coh = \tilde{\mathcal{Z}}_{q,s}^\coh [\tilde{\mathcal{Z}}_{\bar{q},s}^\coh ]^*$, recovering the result for coherent errors~\cite{Venn:2023fp}.

For incoherent errors, the off-diagonal $\mathcal{Z}_{01,s}^\inc =0$, so we take $q=\bar{q}$. The classical effective Hamiltonian~\eqref{eq:hamiltonian} contains terms $\propto \ln \gamma_{vv'} (1-\sigma_v \sigma_{v'}\bar{\sigma}_{v} \bar{\sigma}_{v'})$, which implies that spin configurations with $\sigma_v \sigma_{v'}\bar{\sigma}_{v} \bar{\sigma}_{v'}=-1$ come with an infinite energy cost when $\gamma_{vv'} \to 0$.
We can thus take $\sigma_v = \bar{\sigma}_v$, which recovers the one-spin-species RBIM for incoherent errors~\cite{Dennis:2002ds}.

\section{Quantum circuit}
\label{sec:quantum_circuit}

To evaluate the partition function [Eq.~\eqref{eq:partition_function}], we cannot resort to standards methods like Monte-Carlo sampling due to the complex coefficients in the Hamiltonian~\eqref{eq:hamiltonian}.
Instead, we express the partition function as $\mathcal{Z}_{q\bar{q},s}= \braket{\phi_0| \calM |\phi_0}$ with the transfer matrix
\begin{equation}
 \calM  = \hat{V}_{L}^{(q\bar{q},s)} \hat{H}' \hat{V}_{L-1}^{(q\bar{q},s)} \dots \hat{H} \hat{V}_{1}^{(q\bar{q},s)} ,
\end{equation}
where $\hat{V}_l^{(q\bar{q},s)} = \prod_m \hat{V}_{l,m}^{(q\bar{q},s)}$ denote vertical slices of the transfer matrix, with each $\hat{V}_{l,m}^{(q\bar{q},s)}$ corresponding to one physical qubit at site $j\to (l,m)$.
Because of the geometry of the lattice, we need to include intermediate layers $\hat{H} = \prod_{m=1}^{(M-1)/2} (1+\tau_{2m}^x) (1+\bar{\tau}_{2m}^x)$ and $\hat{H}' = \prod_{m=1}^{(M-1)/2} (1+\tau_{2m-1}^x)(1+ \bar{\tau}_{2m-1}^x)$ between neighboring $\hat{V}_l^{(q\bar{q},s)}$ and $\hat{V}_{l+1}^{(q\bar{q},s)}$ slices, where the Pauli $\tau_m^\mu$ and $\bar{\tau}_m^\mu$ ($\mu \in \{0,x,y,z \}$) act on $(M-1)$-site transfer matrix states $\ket{\{\sigma_m\bar{\sigma}_m\}}$, i.e., in a $4^{M-1}$-dimensional Hilbert space.
When identifying all physical qubits as vertical bonds of a 2D lattice that connect $S_v^X$ stabilizer, as denoted in Fig.~\ref{fig:layout}(a) by dashed lines, stabilizers on neighboring slices of this new 2D lattice are either identical or uncoupled, as shown in Fig.~\ref{fig:layout}(b).
In transfer matrix space, identical spins are horizontally connected via the identity matrix (which gives $\delta_{\sigma\sigma'}$), while uncoupled spins are projected, by $(1+\tau_m^x)(1+\bar{\tau}_m^x)$, onto $2 \ket{++}$, which by $\sqrt{2}\ket{+}=\ket{0}+\ket{1}$ sums over their configurations. 

Apart from boundary terms, the transfer matrices on bonds $(l,m)$ equal
\begin{align}
 \hat{V}_{l,m}^{(q\bar{q},s)}
 =& e^{ J_{l,m}^{(0)} + J_{l,m}^{(1)} \eta_{l,m}^{(q,s)} \tau_m^z \tau_{m+1}^z + J_{l,m}^{(2)} \bar{\eta}_{l,m}^{(\bar{q},s)} \bar{\tau}_m^z \bar{\tau}_{m+1}^z} \nonumber \\
  & \times e^{J_{l,m}^{(3)} \eta_{l,m}^{(q,s)}\bar{\eta}_{l,m}^{(\bar{q},s)} \tau_m^z \tau_{m+1}^z \bar{\tau}_m^z \bar{\tau}_{m+1}^z }
  \label{eq:Vmatrices}
\end{align}
and the boundary state is a product state
\begin{equation}
 \ket{\phi_0} = 2^{M-1} \bigotimes_{m=1}^{M-1} \ket{++}_m .
 \label{eq:boundary_state}
\end{equation}
We discuss the transfer matrix construction for a different square lattice geometry in the Appendix~\ref{sec:quantum_circuit_different}.

\subsection{Spontaneous symmetry breaking and entanglement in transfer matrix space}
\label{sec:entanglement}

The phases of QEC mirror distinct phases of the quantum circuit~\cite{Behrends:2024bs,behrends2024statistical,Bao2024}.
To explore the phases of the quantum circuit in more detail, we use it to define an effective Hamiltonian $\calH$ via the thermal density matrix $\calMdiag \calMdiag^\dagger = \exp (-L \calH)$ with inverse temperature $L$~\cite{Venn:2023fp}.
Since standard measures for QEC, e.g., the logical error rate defined via the minimum infidelity in Eq.~\eqref{eq:logical_error}, depend only on the diagonal elements $\calMdiag$, with identical bond configurations $\{ \eta_{l,m}^{(q,s)} \}$ and $\{ \bar{\eta}_{l,m}^{(\bar{q},s)} \}$, we focus on $\calMdiag$ for this discussion.

The effective Hamiltonian $\calH$ has an approximate $\mathbb{Z}_2 \times \mathbb{Z}_2$ symmetry defined by the operators $W = \prod_{m} \tau_m^x$ and $\bar{W} =\prod_m \bar{\tau}_m^x$: $W$ and $\bar{W}$ commute with the intermediate layers $\hat{H}$ and $\hat{H}'$ and with all bulk $\hat{V}_{l,m}^{(qq,s)}$, but not with the gates $\hat{V}_{l,1}^{(qq,s)}$ and $\hat{V}_{l,M}^{(qq,s)}$ at the top and bottom boundaries [in terms of Fig.~\ref{fig:layout}(b)].

As we now discuss, below the QEC threshold, one of $\calH$ is gapped~\cite{Venn:2023fp,Behrends:2024bs,behrends2024statistical}, and the long-time state of the circuit (i.e., the ground state of this $\calH$)  exhibits spontaneous symmetry breaking and an entanglement area law.
Well below the threshold, i.e., for $p \to 0$, the real part of $J_j^{(1)}$ and $J_j^{(2)}$ tend to infinity, hence the effective Hamiltonian is dominated by the $\propto \tau_m^z \tau_{m+1}^z$ ($\propto \bar{\tau}_m^z \bar{\tau}_{m+1}^z$) terms that contribute to $\hat{V}_{l,m}^{(qq,s)}$ [Eq.~\ref{eq:Vmatrices}].
Domain walls, as detected by $\tau_m^z \tau_{m+1}^z$ (and $\bar{\tau}_m^z \bar{\tau}_{m+1}^z$) are thus costly, signaling an ordered phase spontaneously breaking $\mathbb{Z}_2 \times \mathbb{Z}_2$ symmetry.
The boundary operators $\hat{V}_{l,1}^{(qq,s)}$ and $\hat{V}_{l,M}^{(qq,s)}$ effectively introduce terms $\propto \tau_m^z$ and $\propto \bar{\tau}_m^z$ with $m=1,M$ to $\calH$, i.e., they generate a boundary magnetic field~\cite{behrends2024statistical}.
Assuming that both boundary magnetic fields have the same sign for $\hat{\mathcal{H}}_{0,s}$ (the roles of $\mathcal{Z}_{00,s}$ and $\mathcal{Z}_{11,s}$ in the following are reversed when the sign is opposite), this picks one of the two spin-polarized (i.e., symmetry-breaking) states as the ground state for $\hat{\mathcal{H}}_{0,s}$.
The lowest-energy excitations have a pair of domain walls, thus energy gap $2/\xi$ (i.e., energy $1/\xi$ per domain wall).
Owing to this gap and the short-range correlations of the ordered phase, the ground state---i.e., the long-time state in transfer matrix space---satisfies an entanglement area law~\cite{Brandao:2013fz,Brandao:2015cn,Cho_PhysRevX.8.031009}.

We now relate the gap to QEC through the ratio of $P_{0,s} = \mathcal{Z}_{00,s}$ and $P_{1,s} = \mathcal{Z}_{11,s}$.
Introducing a logical $\logX$ effectively flips one of the boundary magnetic fields by flipping the bonds along $\logX$, e.g., by flipping the $\eta_{l,M}$ and $\bar{\eta}_{l,M}$ at the bottom for $\logX$ shown in Fig.~\ref{fig:layout}(a).
Hence, the lowest-energy states of $\hat{\mathcal{H}}_{1,s}$ must have one domain wall, and there must be approximately $M$ states with similar energies due to the $M$ choices for the position of the domain wall~\cite{behrends2024statistical}.
The energy of such a domain wall state is larger than the ground-state energy  of $\hat{\mathcal{H}}_{0,s}$, roughly by $1/\xi$.
The ratio of probabilities
\begin{equation}
 \frac{P_{1,s}}{P_{0,s}} = \frac{\mathcal{Z}_{11,s}}{\mathcal{Z}_{00,s}} \propto M \exp \left( -\frac{L}{2 \xi} \right)
\end{equation}
thus decreases exponentially to zero, at a rate set by the gap $2/\xi$ of $\hat{\mathcal{H}}_{0,s}$.
The logical error rate, therefore, also decreases exponentially to zero when $\hat{\mathcal{H}}_{0,s}$ is gapped, thus linking the gap to being below the QEC threshold.

The behavior of $\calH$ above threshold is non-universal and depends on the type of error. For incoherent errors, $\calH$ is gapped but disordered~\cite{Dennis:2002ds,Chubb:2021cn,Behrends:2024bs} (and hence the long-time state again satisfies an area law).
For coherent errors, $\calH$ is gapless~\cite{Venn:2023fp} (and hence displays a logarithmic law for the entanglement entropy), resulting in critical behavior of the logical error rate~\cite{behrends2024statistical}.
In Sec.~\ref{sec:FT}, we introduce a phenomenological theory, motivated by the anyon description of $\mathcal{Z}_{00,0}$, that captures these properties.

\subsection{Error string sampling}
\label{sec:error_strings}

All quantities of interest, e.g., the logical error rate, require a sum over all syndromes.
Since the number of possible syndromes scales exponentially with system size, summing over all syndromes becomes unfeasible.
In practice we thus take averages $\sum_s P(s) [\dots ] = \langle \dots \rangle_s$ over syndromes by sampling $s$ from the probability distribution $P(s)$.
For incoherent errors, we can easily sample from $P(s)$ by choosing error strings with $\eta_j=-1$ with probability $p_j$ and $\eta_j=1$ otherwise.
However, for any nonzero coherent contribution $\gamma_j$, nontrivial correlations between different $j$ start forming~\cite{Bravyi:2018ea,behrends2024statistical}.
Hence, we cannot sample directly from $P(s)$.
Here we show how to sample distributions $\{\eta\}$ based on the quantum circuit $\hat{\mathcal{M}}$, building on our approach for fully coherent errors~\cite{behrends2024statistical}.
Since we are interested in probabilities of error strings, we always set $\eta_j = \bar{\eta}_j$ is the following discussion.

The algorithm samples configurations $\{ \eta\}$. Each configuration corresponds to an error string that equals, up to stabilizer products, a string $C_s \logX^q$ with some $s$ and $q$.
Thus, the $\{\eta\}$ configurations span all combinations of $s$ and $q$ and we can directly sample configurations $\{\eta\}$ according to their probabilities $P(\{\eta\}) = \braket{\phi_0| \hat{\mathcal{M}}_{qq,s} |\phi_0}$ (where the corresponding $s$ and $q$ is set by $\{\eta\}$).
To simplify the notation for sampling from $P(\{\eta\})$, we label all sites by $j \in \{1,\dots N\}$ instead of their horizontal and vertical positions, and label the $\hat{V}_{l,m}^{(qq,s)}$ by $\hat{T}^{(\eta_j)}_j$, with $\eta_j \in \{-1,1\}$.
Using this notation, the quantum circuit is
\begin{equation}\label{eq:Tnotation}
 \hat{\mathcal{M}} = \hat{T}^{(\eta_N)}_N \dots \hat{T}^{(\eta_{M+1})}_{M+1} \hat{H} \hat{T}^{(\eta_M)}_M \dots \hat{T}^{(\eta_1)}_1,
\end{equation}
with intermediate gates $\hat{H}$ and $\hat{H}'$  after every $M^{\text{th}}$ site.
The probability of one error string thus equals
\begin{equation}
 P(\{\eta\}) = \braket{\phi_0| \hat{T}^{(\eta_N)}_N \dots \hat{T}_1^{\eta_1} | \phi_0} ,
 \label{eq:full_probability}
\end{equation}
which is, apart from the relabeling of the gates $\hat{V}^{(qq,s)}_{l,m} \to \hat{T}^{(\eta_j)}_j$, identical to the contracted quantum circuit shown in Fig.~\ref{fig:layout}(b).

Now consider the scenario when all $\eta_{j<N}$ are known and only $\eta_N$ needs to be determined.
Then, we can sample $\eta_N$ based on its conditional probability $P_N (\eta_N|\eta_{N-1}\dots \eta_1)$, which equals $P(\{\eta\})$ from Eq.~\eqref{eq:full_probability} divided by the marginal distribution
\begin{align}
 P_{N-1} (\eta_{N-1} \dots \eta_1 ) &= \sum_{\eta_N=\pm 1} \braket{\phi_0| \hat{T}^{(\eta_N)}_N | \phi_{N-1}^{\{ \eta\} } } \nonumber \\
 &= \braket{\omega_N|\phi_{N-1}^{\{\eta\}}},
 \label{eq:marginal}
\end{align}
where $\phi_{N-1}^{\{\eta\}} = \hat{T}^{(\eta_{N-1})}_{N-1} \dots \hat{T}_1^{(\eta_1)} \ket{\phi_0}$ is the initial state evolved by the first $N-1$ qubit gates and intermediate gates $\hat{H}$ and $\hat{H}'$.
The sum over both $\eta_N=\pm1$ defines the state $\ket{\omega_N} = \sum_{\eta_N} [\hat{T}^{(\eta_N)}_N]^\dagger \ket{\phi_0}$.
The conditional probability $P_{N-1} (\eta_{N-1}|\eta_{N-2}\dots \eta_1)$ equals the marginal probability of the first $N-1$ qubits [Eq.~\eqref{eq:marginal}] divided by the marginal probability of the first $N-2$ qubit gates.
For the $j^\text{th}$ qubit, the conditional probability equals $P_j (\eta_j|\eta_{j-1}\dots \eta_1) =  P_{j} (\eta_{j}\dots \eta_1 )/P_{j-1} (\eta_{j-1}\dots \eta_1 )$ but in practice the normalization by $P_{j-1}$ is not necessary since it is independent of $\eta_j$; hence we use
\begin{equation}
 P_j (\eta_j|\eta_{j-1} \dots \eta_1) \propto \braket{\omega_{j+1} | \hat{T}_j^{(\eta_j)} | \phi_{j-1}^{\{\eta\}}} .
 \label{eq:conditional}
\end{equation}
We show a quantum-circuit representation of this expression in Fig.~\ref{fig:layout}(c).

The expression for the $j^\mathrm{th}$ marginal probability contains the state (including horizontal $\hat{H}$ and $\hat{H'}$ gates when $j\le N-M+1$)
\begin{equation}
 \ket{\omega_{j+1}} = \sum_{\eta_{j+1}\dots \eta_N} [\hat{T}^{(\eta_{j+1})}_{j+1}]^\dagger \dots [\hat{T}^{(\eta_N)}_N]^\dagger \ket{\phi_0} .
 \label{eq:omega_sampling}
\end{equation}
We now express $\ket{\omega_j}$ analytically.
Starting with $\ket{\phi_0}$ from Eq.~\eqref{eq:boundary_state}, we first compute $\ket{\omega_N}$.
Choosing $\hat{T}_N^{(\eta_N)}$ such that it acts on the qubit at coordinates $L,M$, taking the sum of both $\hat{T}_N^{(\pm 1)}$ effectively changes the $(M-1)^\mathrm{th}$ site in the transfer matrix state from $2 \ket{++}_{M-1}$ to the Bell state $(\ket{00}+\ket{11})_{M-1}$, which implies
\begin{equation}
 \ket{\omega_N} = 2^{M-2} \ket{++}_1 \otimes \ket{++}_2 \dots (\ket{00}+\ket{11})_{M-1}.
\end{equation}
Next, $\hat{T}_{N-1}^{(\eta_{N-1})}$ at $(L,M-1)$ effectively changes $2 \ket{++}_{M-2} \to (\ket{00}+\ket{11})_{M-2}$ (cf.\ Fig.~\ref{fig:layout}(c) for an example).
This pattern continues until
\begin{equation}
 \ket{\omega_{N-M+2}} = \bigotimes_{m=1}^{M-1} (\ket{00}+\ket{11})_{m}.
 \label{eq:only_Bell_pairs}
\end{equation}
To continue to the next site $j=N-M$, we (after summing over the gates $\hat{T}_{N-M+1}^{(\eta_{N-M+1})}$ that do not change $\ket{\omega_{N-M+2}}$) need to apply the layer $\hat{H'}$; cf.\ Fig~\ref{fig:layout}(b).
The layer $\hat{H'}$ projects the state for all even $m$ onto $\ket{++}$, such that
\begin{equation}
 \ket{\omega_{N-M+1}} = 2^{\frac{M-1}{2}} \bigotimes_{m=1}^{\frac{M-1}{2}} (\ket{00}+\ket{11})_{2m-1} \otimes \ket{++}_{2m} .
\end{equation}
When the sum of $T_{j}^{(\eta_j=\pm 1)}$ acts on this state (with $j< N-M$), the even sites that $\hat{T}_j^{(\eta_j)}$ acts on $j$ are again changed to a Bell state, i.e., $2 \ket{++}_{2m} \to (\ket{00}+\ket{11})_{2m}$ until we arrive at the product state of Bell pairs, $\ket{\omega_{N-2 M+2}}= \ket{\omega_{N-M+2}}$, Eq.~\eqref{eq:only_Bell_pairs}.
Finally, the layer $\hat{H}$ effectively resets the other half of the gates to $\ket{++}$, such that
\begin{equation}
 \ket{\omega_{N-2M+1}} = 2^{\frac{M-1}{2}} \bigotimes_{m=1}^{\frac{M-1}{2}} \ket{++}_{2m-1} \otimes (\ket{00}+\ket{11})_{2m} .
\end{equation}

\begin{figure*}
\includegraphics[scale=1]{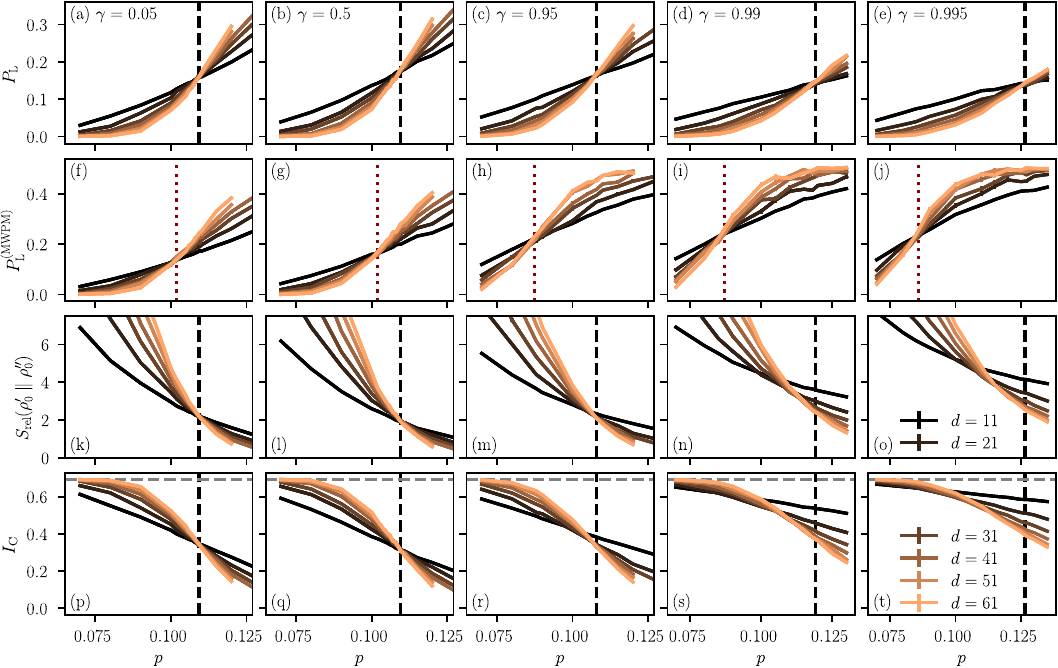}
\caption{(a)--(e) Logical error rate, (f)--(j) MWPM error rate, (k)--(o) quantum relative entropy, and (p)--(t) coherent information as a function of $p$ for fixed $\gamma$, which increases from left to right.
The colors denote different system sizes.
Results are averaged over 1000 to 10000 syndromes, and the error bars show the standard error of the mean.
The black dashed line shows the approximate maximum likelihood threshold, the red dotted line the MWPM threshold, and the gray dashed line in panels (p)--(t) the perfect-recoverability limit $I_\text{C} = \ln 2$ as a guide for the eyes.}
\label{fig:logical_errors}
\end{figure*}

Thanks to its area-law entanglement entropy away from the threshold, we can efficiently represent the evolved state $\ket{\phi_{j}^{\{\eta\}}}$ as a matrix-product state~\cite{Hauschild:2018bp,Cirac:2021gx}.
Using the truncated MPS and the product-state $\ket{\omega_{j+1}}$, we loop through the system and each site $j$ compute the conditional probabilities $P_j (\eta_j|\eta_{j-1} \dots \eta_1)$ [Eq.~\eqref{eq:conditional}, and sample $\eta_j$ using a suitable coin toss.

\section{Numerics: thresholds and information-theoretic measures}
\label{sec:numerics}

We now map out the phase diagram [Fig.~\ref{fig:overview}(a) and (b)] and compute error thresholds.
To this end, we numerically simulate the quantum circuit described in Sec.~\ref{sec:quantum_circuit} using MPS~\cite{Hauschild:2018bp,Cirac:2021gx}.
We compute averages over syndromes by drawing syndromes $s$ from the distribution $P(s)$ using the algorithm from Sec.~\ref{sec:error_strings}, which enables us to compute the logical error rate, quantum relative entropy, and coherent information.
We use the syndromes $s$ to additionally compute the minimum weight perfect matching (MWPM) threshold~\cite{pymatchingv1,pymatchingv2}.
For all results, we use the same uniform $p_j \to p$ and $\gamma_j \to \gamma$ and take a square geometry with code distance $d=L=M$.

The phase diagram is based on the data shown in Fig.~\ref{fig:logical_errors}.
For five different values $\gamma$, we show the logical error rate, the MWPM error rate, quantum relative entropy, and coherent information as a function of $p$ for different code distances $d$.

Fig.~\ref{fig:logical_errors}(a)--(e) shows the logical error rate $P_\text{L}$ as a function of $p$.
For $p<p_\thr$, i.e., below a maximum-likelihood threshold, $P_\text{L}$  decreases exponentially with code distance $L$, and above $p_\thr$, it increases with $L$ to to $P_\text{L} \to 1/2$.
We estimate 
$p_\thr =\{0.109(1), 0.109(1), 0.108(2), 0.119(2), 0.127(3)\}$
for $\gamma = \{0.05, 0.5, 0.95, 0.99, 0.995\}$, respectively.

Fig.~\ref{fig:logical_errors}(f)--(j) shows the MWPM error rate as a function of $p$, which we computed using PyMatching~\cite{pymatchingv1,pymatchingv2}.
For each syndrome $s$ sampled according to $P(s)$, PyMatching chooses a correction operation equivalent for the error string $C_s X^{q^\MWPM}$ (with $q^\MWPM \in \{0,1\}$), whose probability is $P^\MWPM (s) = \mathcal{Z}_{q^\MWPM,q^\MWPM,s}$.
The MWPM error rate is averaged over the syndromes and given by $P_\text{L}^\MWPM = \langle 1 - P^\MWPM (s)/P(s) \rangle_s$.
For $p<p_\thr^\MWPM$, it decreases exponentially to zero with code distance; for $p>p_\thr^\MWPM$, it increases up to $P_\text{L}^\MWPM \to 1/2$.
From the data in Fig.~\ref{fig:logical_errors}(f)--(j), we estimate
$p_\thr^\MWPM =\{0.102(3),0.102(3),0.088(3),0.087(3),0.086(3) \} $
for $\gamma = \{0.05, 0.5, 0.95, 0.99, 0.995\}$, respectively.
Unlike $p_\thr$, we find that $p_\thr^\MWPM$ decreases upon increasing $\gamma$ towards the coherent limit.

Fig.~\ref{fig:logical_errors} also shows the quantum relative entropy $S_\mathrm{rel} (\rho_0' \parallel \rho_0'')$
[panels (k)--(o)] and coherent information $I_\text{C}$ [panels (p)--(t)].
For all simulations we used the layout in Fig.~\ref{fig:layout}(a), hence $\Re [\mathcal{Z}_{01,s}]=0$.
For small $\gamma \lesssim 0.95$, both $S_\mathrm{rel} (\rho_0' \parallel \rho_0'')$ and $I_\text{C}$ signal the transition from the QEC regime to a noncorrecting phase well, even for small code distances. Similar behavior for $I_\text{C}$ was observed in Ref.~\onlinecite{Colmenarez:2024iz} for fully incoherent errors. 
For $\gamma \gtrsim 0.95$ close to the coherent limit, however, both measures become less reliable for small code distance:
For $p$ below threshold, both $S_\mathrm{rel} (\rho_0' \parallel \rho_0'')$ and $I_\text{C}$ decrease for small $d$ with code distance before they increase again [cf.\ Fig.~\ref{fig:overview}(c)], thus displaying the expected subthreshold increase only for sufficiently large $d$.

This behavior can be explained by considering $|\gamma^{(s)}_\text{L}|=|\mathcal{Z}_{10,s}|/\sqrt{\mathcal{Z}_{00,s} \mathcal{Z}_{11,s}}$, the measure,  from Secs.~\ref{sec:error_correction} and \ref{sec:information}, of logical channel coherence.
As shown in Fig.~\ref{fig:overview}(d), $\gamma_\text{L} = \langle|\gamma^{(s)}_\text{L}|\rangle_s$ decreases exponentially to zero with $d$, which holds for all $\gamma<1$ that we considered.
The logical noise thus becomes increasingly incoherent with $d$.
Hence, both quantum relative entropy and coherent information of this channel tend towards the incoherent limit, where they become able to signal the QEC threshold.
In general, both measures can thus be used to distinguish the phases of QEC only for sufficiently large code distances, and only if the error channel has some incoherent component.

\begin{figure*}
\includegraphics[scale=1]{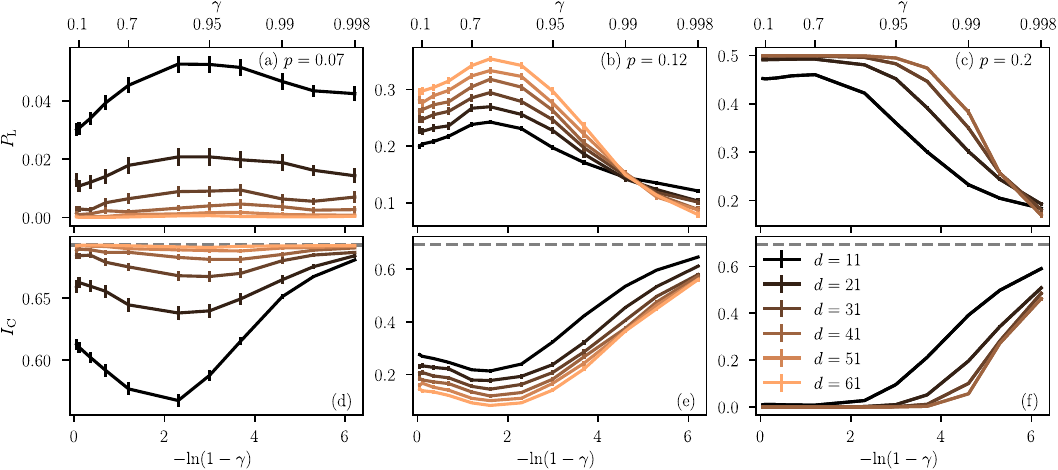}
\caption{(a)--(c) Logical error rate and (d)--(f) coherent information as a function of $-\log(1-\gamma)$ for fixed $p$, which increases from left to right.
Note that we have used an unusual scaling of the $x$-axis with $\gamma$ to highlight features visible for large $\gamma$ close to the coherent limit $\gamma = 1$. ($\gamma$ itself is shown on the second $x$ axis on top of the panel.)
The colors denote different system sizes.
Results are averaged over 1000 to 10000 syndromes, and the error bars show the standard error of the mean.
The gray dashed line in panels (d)--(f) shows $\ln 2$ as a guide for the eyes.}
\label{fig:logical_error_gamma}
\end{figure*}

To visualize the impact of noise coherence, in Fig.~\ref{fig:logical_error_gamma} we show the logical error rate and coherent information as a function of $\gamma$ for three fixed $p$: $p = 0.07$ (below $p_\thr$), $p=0.12$ (around $p_\thr$) and $p=0.2$ (above $p_\thr$).
We plot both quantities as a function of $-\log(1-\gamma)$ to highlight features visible for large values of $\gamma$ close to the coherent limit.
The logical error rate starts with a slow increase with $\gamma$ before reaching a maximum around $\gamma \approx 0.9$, followed by a decrease.
At $p=0.12$, a transition from a non-correcting regime to a QEC regime occurs above $\gamma_\thr \gtrsim 0.99 $.
This transition is well-captured by the logical error rate, however, for the system sizes we can simulate numerically, it is not captured by the coherent information, which continues to decrease with system size.
At $p=0.2$ and $\gamma = 0.998$, the behavior of the logical error is inconclusive. More disorder realizations are necessary to resolve the behavior close to the coherent limit.
For all $p$, the coherent information slowly decreases with $\gamma$ before reaches a minimum around $\gamma \approx 0.9$, and then increases towards $\ln 2$. This is consistent with the observation that the coherent information for even-weight stabilizers and odd-weight logicals always goes to $\ln 2$ in the coherent limit, and thus cannot distinguish the quasi-long-range ordered~\cite{behrends2024statistical} above-threshold behavior from the error-correcting subthreshold regime.

\begin{figure*}
\includegraphics[scale=1]{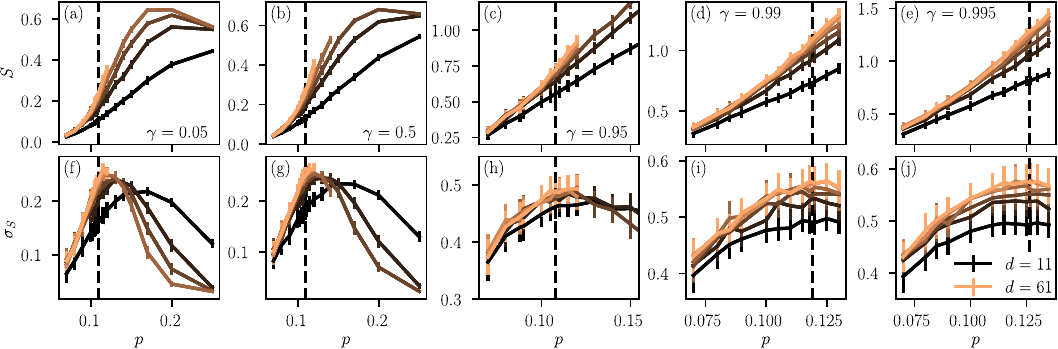}
\caption{(a)--(e) Entanglement entropy and (f)--(j) sample standard deviation of half-system entanglement entropy of the quantum circuit.
The colors denote different system sizes, and the black dashed line the maximum likelihood threshold.
Results are averaged over 1000 to 10000 syndromes, and the error bars show the standard error of the mean.}
\label{fig:entanglement}
\end{figure*}

In Fig.~\ref{fig:entanglement}, we show the half-system entanglement entropy $S$ of the final state $\ket{\phi_N^{\{\eta\}}}$ of the circuit, and its sample standard deviation $\sigma_S$ as a function of $p$.
Both phases of QEC above and below threshold are characterized by an area law.
At the transition, $\sigma_S$ exhibits a maximum that becomes sharper with increasing system size.
Due to the geometry we chose, in particular due to the intermediate transfer matrix layers $\hat{H}$ and $\hat{H}'$, the entanglement entropy for a bipartition of the final state $\ket{\phi_N^{\{\eta\}}}$ is not a smooth function of the position of the bipartition
---since the gates $\hat{H}$ and $\hat{H}'$ in the penultimate layer project half of the sites onto $\ket{++}$ and thereby reduce the entanglement.
We suspect the large error bars of $\sigma_S$ in Fig.~\ref{fig:entanglement}(h)-(j) are related to the sensitivity of the entanglement entropy on the position of the bipartition.

\begin{figure*}
\includegraphics[width=0.9\textwidth]{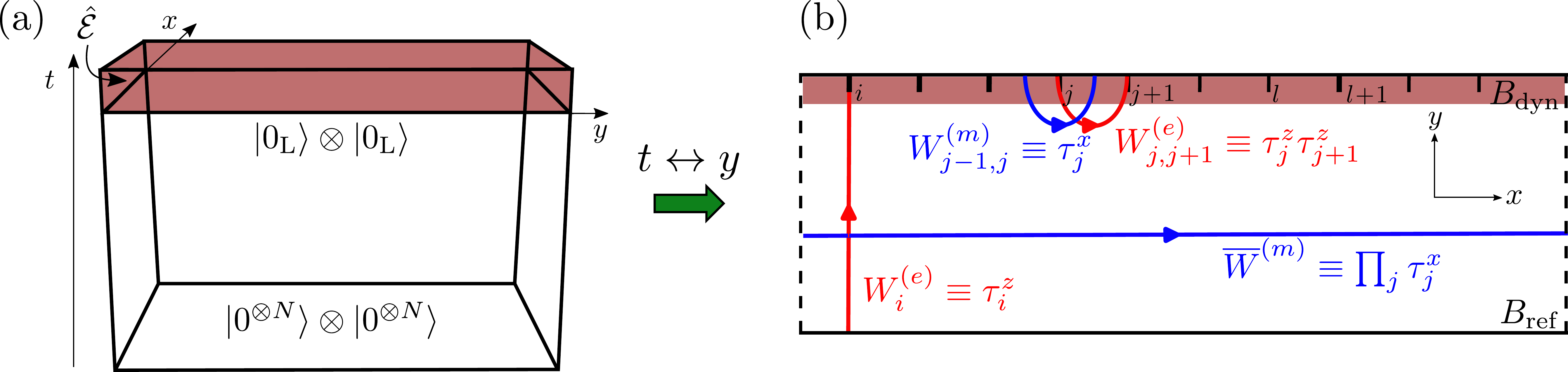}
\caption{Panel (a) shows $\ket{0_\text{L}}\times\ket{0_\text{L}}$ as arising from imaginary-time evolution with a surface code Hamiltonian (starting from $\ket{0^{\otimes N}}\otimes\ket{0^{\otimes N}}$ in the distant past), followed by the action of $\hat{\mathcal{E}}$ in the final time slice (shown red shaded). Panel (b) shows how, upon exchanging $t\leftrightarrow y$, this final time slice becomes a spatial boundary $B_\text{dyn}$; the initial $\ket{0^{\otimes N}}\otimes\ket{0^{\otimes N}}$ is now at the other boundary $B_\text{ref}$. Also shown is the string operator $W^{(e)}_{j,j+1}$ creating an $e$-pair; the corresponding spin-chain operator is $\tau_j^z \tau_{j+1}^z$. Similarly, we have $W^{(m)}_{j-1,j}\equiv \tau_j^x$ for $m$-pairs. We also show the $\mathbb{Z}_2$ symmetry $\overline{W}^{(m)}\equiv \prod_j \tau_j^x$ and the order parameter $W^{(e)}_i\equiv \tau_i^z$. The bra sector also has all the corresponding operators (not shown). 
}
\label{fig:SymTFT}
\end{figure*}

One important feature of the resulting phase diagram is the absence of an extended phase with logarithmic entanglement:
From our numerics, we cannot find indications of an such a critical regime away from $\gamma = 1$, i.e., fully coherent errors, where such a phase was previously observed~\cite{Venn:2023fp,Behrends:2024bs,behrends2024statistical}.
Such a critical phase displays logical error rate decaying as a power-law with code distance to a nonzero value~\cite{Venn:2023fp}, and by an approximately logarithmically increasing entanglement entropy~\cite{Behrends:2024bs,behrends2024statistical}.
Our numerics thus suggests that this extended critical above-threshold phase is special to coherent errors.

\section{Errorfield double, 1D Ising, and field theory phenomenology}
\label{sec:FT}

We now introduce a phenomenological model for the structure of
the phase diagram (Fig.~\ref{fig:overview}). Our approach is motivated
by Ref.~\onlinecite{Bao2023}, but uses a slightly different construction afforded by having only $X$ errors. To our knowledge, our phenomenological model gives a novel description already for the fully coherent limit, and will be seen to 
capture several key features we found in Sec.~\ref{sec:numerics}. 

We first introduce the setup and show how a holographic SymTFT correspondence~\cite{WenWei15,aasen2016topological, JiWen2020categorical,lichtman2020bulk, ChatterjeeWen23,FreedMooreTeleman22,TH23,Bhardwaj_PtII, Lootens_PRXQuantum.4.020357,Lootens_PRXQuantum.5.010338,Bhardwaj_PhysRevLett.133.161601,Fechisin_PhysRevX.15.011058,Barbar_PhysRevLett.134.151603,Sun_PRXQuantum.6.020333,Bottini_PhysRevLett.134.191602,Seifnashri_PhysRevLett.133.116601,Okada_PhysRevLett.133.191602,SymTFT_TC} allows us to link the problem to a non-Hermitian quantum Ising chain. This will already allow us to establish some essential features in the fully coherent and incoherent limits. We then move to the field theory to establish how incoherent-error perturbations impact the system.

We use the fidelity $F(\rho_{0},\mathcal{E}[\rho_{0}])=\mathcal{Z}_{00,0}$ to motivate our model, however our goal is not to capture the precise behavior of $\mathcal{Z}_{00,0}$, but to deduce structural ingredients using which to describe the phenomenology of the typical $\mathcal{Z}_{q\bar{q},s}$. In particular, the typical $\mathcal{Z}_{00,s}$ will be qualitatively the same as $\mathcal{Z}_{00,0}$ deep in the QEC phase.

Using  $\ket{0_{\text{L}}}=2^{-N_{v}/2}\prod_{v}(\mathbb{1}+S_{v}^{X})\ket{0^{\otimes N}}$,
where $N_{v}$ is the number of $S_{v}^{X}$
operators, 
\begin{align}
\mathcal{Z}_{00,0}=2^{N_{v}}\mathcal{Z}_\text{F}, & &
\mathcal{Z}_\text{F}=\bra{0^{\otimes N}}\mathcal{E}[\rho_{0}]\ket{0^{\otimes N}}.\label{eq:FZF}
\end{align}
Deep in the error-correcting phase $\mathcal{Z}_\text{F}$ is near maximal,
while it is suppressed in a non-correcting phase.

Using the mapping $\bra{\psi}A\to A^T\ket{\psi^*}$ for any operator $A$, we vectorize the density matrix on a space with ket and bra sectors~\cite{Baranger_PhysRev.111.494,bengtsson20006geometry,Gilchrist_vec}; on this space $\mathcal{E}$ acts as 
\begin{multline}
\hat{\mathcal{E}}=\prod_j\left[(1-p)\openone+i\gamma\sqrt{(1-p)p}(X_j-\bar{X}_j)+pX_j\bar{X}_j\right] \\ =\prod_{j}\exp\left[\kappa_{0}+\kappa_{1}X_{j}+\kappa_{2}\bar{X}_{j}+\kappa_{3}X_{j}\bar{X}_{j}\right],\label{eq:hatE}
\end{multline}
where $X_j$ act in ket and $\bar{X}_j$ in bra sectors, and
\begin{align}
\kappa_{0}=-\kappa_3&=\frac{1}{4}\log\left[1-4(1-\gamma^{2})p(1-p)\right],\label{eq:kappa0} \\
\kappa_{1}=-\kappa_2&=\frac{i}{2}\arctan\left(\frac{2\gamma\sqrt{p(1-p)}}{1-2p}\right).\label{eq:kappa1}
\end{align}
Note that $\kappa_{3}>0$ and $\kappa_1=-\kappa_2$ are imaginary. With vectorization we have
\begin{equation}
\mathcal{Z}_\text{F}=\langle\!\bra{0^{\otimes N}}\hat{\mathcal{E}}\ket{0_{\text{L}}\rangle\!},\label{eq:ZF}
\end{equation}
where $\ket{\psi}\!\rangle=\ket{\psi}\otimes\ket{\psi^*}$ is the product of bra and ket factors. (Here $\ket{\psi}\in\{\ket{0_\text{L}}, \ket{0^{\otimes N}}\}$.)

The state $\ket{0_{\text{L}}\rangle\!}$ is two copies of the surface code and $\hat{\mathcal{E}}\ket{0_{\text{L}}\rangle\!}$ is the ``errorfield double''~\cite{Bao2023}. 
In terms of surface-code anyons $e$ and $m$, at flipped
$S_{v}^{X}$ and $S_{p}^{Z}$, respectively, $X_{j}$ creates a pair
of  $m$ and $\bar{X}_{j}$ a pair of $\bar{m}$ in the first and second copy, respectively. The state $\ket{0^{N}\rangle\!}$ is invariant under $e$ pair (and $\bar{e}$ pair) creation by $Z_j$ (and $\bar{Z}_j$) hence condenses $e$  and $\bar{e}$ and confines $m$ and 
$\bar{m}$~\cite{baisCondensateinduced2009,kongAnyon2014,burnellAnyon2018,kesselringAnyon2024}.

\subsection{From 2D surface code to 1D Ising via SymTFT}
\label{sec:1DIsing}

As illustrated in Fig.~\ref{fig:SymTFT}(a), one can view $\ket{0_{\text{L}}\rangle\!}$ as arising from imaginary
time evolution with a surface code Hamiltonian followed by 
$\hat{\mathcal{E}}$ in the final time-slice. Upon a spacetime rotation, i.e., exchanging a spatial and the imaginary time direction~\cite{Nayak_RevModPhys.80.1083,Bao2023}, cf.~Fig.~\ref{fig:SymTFT}(b), this time-slice can also be viewed as as a spatial boundary, henceforth denoted by $B_\text{dyn}$. As the nontrivial physics is at this boundary, we get a (1+1)D theory.
To qualitatively describe $\mathcal{Z}_\text{F}$, this theory should
condense $e$ and $\bar{e}$ in the QEC phase and see errors competing with this by creating $m$ (and $\bar{m}$) anyon pairs.

As we now explain, this boundary physics is an interacting quantum Ising chain in an imaginary transverse field, matching the Ising nature and the types of gates in our (1+1)D quantum circuits in Sec.~\ref{sec:quantum_circuit} and App.~\ref{sec:quantum_circuit_different}.

We take $m$ anyons of $B_\text{dyn}$ to live at the links of a 1D lattice, while $e$ anyons live on the sites of this lattice, see Fig.~\ref{fig:SymTFT}(b). This alternating assignment captures the pattern imposed by the surface code [cf.~Fig.~\ref{fig:layout}(a)].
A pair of $m$-anyons adjacent to $j\in B_\text{dyn}$ is created (or annihilated) by Pauli-$X$ string $W^{(m)}_{j-1,j}$ with endpoints at links $(j-1,j)$ and $(j,j+1)$ of $B_\text{dyn}$. [More precisely, $W^{(m)}_{j-1,j}$ can be any member of the equivalence class of such Pauli-$X$ strings under endpoint-preserving path deformations, i.e.,  multiplication by $S_{v}^{X}$ in the system of Fig.~\ref{fig:SymTFT}(b).]
Similarly, a pair of $e$ anyons, one at site $j$ and another at site $j+1$ of $B_\text{dyn}$ is created by Pauli-$Z$ string $W^{(e)}_{j,j+1}$ (or any of its deformations) with endpoints at those sites. 
We also have the corresponding operators for $\bar{m}$ and $\bar{e}$ pair creation in the bra sector. 

As a consequence of the fusion rules $e^2=m^2=\bm{1}$ (corresponding to Pauli strings squaring to the identity) and the mutual semionic nature of $e$ and $m$ anyons (corresponding to the anticommutation of Pauli $X$ and $Z$ strings sharing an odd number of qubits), the operators $W^{(m)}_{i-1,i}$  and $W^{(e)}_{j,j+1}$, regardless of their details apart from their endpoints, satisfy the same algebraic properties as Pauli operators $\tau_i^x$, $\tau_j^z \tau_{j+1}^z$~\cite{WenWei15,aasen2016topological, JiWen2020categorical}. We hence identify
\begin{equation}\label{eq:Ising_gen}
W^{(m)}_{j-1,j} \equiv \tau_j^x\quad
W^{(e)}_{j,j+1} \equiv \tau_j^z \tau_{j+1}^z,
\end{equation}
where $\tau^\alpha_j$ are Pauli operators for a spin chain at $B_\text{dyn}$.
We can also embed $\tau_j^{z}$ in this approach noting that under the spacetime rotation, the initial state from which we imaginary-time evolve resides at another boundary, $B_\text{ref}$, cf.~Fig.~\ref{fig:SymTFT}(b). 
Using $\ket{0_{\text{L}}}\propto\prod_{v}(\mathbb{1}+S_{v}^{X})\ket{0^{\otimes N}}$, we take this initial state to be $\ket{0^{\otimes N}}\!\rangle$. Hence, $B_\text{ref}$ condenses $e$. Using this, we can deform $W^{(e)}_{j,j+1}$ to $W^{(e)}_{j}W^{(e)}_{j+1}$,  where each $W^{(e)}_{i}$ connects $B_\text{ref}$ and $B_\text{dyn}$, cf.~Fig.~\ref{fig:SymTFT}(b). [Here we used that $e$-condensation allows us to absorb a path segment from $W^{(e)}_{j,j+1}$ at $B_\text{ref}$.] We thus take $W^{(e)}_{j}\equiv \tau^z_j$. Owing to the single intersection of the corresponding strings, this anticommutes with $W^{(m)}_{j-1,j} \equiv \tau_j^x$ as it should.
With analogous considerations, we also introduce $\bar{\tau}_j^\alpha$ by considering $\bar{m}$ and $\bar{e}$ anyons.

The $e$-condensing $B_\text{dyn}$ we have without errors gives a state invariant under  $W^{(e)}_{j,j+1}$; for the spin chain, this is $\ket{\psi}_\text{1D}$ with $\tau_j^{z}\tau_{j+1}^{z}\ket{\psi}_\text{1D}=\ket{\psi}_\text{1D}$: a ferromagnetic state. Similarly, we have the corresponding ferromagnetic state for the bra sector, from $\bar{e}$ condensation.

In terms of these ingredients, and motivated by Eq.~\eqref{eq:hatE}, we seek to  capture the essential features of $\mathcal{Z}_\text{F}$'s phenomenology by  studying the imaginary-time evolution $\exp(-Ht)$ for large $t>0$ for the spin chain, with Hamiltonian $H=\sum_j H_j$ where
\begin{equation}
 H_j= -J \tau_j^{z}\tau_{j+1}^z - \bar{J} \bar{\tau}_j^{z}\bar{\tau}_{j+1}^z + h_x \tau_j^x +\bar{h}_x \bar{\tau}_j^x + h_\text{int} \tau_j^x\bar{\tau}_j^x.
 \end{equation}
The Hamiltonian has a $\mathbb{Z}_2\times \mathbb{Z}_2$ symmetry, generated by operators $W^{(m)}=\prod_k \tau_k^x$ and $\smash{\overline{W}^{(m)}}=\prod_k \bar{\tau}_k^x$. This captures a key feature of the quantum circuit, cf.~Sec.~\ref{sec:entanglement} (here we focus on the bulk physics, thus ignoring boundaries where the symmetry may be broken). Upon requiring $H$ to continue to describe vectorized dynamics, but now for the spin-chain, the Hermiticity of the spin-chain density matrix imposes an antiunitary $\mathbb{Z}_{2}^{H}$ symmetry swapping ket and bra sectors~\cite{bengtsson20006geometry}. As a result, $\bar{J}=J^*$, $\bar{h}_x=h_x^*$, and $h_\text{int}$ is real. These are again features shared with the quantum circuit; this is most apparent for the geometry discussed in Appendix~\ref{sec:quantum_circuit_different}.

We now further specify the couplings using Eq.~\eqref{eq:hatE} and phenomenological considerations. In the absence of errors ($h_x = \bar{h}_x=h_\text{int}=0$), since we spacetime rotate a topological theory in Euclidean spacetime, we have imaginary-time evolution with a Hermitian $H$~\cite{Nayak_RevModPhys.80.1083}, hence $J=\bar{J}$ is real; to imprint the ferromagnetic order we take $J>0$. For pure bit-flip errors, we have $\kappa_{1,2}=0$ in Eq.~\eqref{eq:hatE} and $\kappa_3$ drives the system towards correlated $m$-$\bar{m}$ condensation. We imprint this by taking $h_\text{int}<0$, to promote tendency towards $\tau_j^x \bar{\tau}_j^x\to 1$, corresponding to this condensation. Finally, motivated by the imaginary $\kappa_{1,2}$ with coherent errors, we take $h_x = i\mu$ with $\mu$ real [see also Eqs.~\eqref{eq:circkappaApp} for the quantum circuit of App.~\ref{sec:quantum_circuit_different}]. 

Overall, we thus take $H=\sum_j H_j$ with 
\begin{equation}\label{eq:HIsing}
H_j=-J(\tau_j^{z}\tau_{j+1}^z +\bar{\tau}_j^{z}\bar{\tau}_{j+1}^z)+i\mu (\tau_j^x - \bar{\tau}_j^x)+h_\text{int} \tau_j^x\bar{\tau}_j^x,
\end{equation}
with $J>0$, $h_\text{int}<0$, and $\mu$ real.  
This Hamiltonian captures key qualitative features both for the purely coherent and purely bit-flip limits. 

For pure bit-flip errors ($\mu=0$), $H$ commutes with $S_j=\tau_j^{z}\tau_{j+1}^z\bar{\tau}_j^{z}\bar{\tau}_{j+1}^z$ for all $j$ and hence is block diagonal in the corresponding sectors. For the long-time limit we are interested in, we focus on the $S_j= 1$ sector: the QEC phase ($J\gg |h_\text{int}|$) is smoothly connected to the $e$- and $\bar{e}$-condensing phase with $\tau_j^{z}\tau_{j+1}^z\to 1$, $\bar{\tau}_j^{z}\bar{\tau}_{j+1}^z\to 1$, while for the non-QEC phase ($J\ll |h_\text{int}|$), through $-(J^2/|h_\text{int}|)S_j$ terms arising in perturbation theory, the system also approaches the $S_j =  1$ subspace. [Working in this sector for bit-flips also captures a key feature of the quantum circuit; there $S_j\to 1$ arises as a constraint for $\gamma\to 0$, through the $J^{(3)}_j$ terms.] In the $S_j= 1$ subspace, we effectively have $\tau_j^{z}\tau_{j+1}^z=\bar{\tau}_j^{z}\bar{\tau}_{j+1}^z\equiv \sigma_j^{z}\sigma_{j+1}^z$ and $\tau_j^x\bar{\tau}_j^x\equiv \sigma_j^x$ in terms of Pauli operators $\sigma_j^\alpha$, hence $H$ becomes the familiar transverse-field Ising model. This has ferromagnetic and paramagnetic phases, both with area-law entanglement in the ground state~\cite{sachdev_2011,fradkin2013field,Cirac:2021gx}. This phase diagram captures the essence of the one arising from the quantum circuit for purely bit-flip errors~\cite{Bravyi:2014ja,Behrends:2024bs}. 

For purely coherent errors ($h_\text{int}=0$), the ket and bra sectors decouple and each has an Ising chain in an imaginary transverse field $\pm i\mu$. Such Ising chains were studied recently in the context of monitored quantum systems~\cite{Lee_PhysRevX.4.041001,Biella2021manybodyquantumzeno,Turkeshi_PhysRevB.103.224210,yan2024dissipativedynamicalphasetransition} where it was found that for $J\gg |\mu|$ (corresponding to the QEC phase) the system again exhibits an area law ferromagnetic phase, and for  $|\mu|\gg J$ (i.e., the non-QEC phase) this gives way to a logarithmic entanglement phase. This again captures essential features from the quantum circuit description of QEC~\cite{Venn:2023fp,Behrends:2024bs,behrends2024statistical}.

We next describe the system using phenomenological field theory. This theory will be a bosonized formulation of a model closely related to Eq.~\eqref{eq:HIsing}. This will allow us to use the renormalization group (RG) to capture the behavior away from the fully coherent and fully incoherent limits, by considering the physics arising when incoherent noise enters as a perturbation. 
For background on the field theory at the boundary of topological order and on bosonization, we refer the reader to Refs.~\cite{WenZee,wen1995topological,Haldane_PhysRevLett.74.2090,wen2004quantum,Giamarchi}.

\subsection{Field theory ingredients}

The boundary admits a description as a compact boson field theory~\cite{WenZee,wen1995topological,Haldane_PhysRevLett.74.2090,wen2004quantum} in terms of slowly varying fields $\phi_{e}$
and $\phi_{m}$ in the ket sector and $\bar{\phi}_{e,m}$ in the bra
sector. As operators, bra and ket fields commute and 
\begin{align}
[\phi_{m}(x),\phi_{e}(y)]&=i\pi\Theta(x-y), \label{eq:emcomm1}\\ 
\quad[\bar{\phi}_{m}(x),\bar{\phi}_{e}(y)]&=-i\pi\Theta(x-y),\label{eq:emcomm2}
\end{align}
where $\Theta$ is the Heaviside function with $\Theta(0)=1/2$. The antiunitary $\mathbb{Z}_{2}^{H}$ symmetry exchanges $\phi_{j}\leftrightarrow\bar{\phi}_{j}$, resulting in the
sign difference in Eqs.~\eqref{eq:emcomm1} and \eqref{eq:emcomm2}. This corresponds to ket and bra sectors forming conjugate topological orders, another consequence of the $\mathbb{Z}_{2}^{H}$ symmetry~\cite{Bao2023}.
The field $\psi_{e}(x)=\exp[i\phi_{e}(x))]$ creates an $e$
and $\psi_{m}(x)=\exp[i\phi_{m}(x)]$ an $m$ anyon
at position $x$; the creation of $\bar{e}$ and $\bar{m}$ works similarly via the $\bar{\phi}_j$ fields.

\subsection{Coherent error limit}
\label{subsec:cohFT}

We start with fully coherent errors where $\phi_j$ and $\bar{\phi}_j$   decouple;
we focus on the  $\phi_j$ sector. As local processes
can create  $e$ and $m$ anyons each in pairs, a generic boundary
Hamiltonian $H=\int dx\,h (x)$ has structure~\cite{WenZee,wen1995topological,Haldane_PhysRevLett.74.2090,wen2004quantum}  
\begin{multline}
h=\frac{1}{2\pi}\sum_{\alpha,\beta\in\{e,m\}}V_{\alpha\beta}(\partial_{x}\phi_{\alpha})(\partial_{x}\phi_{\beta})+ \\ +\sum_{n_{e},n_{m}\in\mathbb{Z}}C_{n_{e},n_{m}}\psi_{e}^{2n_{e}}\psi_{m}^{2n_{m}},\label{eq:hgeneric}
\end{multline}
with real positive definite $V_{\alpha\beta}$ in the kinetic energy term. 
By $\kappa_{1,2}$ being imaginary, we allow non-Hermitian
$C_{n_{e},n_{m}}$ terms, hence we omit the Hermitian conjugate. We next constrain $h$ through the structure
and symmetries of $\hat{\mathcal{E}}$ in Eq.~\eqref{eq:hatE} and the requirement of an $e$-condensing QEC phase and logarithmic entanglement above threshold.

To implement $e$ condensation, the $C_{n_{e},n_{m}}$ terms include $\cos\left[2(k_{e}x+\phi_{e})\right]$
with a real coefficient $\Delta$; we allow a nonzero momentum transfer $2k_e$ for $e$-pair creation. (Here we use that $\hat{\mathcal{E}}$ is lattice translation invariant; for the typical  $\mathcal{Z}_{00,s}$ this holds due the underlying syndrome average.) To ensure that this term survives the integration in $H=\int dx\,h$ we take $2k_{e}a=0$ ($\text{mod }2\pi$, with $a$ the lattice spacing). In the $e$-condensing phase this term dominates and hence the $e$ phase operator $\phi_{e}$ takes a definite value. This feature of the condensate suggests $e$ current $\sim\partial_x \phi_e$ and the ($k=0$ part of the) $e$ density to be $\sim\partial_x \phi_m$. This suggests $(\partial_{x}\phi_{e})(x)\to-(\partial_{x}\phi_{e})(-x)$ and $(\partial_{x}\phi_{m})(x)\to(\partial_{x}\phi_{m})(-x)$ under spatial reflections. By reflection symmetry (for the typical $\mathcal{Z}_{00,s}$ this holds by the underlying syndrome average), odd powers of $\partial_x \phi_e$ are absent from $h$. In particular $V_{em}=0$.

Of the other $C_{n_{e},n_{m}}$ terms, we focus on two-anyon processes; higher-orders are expected to be less relevant in the renormalization group (RG). This leaves the $m$ terms to consider. From the $\kappa_{1}$
term, the $m$-anyons enter through a Hermitian term times the imaginary unit,
\begin{equation}
 h_{m}= i\mu\partial_{x}\phi_{m}+i\mu'\cos\left[2(k_{m}x+\phi_{m})\right],
 \label{eq:mpert}
\end{equation}
with the first term from $\psi_{m}^{\dagger}\psi_{m}$ and
the second from $\psi_{m}^{\dagger2}+\psi_{m}^{2}$ and where $\mu,\mu'$ are real. Symmetries do
not constrain $k_{m}$ and the logarithmic entanglement
phase will require the cosine to be inoperative; since generically $\mu'\neq0$
we set $2k_{m}a\neq 0$ $\text{mod }2\pi$.  We therefore drop the $\mu'$ term (but comment
on its effect later). Hence
\begin{equation}
h=\frac{v}{2\pi}\left[g(\partial_{x}\phi_{e})^{2}+g^{-1}(\partial_{x}\phi_{m})^{2}\right]+\Delta\cos(2\phi_{e})+i\mu\partial_{x}\phi_{m},\label{eq:hcoherent}
\end{equation}
where $vg=V_{ee}$, $v/g=V_{mm}$, with Luttinger
parameter $g$ and velocity $v$. The bra sector has identical Hamiltonian $\bar{h}$ (with fields $\bar{\phi}_j$),
except for $\mu\to-\mu$ by
$\mathbb{Z}_{2}^{H}$ symmetry.

We now relate Eq.~\eqref{eq:hcoherent} to the 1D Ising picture. For this, we describe the Ising chain's ferromagnetic order starting from a ZY chain with $H^{(ZY)}_j = -J (\tau_j^{z}\tau_{j+1}^z + \tau_j^{y}\tau_{j+1}^y)$, with $J>0$, and explicitly breaking its $U(1)$ symmetry to $\mathbb{Z}_2$ (i.e., Ising) by reducing the $\tau_j^{y}\tau_{j+1}^y$ coupling through adding $J_y \tau_j^{y}\tau_{j+1}^y$, with $0<J_y\leq J$. [Hence, $J_y$ sets by how much $\tau_j^{z}\tau_{j+1}^z$ dominates over $\tau_j^{y}\tau_{j+1}^y$; while $J_y=J$ in the Ising limit (i.e., without $\tau_j^{y}\tau_{j+1}^y$ terms), for the field theory we take $0<J_y\ll J$; this already captures the requisite ferromagnetic order at low energies.] Upon Jordan-Wigner transformation, this gives fermions on a 1D chain, with nearest neighbor hopping amplitude $2J-J_y$ and pairing [i.e., $U(1)\to\mathbb{Z}_2$ breaking] amplitude $J_y$. At low energies the hopping terms yield left and right moving fermion branches at Fermi momenta $\pm k_\text{F}=\pm \pi/(2a)$. Upon bosonization (with $\varphi=2\phi_{m}$, $\theta=2\phi_{e}$ the standard bosonization fields~\cite{Giamarchi}), the hopping term gives the derivative terms in Eq.~\eqref{eq:hcoherent} with $g=1$ owing to the free-fermion nature of the theory. 

The pairing terms bosonize to $\sim J_y\cos\left[2(k_{e}x+\phi_{e})\right]$; they thus correspond to $e$-pair creation with $\Delta \sim J_y$. This is what imprints the Ising order (the dominance of $\tau_j^{z}\tau_{j+1}^z$ in our approach), matching the Ising chain's phenomenology. As pairing involves a left- and right-moving fermion, it has zero momentum transfer, $k_e=0$. In turn, $\tau_j^x$ corresponds to the fermion density; this has $k=0$ and $k=2k_\text{F}$ components that lead to the terms in Eq.~\eqref{eq:mpert} with $k_m=k_\text{F}=\pi/(2a)$ (thus indeed satisfying $2k_{m}a\neq 0$ $\text{mod }2\pi$).
The ZY-deformed formulation is particularly well suited for capturing the logarithmic phase that arises for the Ising model Eq.~\eqref{eq:HIsing} for $|\mu|\gg J$ since that also involves gapless modes with momenta $k=\pm \pi/(2a)$~\cite{Lee_PhysRevX.4.041001,Biella2021manybodyquantumzeno,Turkeshi_PhysRevB.103.224210,yan2024dissipativedynamicalphasetransition} and the ZY deformation supplies a low-energy theory precisely around these momenta. [It can also describe the standard Ising transition for real $h_x$ if we precede the ZY deformation by a unitary transformation, taking $\tau_j^x\to (-1)^j \tau_j^x$. This picks out the $ka=\pi$ component of the density thus now retaining $\cos(2\phi_{m})$ instead of $\partial_x \phi_m$. This gives a sine-Gordon formulation that was used to study Ising criticality~\cite{lecheminant2002criticality}.]

We now show how evolution with
$\exp(-Ht)$ for large $t>0$ leads to an area-law in the $e$-condensing phase ($\Delta\gg|\mu|$)
and logarithmic entanglement for $|\mu|\gg\Delta$. 
To this end, we set $g=1$, as befits the free fermions of the coherent-error RBIM, or the 1D Ising model, and fermionize $h$. This amounts to reversing the bosonization identities~\cite{Giamarchi} for $\varphi=2\phi_{m}$
and $\theta=2\phi_{e}$ to introduce left and right moving fermions
$\chi_{+}$, $\chi_{-}$, respectively. (These are the left- and right-movers for the ZY-deformed Ising chain, i.e., the slowly-varying parts of the Jordan-Wigner fermions before bosonization.) In terms of $\boldsymbol{\chi}_{\pm}=(\chi_{\pm},\chi_{\mp}^{\dagger})$
we have, up to
a constant,
\begin{align}
h=\frac{1}{2}\sum_{p\in\pm}\boldsymbol{\chi}_{p}^{\dagger}\mathcal{H}_{\text{BdG}}^{(p)}\boldsymbol{\chi}_{p}, &\ 
\mathcal{H}_{\text{BdG}}^{(\pm)}=(\pm iv\partial_{x}+i\mu)\Sigma_{3}\pm\Delta\Sigma_{1},
\label{eq:hfermionize}
\end{align}
where $\Sigma_{j}$ are the Pauli matrices in Nambu space and we absorbed
the short distance cutoff $a$ in $\Delta$. The spectrum of $\mathcal{H}_{\text{BdG}}^{(\pm)}$
is $\pm\varepsilon(k)=\pm\sqrt{(\pm vk+i\mu)^{2}+\Delta^{2}}$.
For small $k$, this is $\varepsilon(k)\approx\sqrt{\Delta^{2}-\mu^{2}}\pm ivk\mu/\sqrt{\Delta^{2}-\mu^{2}}$,
hence $\Re \varepsilon(k)$ is gapped for $\Delta^{2}>\mu^{2}$
but $\Re \varepsilon(k)\propto vk$ for $\mu^{2}>\Delta^{2}$.
Thus, for $\Delta^{2}>\mu^{2}$, the $\exp(-Ht)$ evolution leads to a state
akin to the ground state of a gapped system, suggesting an area law.
For $\mu^{2}>\Delta^{2}$, the system evolves to a state akin to a
CFT ground state, suggesting logarithmic
entanglement.
{[}From the $\mu'$ term with $k_{m}=0$,
$\mathcal{H}_{\text{BdG}}$ would include $i\mu'(\chi_{+}^{\dagger}\chi_{-}+\text{h.c.})$,
replacing $\Re \varepsilon(k)\propto vk$ with a $\sim\sqrt{(vk)^{2}-\mu'^{2}}$ contribution.{]}

The CFT becomes concrete for $\mu^{2}\gg\Delta^{2}$;
here $\mathcal{H}_{\text{BdG}}^{(\pm)}$ is dominated by the $\Sigma_{3}$
term and its eigenvectors approach $(1,0)^{T}$ and $(0,1)^{T}$ as
$|\mu|\to\infty$. Hence $\exp(-Ht)$ evolves to the same state as
for $\mu=\Delta=0$, the free-fermion (or in terms of $\phi_{e,m}$
the free-boson) CFT ground state. The reason this can remain gapless in the presence of the pairing perturbation $\Delta\chi_{+}\chi_{-}$ is that this now does not connect degenerate
states for $k=0$ (as it would in the Hermitian case, hence opening
a gap) but states that differ in energy by $2i\mu$. Hence $\Delta\chi_{+}\chi_{-}$
is a ``high-energy'' process that can enter the long-distance theory
only in second order in perturbation theory. 
The free-fermion CFT state implies, for $\mu^{2}\gg\Delta^{2}$,  entanglement entropy $S\propto 1/6 \log(\ell)$ for interval $[0,\ell]$ in a half-infinite system on $[0,\infty]$\cite{Calabrese:2009dx,calabrese2004entanglement}, in perfect agreement with features seen in numerics for the monitored Ising chain~\cite{Turkeshi_PhysRevB.103.224210}.

\subsection{Effect of incoherent noise}
\label{subsec:incohFT}

Armed with the picture for the coherent case, we can now turn to $\kappa_{3}\neq0$. The direct consequence is the appearance of correlated $m$-pair--$\bar{m}$-pair processes,
\begin{equation}
\sum_{n_{m}\in\mathbb{Z}}C_{n_{m}}\psi_{m}^{2n_{m}}\bar{\psi}_{m}^{2n_{m}}+C'_{n_{m}}\psi_{m}^{2n_{m}}\bar{\psi}_{m}^{-2n_{m}}+ \text{h.c.}
\label{eq:incohpert}
\end{equation}
We require
the terms to be Hermitian motivated by $\kappa_{3}$ being real [cf.~also the real $h_\text{int}$ in Eq.~\eqref{eq:HIsing}]. In terms
of $\phi_j$ and $\bar{\phi}_j$, focusing on $n_{m}\leq1$, we get
\begin{equation}
\lambda_{0}(\partial_{x}\phi_{m})(\partial_{x}\bar{\phi}_{m})+\lambda_{1}\cos[2(\phi_{m}+\bar{\phi}_{m})]+\lambda_{2}\cos[2(\phi_{m}-\bar{\phi}_{m})].\label{eq:incoh_m}
\end{equation}
By the $\mathbb{Z}_2^H$ symmetry, $\phi_m$ and $\bar{\phi}_m$ in the cosines are attached the same momentum $k_m$, that in Eq.~\eqref{eq:mpert}; this cancels in the $\lambda_2$ term and we took the Ising value $k_ma=\pi/2$ in the [later RG-irrelevant, Eq.~\eqref{eq:scalingdims}] $\lambda_1$ term.

An indirect consequence is that the coupling between bra and ket sectors
allows correlated $e$ terms to appear in perturbation theory (cf.~the $\tau_j^{z}\tau_{j+1}^z\bar{\tau}_j^{z}\bar{\tau}_{j+1}^z$ terms in Sec.~\ref{sec:1DIsing}). Focusing henceforth on the $|\mu| \gg \Delta$ regime, to
second order in $\Delta/\mu$ we get
\begin{equation}
\lambda_{3}\cos[2(\phi_{e}+\bar{\phi}_{e})]+\lambda_{4}\cos[2(\phi_{e}-\bar{\phi}_{e})],\label{eq:incoh_e}
\end{equation}
where we allowed for $\lambda_{3}\neq\lambda_{4}$ anticipating the
different RG flow of the two couplings.

We now make contact with spinful Luttinger liquids and introduce
$\varphi_{\uparrow}=2\phi_{m}$, $\theta_{\uparrow}=2\phi_{e}$, $\varphi_{\downarrow}=2\bar{\phi}_{m}$,
$\theta_{\downarrow}=-2\bar{\phi}_{e}$, which obey the standard commutation relation 
\begin{equation}
[\varphi_{\sigma}(x),\theta_{\sigma'}(y)]=4\pi \delta_{\sigma\sigma'}i\Theta(x-y),\quad \sigma,\sigma'\in\{\uparrow,\downarrow\}.\label{eq:comm_standard}
\end{equation}
Introducing ``spin'' $A_\mathsf{s}=(A_{\uparrow}-A_{\downarrow})/\sqrt{2}$ and ``charge'' $A_\mathsf{c}=(A_{\uparrow}+A_{\downarrow})/\sqrt{2}$ fields, for $A\in\{\varphi,\theta\}$, the problem decouples; the low-energy (i.e., long-time) physics is governed by $H=\int dx\,(h_\mathsf{c}+h_\mathsf{s})$ with
\begin{multline}
 h_\mathsf{s} = \frac{u_\mathsf{s}}{8\pi}\left[g_\mathsf{s} (\partial_{x}\theta_\mathsf{s} )^{2}+g_\mathsf{s}^{-1}(\partial_{x}\varphi_\mathsf{s})^{2}\right]+i\mu\partial_{x}\varphi_\mathsf{s}/\sqrt{2}+\\
 +\lambda_{2}\cos(\sqrt{2}\varphi_\mathsf{s})+\lambda_{3}\cos(\sqrt{2}\theta_\mathsf{s}),\label{eq:spincharge_spin}
\end{multline}
and
\begin{multline}
h_\mathsf{c} = \frac{u_\mathsf{c}}{8\pi}\left[g_\mathsf{c} (\partial_{x}\theta_\mathsf{c})^{2}+g_\mathsf{c}^{-1}(\partial_{x}\varphi_\mathsf{c})^{2}\right]+\\
+\lambda_{1}\cos(\sqrt{2}\varphi_\mathsf{c})+\lambda_{4}\cos(\sqrt{2}\theta_\mathsf{c}),\label{eq:spincharge_charge}
\end{multline}
where, assuming $g=1$ for $\kappa_{3}=0$, we have $u_\mathsf{c} g_\mathsf{c} = u_\mathsf{s} g_\mathsf{s} = v$ with $g_\mathsf{c}=(1+r\lambda_{0})^{-1/2}$ and $g_\mathsf{s}=(1-r\lambda_{0})^{-1/2}$, with $r>0$ constant.
Eqs.~\eqref{eq:spincharge_spin} and \eqref{eq:spincharge_charge} are reminiscent of the Hubbard model in an imaginary magnetic field $\mu$ and both backscattering and pairing terms in each of the charge and spin sectors.
Since $\kappa_{3}>0$, the corresponding coupling in $\exp(-Ht)$ is negative, hence $\lambda_{0}<0$ [cf.~$h_\text{int}<0$ in Eq.~\eqref{eq:HIsing}] and $g_\mathsf{c}>1$, $g_\mathsf{s}<1$, giving an attractive Hubbard model. 

Under RG, to first order in the couplings, the cosines flow via $d\lambda/d\ell=(2-\Delta_{\lambda})\lambda$ with the scale parameter $\ell$.
The scaling dimensions of the cosines are
\begin{equation}
 \Delta_{\lambda_{1}}=2g_\mathsf{c},\ \Delta_{\lambda_{2}}=2g_\mathsf{s},\ \Delta_{\lambda_{3}}=\frac{2}{g_\mathsf{s}},\ \Delta_{\lambda_{4}}=\frac{2}{g_\mathsf{c}}.
 \label{eq:scalingdims}
\end{equation}
Hence, $\lambda_{1}$ and $\lambda_{3}$ are irrelevant, while $\lambda_{2}$ and $\lambda_{4}$ are relevant under RG.
As the charge sector has a Hermitian Hamiltonian, this has the standard interpretation of a gap opening in that sector.
In the spin sector we must be more careful because of the large $\mu$ term.
To this end, we inspect terms, first by fermionizing at $g_\mathsf{s} = 1$.
In terms of spin-sector fermions $\chi_{\mathsf{s}+}$, $\chi_{\mathsf{s}-}$, the $\lambda_{2}$ term has $\chi_{\mathsf{s}-}^{\dagger}\chi_{\mathsf{s}-}^{\dagger}\chi_{\mathsf{s}+}\chi_{\mathsf{s}+}$ umklapp terms, while the $\lambda_{3}$ term has $\chi_{\mathsf{s}-}\chi_{\mathsf{s}-}\chi_{\mathsf{s}+}\chi_{\mathsf{s}+}$ correlated pairing.
As the latter connects states differing in energy by $4i\mu$, it is not a low-energy process, consistently with its RG irrelevance.
{[}A high-energy process from $\cos(2\phi_{e})\cos(2\bar{\phi}_{e})$ behind Eq.~\eqref{eq:incoh_e} is expected since each of the $\cos(2\phi_{e})$ and $\cos(2\bar{\phi}_{e})$ are sums of high-energy processes with opposite energies.{]}
The $\lambda_{2}$ term connects degenerate states.
To substantiate this opening a gap, we set $\lambda_{3}=0$ and refermionize at $g_\mathsf{s} =1/2$~\cite{LutherEmery_PhysRevLett.33.589,Giamarchi}: $\tilde{\varphi}=\sqrt{2}\varphi_\mathsf{s}$ and $\tilde{\theta}=\theta_\mathsf{s}/\sqrt{2}$ obey Eq.~\eqref{eq:comm_standard}, and $h_\mathsf{s}$ becomes Eq.~\eqref{eq:hcoherent} with $g=1$, $v=u_\mathsf{s}$, and with an $i\mu'$ term [from Eq.~\eqref{eq:mpert}] instead of the $\Delta$ term with $i\mu'\to\lambda_{2}$ and $k_m=0$.  
This maps to spinless fermions in an imaginary chemical potential $i\mu$ and backscattering $\lambda_{2}$, hence the spectrum is $\varepsilon(k)=\sqrt{(vk)^{2}+\lambda_{2}^{2}}+i\mu$ with gap $\lambda_{2}$ in the real part. 

The field theory thus shows that the logarithmic entanglement for purely coherent errors is fragile; it gives way to an area law (from the gapped phase) upon introducing incoherent noise even as perturbation. 
This is consistent with our numerical observation of the incoherent parts of the error being dominant. 
This is also suggestive of the phase boundary approaching, as in Fig.~\ref{fig:overview}(b), the incoherent $p$ value upon increasing  incoherent noise, but we leave analyzing this in our field theory to future work. 

As the field theory also captures, through $e$ (and $\bar{e}$) condensation, the correct behavior in the QEC phase, it provides a qualitative description of the phase diagram in Fig.~\ref{fig:overview}. 
This is further supported by the behavior of the correlation functions phenomenologically corresponding to the typical $\mathcal{Z}_{11,s}/\mathcal{Z}_{00,s}$; the insertion of the $q=\bar{q}$ logical operator $X_\text{L}\bar{X}_\text{L}$ corresponds to a correlation function with $\exp[i(\phi_m(x)\pm \bar{\phi}_m(x))]$ inserting an $m$-$\bar{m}$ pair at time $0$ which is then removed at time $L$. By creating kinks of energy $\sim \Delta$ in the $\phi_e$ and $\bar{\phi}_e$ fields, this decays exponentially with $L$ in the QEC phase; by the emergent CFT it decays as a power law in the coherent non-correcting phase; and, by including a $\varphi_\mathsf{s}$ exponential (with  $\varphi_\mathsf{s}$ locked to a cosine minimum by the $\lambda_2$ term and commuting with the $\lambda_4$-locked $\theta_\mathsf{c}$) it has a non-decaying part in the incoherent non-correcting phase. 

The theory also captures the phenomenology of the typical logical subspace coherence $\gamma_\text{L}^{(s)}=(\mathcal{Z}_{10,s}/\mathcal{Z}_{00,s})/\sqrt{\mathcal{Z}_{11,s}/\mathcal{Z}_{00,s}}$ which corresponds to the ratio of two correlation functions: one with a $X_\text{L}$ and the square root of one with a $X_\text{L}\bar{X}_\text{L}$ insertion. 
In the purely coherent limit, this correlator ratio has unit magnitude due to the decoupling of ket and bra sectors.
In the non-correcting phase with incoherent noise, the correlator for $\sqrt{\mathcal{Z}_{11,s}/\mathcal{Z}_{00,s}}$ is non-decaying as noted above, but $\mathcal{Z}_{10,s}/\mathcal{Z}_{00,s}$ corresponds to an $\exp(\pm i \phi_m)$ correlator featuring a kink in the gapping $\lambda_4$ term, leading to an exponential decay. In the QEC phase with incoherent noise, $e$ and $\bar{e}$ condensation allows the $\lambda_{3}$ and $\lambda_{4}$ terms to be present and give additional energy penalty to a kink only in the ket sector, corresponding to $\mathcal{Z}_{10,s}/\mathcal{Z}_{00,s}$, compared with half of the energy of the correlated kink-pair  corresponding to $\sqrt{\mathcal{Z}_{11,s}/\mathcal{Z}_{00,s}}$.
This leads to an exponential decay with $L$ for the typical $\gamma_\text{L}^{(s)}$, capturing the phenomenology shown in Fig.~\ref{fig:overview}(d).

\section{Conclusion and Outlook}
\label{sec:conclusion}

Considering the surface code under the most general single-qubit $X$-errors, we have investigated the ensemble $\rho_0'$ of error-corrupted post-syndrome-measurement states, Eq.~\eqref{eq:ensembleZ}, and studied associated information-theoretic quantities and their relation to QEC. 

To this end, we have developed a statistical mechanics mapping and showed that $\rho_0'$, and the information-theoretic quantities, as well as those for QEC (e.g., the logical error rate), can be expressed using partition functions of classical interacting random-bond Ising models (RBIMs) with complex couplings.

Our statistical-mechanics mapping gives access to the full ensemble of post-measurement states in geometries with even-weight $X$-stabilizers and odd-weight
$\logX$ (and to the full ensemble of suitable Bloch-sphere-averaged quantities in more general geometries).
This facilitates the computation of information-theoretic quantities that are nonlinear in $\rho_0'$, including the coherent information $I_\text{C}$ and quantum relative entropy $S_\text{rel}$.
We showed that $I_\text{C}$, for fully coherent errors, always goes to $\ln 2$; this conclusion also holds for generic $\text{SU}(2)$ single-qubit coherent errors. 
For generic $X$ errors close to the coherent limit, both  $I_\text{C}$ and $S_\text{rel}$ suffer from finite size effects. 

To explain this, we studied the coherence $\gamma_\text{L}$ of the logical noise. We found that it decreases with code distance for any nonzero incoherent noise component [Fig.~\ref{fig:overview}(d)]. This suggests that the suppression of finite-size effects for $I_\text{C}$ and $S_\text{rel}$ occurs when the code distance is sufficiently large for $\gamma_\text{L}$ to become sufficiently suppressed.

To establish the link to QEC, we have used our mapping to compute maximum likelihood thresholds and logical error rates. We found that a coherent contribution of the $X$-error channel does not significantly affect the maximum-likelihood threshold and logical error rate as long as the channel is not near the fully coherent limit, cf.\ Fig.~\ref{fig:logical_errors}.
The logical error rate as a function of $\gamma$ [characterizing the coherent noise component in Eq.~\eqref{eq:error_channel}] initially slightly increases before it decreases again close to the coherent limit, cf.\ Fig.~\ref{fig:logical_error_gamma}.
This decreasing logical error rate results in an increased maximum-likelihood threshold close to the coherent limit.

We have also developed a phenomenological errorfield double field theory that can account for some of these qualitative features, including the behavior of the logical error rate and the suppression of $\gamma_\text{L}$. This field theory gives analytical support to the observed fragility of the quasi-long-range ordered above-threshold behavior for fully coherent errors by showing that the incoherent noise component enters as a relevant perturbation to the critical above-threshold regime of the coherent case.

A key ingredient for our results is our syndrome sampling algorithm.
This is based on evaluating the quantum circuit (i.e., transfer matrix) evolution using matrix-product states, hence it samples syndromes according to their approximate probability.
This algorithm can be used with any decoder as we illustrated by computing thresholds for minimum weight perfect matching. This threshold---unlike the maximum-likelihood-threshold---decreases with increasing coherence $\gamma$.

Our work forms the basis for statistical-mechanics mappings and transfer matrix calculations of generic single-qubit error channels, as we sketch in the Appendix~\ref{sec:generic_errors}.
We expect that the large parameter space of such generic channels can be explored using similar techniques to those in this work since we expect the below-threshold regime to be generally characterized by an area-law entanglement entropy, and hence to be efficiently simulable using matrix product states.

Studying the impact of generic single-qubit error channels is not only relevant for a quantum memory, but also for state preparation~\cite{Zhu:2023be,Cheng2024} and teleportation~\cite{Eckstein:2024ev}, and decoder-encoder problems~\cite{Turkeshi:2024dd}. We expect that our work can be extended to these interesting areas.

\paragraph*{Note added.} During the completion of this manuscript, we became aware of a related independent work on sampling and maximum likelihood decoding in the surface code under local noise~\cite{Anand2025}.

\begin{acknowledgments}
We thank Florian Venn for collaboration on related topics, Max McGinley for input on information-theoretic measures, Claudio Castelnovo for providing additional computational resources, and Xhek Turkeshi for helpful discussions.
This work was supported by EPSRC Grant No.\ EP/V062654/1, a Leverhulme Early Career Fellowship and the Newton Trust of the University of Cambridge.
Our simulations used resources at the Cambridge Service for Data Driven Discovery operated by the University of Cambridge Research Computing Service (\href{https://www.csd3.cam.ac.uk}{www.csd3.cam.ac.uk}), provided by Dell EMC and Intel using EPSRC Tier-2 funding via grant EP/T022159/1, and STFC DiRAC funding (\href{https://www.dirac.ac.uk}{www.dirac.ac.uk}).

\end{acknowledgments}

\appendix

\section{Coefficient phase relation}
\label{sec:phase_relation}

Here we show that the off-diagonal coefficients $\mathcal{Z}_{01,s}$ of the channel $D_s[\rho]$ are purely imaginary in surface code geometries with only even-weight stabilizers and odd-weight logical operators, which generalizes previous argument for coherent errors~\cite{Bravyi:2018ea} to error channels of the form~\eqref{eq:error_channel}.

We start by expressing the post-error state in the form of Eq.~\eqref{eq:error_string_configurations}, i.e., as a sum over configurations $\{ \xsf \}$ and $\{ \xsfb \}$ of Pauli strings that act from the left and the right on the initial state $\rho$ via $\mathcal{P}(\{\xsf\}) \rho \mathcal{P}(\{\xsfb\})$.
Each coefficient $e^{\sum_j (J_j^{(0)} + J_j^{(1)} \xsf + J_j^{(2)} \xsfb + J_j^{(3)} \xsf \xsfb)}$ in this expansion must be either purely real or purely imaginary since the coefficients of the channel~\eqref{eq:error_channel} are also purely real ($p$ and $1-p$) and purely imaginary ($\pm i \gamma$).
More precisely, the coefficients are real when the difference between the number of $\xsf = -1$ and the number of $\xsfb = -1$ is even; conversely, the coefficients are imaginary when the difference is odd.

As argued in the main text, the only Pauli strings that contribute to $\calZ$ are those where $\mathcal{P}(\{\eta_j\} \mathcal{P} (\{ \xsf \} )$ and  $\mathcal{P}(\{\bar{\eta}_j\}) \mathcal{P} (\{ \xsfb \} )$ form products of $S_v^X$ stabilizers.
Now consider geometries with only even-weight $S_v^X$ stabilizers and odd-weight $\logX$: 
For coefficients $\mathcal{Z}_{01,s}$, the configurations $\{ \eta_j\}$ and $\{\bar{\eta}_j\}$ differ by a logical operator.
Hence, when $\{ \eta_j \}$ contains an even number of elements with $\eta_j = -1$, $\{ \bar{\eta}_j \}$ must contain an odd number of elements with $\bar{\eta}_j = -1$ and vice versa.
To form products of $S_v^X$ stabilizers, $\mathcal{P} (\{ \xsf \} )$ must be an even-weight operator for an even number of $\eta_j = -1$, and odd-weight for an odd number of $\eta_j = -1$; accordingly $\mathcal{P} (\{ \xsfb \} )$ must be odd-weight for an even number of $\eta_j = -1$, and even-weight for an odd number of $\eta_j = -1$.
Thus the difference of the number of $\xsf = -1$ and the number of $\xsfb = -1$ is odd for all strings contributing to $\mathcal{Z}_{01,s}$, and $\mathcal{Z}_{01,s}$ must be imaginary as it is the sum of imaginary terms.

\section{Quantum circuit in different geometry}
\label{sec:quantum_circuit_different}

\begin{figure}
\includegraphics[scale=1]{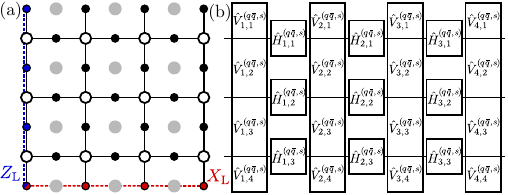}
\caption{(a) Surface code ($L=M=4$) with rough and smooth boundaries on a square lattice, with physical qubits (black discs) on its links, $S^X_v$ stabilizers on its vertices (open discs), and $S_p^Z$ on its faces (gray discs).
The logical $\logX$ and $\logZ$ are denoted by red and blue dashed lines, respectively.
(b) The quantum circuit consists of $\hat{V}_{l,m}^{(q\bar{q},s)}$ gates on vertical links and $\hat{H}_{l,m}^{(q,s)}$ on horizontal links.}
\label{fig:smooth_rough}
\end{figure}

In the main text, we chose a geometry with qubits on vertices of lattice, a ``rotated'' surface code~\cite{Bombin:2007ed,Horsman:2012ga}.
Here we show for completeness the construction of the quantum circuit for another geometry of surface code with ``rough'' and ``smooth'' boundaries~\cite{Bravyi1998quantum}.

In this geometry, qubits are located on the links of the lattice, and the $S_v^X$ and $S_p^Z$ stabilizers on the vertices and plaquettes of the lattice, respectively; cf.\ Fig.~\ref{fig:smooth_rough}(a).
The mapping to a complex RBIM Ising model is analog to the main text, but now Ising spins $\sigma_v,\bar{\sigma}_v$ are located on the vertices of the lattice.
To construct the quantum circuit, we distinguish horizontal and vertical couplings between these spins, analogous to the transfer matrix construction for the noninteracting RBIM with spins on vertices of a 2D square lattice~\cite{Schultz:1964fv}; cf.\ Fig.~\ref{fig:smooth_rough}(b).

The $\hat{V}_{l,m}$ matrices on the vertical bonds have the same form as given in Eq.~\eqref{eq:Vmatrices}.
The matrices on the horizontal bonds factorize for the $l^\mathrm{th}$ layer as $\hat{H}_l = \prod_m \hat{H}_{l,m}$ with the matrices
\begin{equation}
 \hat{H}_{l,m}^{(q\bar{q},s)} = \exp\left( \kappa_{l,m}^{(0)} + \kappa_{l,m}^{(1)} \tau_m^x  + \kappa_{l,m}^{(2)} \bar{\tau}_m^x + \kappa_{l,m}^{(3)} \tau_m^x \bar{\tau}_m^x \right) .
\end{equation}
The couplings are given by
\begin{align}
 \kappa_{l,m}^{(0)} =& \frac{1}{4} \ln \left[ 1-4 \left(1-\gamma^2\right) p (1-p) \right] + \frac{i\pi}{4}(\eta_{l,m}^{(h)}-\bar{\eta}_{l,m}^{(h)}), \nonumber \\
 \kappa_{l,m}^{(1)} =& \frac{i}{2} \arctan \left(\frac{2 \gamma \sqrt{p(1-p)} }{1-2 p}\right) + \frac{i\pi}{4} (1-\eta_{l,m}^{(h)}), \nonumber \\
 \kappa_{l,m}^{(2)} =&-\frac{i}{2} \arctan \left(\frac{2 \gamma \sqrt{p(1-p)} }{1-2 p}\right) - \frac{i\pi}{4} (1-\bar{\eta}_{l,m}^{(h)}), \nonumber \\
 \kappa_{l,m}^{(3)} =&-\frac{1}{4} \ln \left[ 1-4 \left(1-\gamma^2\right) p (1-p) \right],\label{eq:circkappaApp}
\end{align}
where we introduced the superscript $h$ to indicate horizontal bonds.
For $\eta=\bar{\eta}=1$, these couplings reduce to $\kappa_j$ in Eqs.~\eqref{eq:kappa0} and \eqref{eq:kappa1}.

\section{Generic error channel}
\label{sec:generic_errors}

The construction we introduced in this work can be extended to a more generic local error channel
\begin{align}
 \mathcal{E}_j^{\mathrm{(f)}} [\rho] 
 &= \sum_{\mu\nu} \varepsilon_{\mu\nu} \mathsf{O}_j^{\mu} \rho \mathsf{O}_j^{\nu} \\
 &=\sum_{\substack{ \xsf,\xsfb \\ \mathsf{z}_j,\bar{\mathsf{z}}_j}} e^{h(\{ \mathsf{x}_j, \mathsf{z}_j, \bar{\mathsf{x}}_j, \bar{\mathsf{z}}_j \})} X_j^{\frac{1-\xsf}{2}} Z_j^{\frac{1-\mathsf{z}_j}{2}} \rho Z_j^{\frac{1-\bar{\mathsf{z}}_j}{2}} X_j^{\frac{1-\xsfb}{2}} \nonumber
\end{align}
involving all on-site Pauli operators $\mathsf{O}_j^{\mu} = (\mathbb{1}_j,X_j,Y_j,Z_j)$ for $\mu=\{0,1,2,3\}$.
In this expansion, we express the 16 parameters~\footnote{The total number of parameters reduces to 12 when taking into account that $\mathcal{E}_j^\mathrm{(f)}$ is traceless~\cite{Nielsen:2010ga}.} $\varepsilon_{\mu\nu}$ as
\begin{align}
h =& J_j^{(0)}
 + J_j^{(\mathsf{x})} \mathsf{x}_j + J_j^{(\mathsf{y})} \mathsf{x}_j \mathsf{z}_j + J_j^{(\mathsf{z})} \mathsf{z}_j
 + J_j^{(\bar{\mathsf{x}})} \bar{\mathsf{x}}_j + J_j^{(\bar{\mathsf{y}})} \bar{\mathsf{x}}_j \bar{\mathsf{z}}_j  \nonumber \\
 &+ J_j^{(\bar{\mathsf{z}})} \bar{\mathsf{z}}_j+ J_j^{(\mathsf{x}\bar{\mathsf{x}})} \mathsf{x}_j \bar{\mathsf{x}}_j
 + J_j^{(\mathsf{y}\bar{\mathsf{y}})} \mathsf{x}_j \mathsf{z}_j \bar{\mathsf{x}}_j \bar{\mathsf{z}}_j 
 + J_j^{(\mathsf{z}\bar{\mathsf{z}})} \mathsf{z}_j \bar{\mathsf{z}}_j \nonumber \\
 &+ J_j^{(\mathsf{x}\bar{\mathsf{y}})} \mathsf{x}_j \bar{\mathsf{x}}_j \bar{\mathsf{z}}_j
  + J_j^{(\mathsf{y}\bar{\mathsf{x}})} \mathsf{x}_j \mathsf{z}_j \bar{\mathsf{x}}_j
  + J_j^{(\mathsf{y}\bar{\mathsf{z}})} \mathsf{x}_j \mathsf{z}_j \bar{\mathsf{z}}_j \nonumber \\
 &+ J_j^{(\mathsf{z}\bar{\mathsf{y}})} \mathsf{z}_j \bar{\mathsf{x}}_j \bar{\mathsf{z}}_j
  + J_j^{(\mathsf{z}\bar{\mathsf{x}})} \mathsf{z}_j \bar{\mathsf{x}}_j
  + J_j^{(\mathsf{x}\bar{\mathsf{z}})} \mathsf{x}_j \bar{\mathsf{z}}_j .
\end{align}
Analogous to the error channel discussed in the main text, the total error channel $\mathcal{E}^{\mathrm{(f)}}=\bigotimes_j \mathcal{E}_j^{\mathrm{(f)}}$ projected onto a stabilizer measurement can be expressed using complex coefficients $\mathcal{Z}_{\mu\nu,s}^\mathrm{(f)}$ that equal partition functions of interacting Ising models.
The Ising model will host two spin species on both direct and dual lattice. Interaction terms couple both spin species within each lattice, and direct and dual lattice via four-spin, eight-spin, and sixteen-spin interaction terms.

Since the error-correcting phase must be characterized by an area law in the transfer matrix space~\cite{Venn:2023fp,Behrends:2024bs}, we expect that efficient simulations using matrix product states are possible.
We also expect that the sampling of error strings is possible using a variant of the algorithm discussed in Sec.~\ref{sec:error_strings}.
This algorithm relies on an analytical expression of the state $\ket{\omega_j}$ [Eq.~\eqref{eq:omega_sampling}].
If the sum over all four~\cite{behrends2024statistical} configurations of $[\hat{T}^{(\eta_j)}_j]^\dagger \ket{\omega_{j+1}}$ can be expressed as a product state, the other steps of the algorithm follow through. We leave investigating this possibility to future works.

\bibliography{references}

\begin{thebibliography}{135}%
\makeatletter
\providecommand \@ifxundefined [1]{%
 \@ifx{#1\undefined}
}%
\providecommand \@ifnum [1]{%
 \ifnum #1\expandafter \@firstoftwo
 \else \expandafter \@secondoftwo
 \fi
}%
\providecommand \@ifx [1]{%
 \ifx #1\expandafter \@firstoftwo
 \else \expandafter \@secondoftwo
 \fi
}%
\providecommand \natexlab [1]{#1}%
\providecommand \enquote  [1]{``#1''}%
\providecommand \bibnamefont  [1]{#1}%
\providecommand \bibfnamefont [1]{#1}%
\providecommand \citenamefont [1]{#1}%
\providecommand \href@noop [0]{\@secondoftwo}%
\providecommand \href [0]{\begingroup \@sanitize@url \@href}%
\providecommand \@href[1]{\@@startlink{#1}\@@href}%
\providecommand \@@href[1]{\endgroup#1\@@endlink}%
\providecommand \@sanitize@url [0]{\catcode `\\12\catcode `\$12\catcode
  `\&12\catcode `\#12\catcode `\^12\catcode `\_12\catcode `\%12\relax}%
\providecommand \@@startlink[1]{}%
\providecommand \@@endlink[0]{}%
\providecommand \url  [0]{\begingroup\@sanitize@url \@url }%
\providecommand \@url [1]{\endgroup\@href {#1}{\urlprefix }}%
\providecommand \urlprefix  [0]{URL }%
\providecommand \Eprint [0]{\href }%
\providecommand \doibase [0]{https://doi.org/}%
\providecommand \selectlanguage [0]{\@gobble}%
\providecommand \bibinfo  [0]{\@secondoftwo}%
\providecommand \bibfield  [0]{\@secondoftwo}%
\providecommand \translation [1]{[#1]}%
\providecommand \BibitemOpen [0]{}%
\providecommand \bibitemStop [0]{}%
\providecommand \bibitemNoStop [0]{.\EOS\space}%
\providecommand \EOS [0]{\spacefactor3000\relax}%
\providecommand \BibitemShut  [1]{\csname bibitem#1\endcsname}%
\let\auto@bib@innerbib\@empty
%</preamble>
\bibitem [{\citenamefont {Shor}(1995)}]{Shor:1995fj}%
  \BibitemOpen
  \bibfield  {author} {\bibinfo {author} {\bibfnamefont {P.~W.}\ \bibnamefont
  {Shor}},\ }\bibfield  {title} {\bibinfo {title} {{Scheme for reducing
  decoherence in quantum computer memory}},\ }\href
  {https://doi.org/10.1103/PhysRevA.52.R2493} {\bibfield  {journal} {\bibinfo
  {journal} {Physical Review A}\ }\textbf {\bibinfo {volume} {52}},\ \bibinfo
  {pages} {R2493} (\bibinfo {year} {1995})}\BibitemShut {NoStop}%
\bibitem [{\citenamefont {Calderbank}\ and\ \citenamefont
  {Shor}(1996)}]{Calderbank:1996ja}%
  \BibitemOpen
  \bibfield  {author} {\bibinfo {author} {\bibfnamefont {A.~R.}\ \bibnamefont
  {Calderbank}}\ and\ \bibinfo {author} {\bibfnamefont {P.~W.}\ \bibnamefont
  {Shor}},\ }\bibfield  {title} {\bibinfo {title} {{Good quantum
  error-correcting codes exist}},\ }\href
  {https://doi.org/10.1103/PhysRevA.54.1098} {\bibfield  {journal} {\bibinfo
  {journal} {Physical Review A}\ }\textbf {\bibinfo {volume} {54}},\ \bibinfo
  {pages} {1098} (\bibinfo {year} {1996})}\BibitemShut {NoStop}%
\bibitem [{\citenamefont {Steane}(1996)}]{Steane:1996kg}%
  \BibitemOpen
  \bibfield  {author} {\bibinfo {author} {\bibfnamefont {A.~M.}\ \bibnamefont
  {Steane}},\ }\bibfield  {title} {\bibinfo {title} {{Error Correcting Codes in
  Quantum Theory}},\ }\href {https://doi.org/10.1103/PhysRevLett.77.793}
  {\bibfield  {journal} {\bibinfo  {journal} {Physical Review Letters}\
  }\textbf {\bibinfo {volume} {77}},\ \bibinfo {pages} {793} (\bibinfo {year}
  {1996})}\BibitemShut {NoStop}%
\bibitem [{\citenamefont {Preskill}(2018)}]{Preskill:2018gt}%
  \BibitemOpen
  \bibfield  {author} {\bibinfo {author} {\bibfnamefont {J.}~\bibnamefont
  {Preskill}},\ }\bibfield  {title} {\bibinfo {title} {{Quantum Computing in
  the NISQ era and beyond}},\ }\href {https://doi.org/10.22331/q-2018-08-06-79}
  {\bibfield  {journal} {\bibinfo  {journal} {Quantum}\ }\textbf {\bibinfo
  {volume} {2}},\ \bibinfo {pages} {79} (\bibinfo {year} {2018})}\BibitemShut
  {NoStop}%
\bibitem [{\citenamefont {Dalzell}\ \emph {et~al.}()\citenamefont {Dalzell},
  \citenamefont {Hunter-jones},\ and\ \citenamefont
  {Brand{\~{a}}o}}]{Dalzell2021}%
  \BibitemOpen
  \bibfield  {author} {\bibinfo {author} {\bibfnamefont {A.~M.}\ \bibnamefont
  {Dalzell}}, \bibinfo {author} {\bibfnamefont {N.}~\bibnamefont
  {Hunter-jones}},\ and\ \bibinfo {author} {\bibfnamefont {F.~G. S.~L.}\
  \bibnamefont {Brand{\~{a}}o}},\ }\bibfield  {title} {\bibinfo {title}
  {{Random quantum circuits transform local noise into global white noise}},\
  }\Eprint {https://arxiv.org/abs/2111.14907} {arXiv:2111.14907} \BibitemShut
  {NoStop}%
\bibitem [{\citenamefont {{Stilck Fran{\c{c}}a}}\ and\ \citenamefont
  {Garc{\'{i}}a-Patr{\'{o}}n}(2021)}]{StilckFranca:2021cb}%
  \BibitemOpen
  \bibfield  {author} {\bibinfo {author} {\bibfnamefont {D.}~\bibnamefont
  {{Stilck Fran{\c{c}}a}}}\ and\ \bibinfo {author} {\bibfnamefont
  {R.}~\bibnamefont {Garc{\'{i}}a-Patr{\'{o}}n}},\ }\bibfield  {title}
  {\bibinfo {title} {{Limitations of optimization algorithms on noisy quantum
  devices}},\ }\href {https://doi.org/10.1038/s41567-021-01356-3} {\bibfield
  {journal} {\bibinfo  {journal} {Nature Physics}\ }\textbf {\bibinfo {volume}
  {17}},\ \bibinfo {pages} {1221} (\bibinfo {year} {2021})}\BibitemShut
  {NoStop}%
\bibitem [{\citenamefont {Hangleiter}\ and\ \citenamefont
  {Eisert}(2023)}]{Hangleiter:2023cq}%
  \BibitemOpen
  \bibfield  {author} {\bibinfo {author} {\bibfnamefont {D.}~\bibnamefont
  {Hangleiter}}\ and\ \bibinfo {author} {\bibfnamefont {J.}~\bibnamefont
  {Eisert}},\ }\bibfield  {title} {\bibinfo {title} {{Computational advantage
  of quantum random sampling}},\ }\href
  {https://doi.org/10.1103/RevModPhys.95.035001} {\bibfield  {journal}
  {\bibinfo  {journal} {Reviews of Modern Physics}\ }\textbf {\bibinfo {volume}
  {95}},\ \bibinfo {pages} {035001} (\bibinfo {year} {2023})}\BibitemShut
  {NoStop}%
\bibitem [{\citenamefont {Zurek}(2003)}]{Zurek:2003fm}%
  \BibitemOpen
  \bibfield  {author} {\bibinfo {author} {\bibfnamefont {W.~H.}\ \bibnamefont
  {Zurek}},\ }\bibfield  {title} {\bibinfo {title} {{Decoherence, einselection,
  and the quantum origins of the classical}},\ }\href
  {https://doi.org/10.1103/RevModPhys.75.715} {\bibfield  {journal} {\bibinfo
  {journal} {Reviews of Modern Physics}\ }\textbf {\bibinfo {volume} {75}},\
  \bibinfo {pages} {715} (\bibinfo {year} {2003})}\BibitemShut {NoStop}%
\bibitem [{\citenamefont {Nielsen}\ and\ \citenamefont
  {Chuang}(2010)}]{Nielsen:2010ga}%
  \BibitemOpen
  \bibfield  {author} {\bibinfo {author} {\bibfnamefont {M.~A.}\ \bibnamefont
  {Nielsen}}\ and\ \bibinfo {author} {\bibfnamefont {I.~L.}\ \bibnamefont
  {Chuang}},\ }\href {https://doi.org/10.1017/CBO9780511976667} {\emph
  {\bibinfo {title} {Quantum Computation and Quantum Information}}}\ (\bibinfo
  {publisher} {Cambridge University Press},\ \bibinfo {address} {Cambridge,
  U.K.},\ \bibinfo {year} {2010})\BibitemShut {NoStop}%
\bibitem [{\citenamefont {Chamberland}\ \emph {et~al.}(2017)\citenamefont
  {Chamberland}, \citenamefont {Wallman}, \citenamefont {Beale},\ and\
  \citenamefont {Laflamme}}]{Chamberland:2017dm}%
  \BibitemOpen
  \bibfield  {author} {\bibinfo {author} {\bibfnamefont {C.}~\bibnamefont
  {Chamberland}}, \bibinfo {author} {\bibfnamefont {J.}~\bibnamefont
  {Wallman}}, \bibinfo {author} {\bibfnamefont {S.}~\bibnamefont {Beale}},\
  and\ \bibinfo {author} {\bibfnamefont {R.}~\bibnamefont {Laflamme}},\
  }\bibfield  {title} {\bibinfo {title} {{Hard decoding algorithm for
  optimizing thresholds under general Markovian noise}},\ }\href
  {https://doi.org/10.1103/PhysRevA.95.042332} {\bibfield  {journal} {\bibinfo
  {journal} {Physical Review A}\ }\textbf {\bibinfo {volume} {95}},\ \bibinfo
  {pages} {042332} (\bibinfo {year} {2017})}\BibitemShut {NoStop}%
\bibitem [{\citenamefont {Bravyi}\ \emph {et~al.}(2018)\citenamefont {Bravyi},
  \citenamefont {Englbrecht}, \citenamefont {K{\"{o}}nig},\ and\ \citenamefont
  {Peard}}]{Bravyi:2018ea}%
  \BibitemOpen
  \bibfield  {author} {\bibinfo {author} {\bibfnamefont {S.}~\bibnamefont
  {Bravyi}}, \bibinfo {author} {\bibfnamefont {M.}~\bibnamefont {Englbrecht}},
  \bibinfo {author} {\bibfnamefont {R.}~\bibnamefont {K{\"{o}}nig}},\ and\
  \bibinfo {author} {\bibfnamefont {N.}~\bibnamefont {Peard}},\ }\bibfield
  {title} {\bibinfo {title} {{Correcting coherent errors with surface codes}},\
  }\href {https://doi.org/10.1038/s41534-018-0106-y} {\bibfield  {journal}
  {\bibinfo  {journal} {npj Quantum Inf.}\ }\textbf {\bibinfo {volume} {4}},\
  \bibinfo {pages} {55} (\bibinfo {year} {2018})}\BibitemShut {NoStop}%
\bibitem [{\citenamefont {Gottesman}()}]{Gottesman2019}%
  \BibitemOpen
  \bibfield  {author} {\bibinfo {author} {\bibfnamefont {D.}~\bibnamefont
  {Gottesman}},\ }\bibfield  {title} {\bibinfo {title} {{Maximally Sensitive
  Sets of States}},\ }\Eprint {https://arxiv.org/abs/1907.05950}
  {arXiv:1907.05950} \BibitemShut {NoStop}%
\bibitem [{\citenamefont {Guti{\'{e}}rrez}\ \emph {et~al.}(2016)\citenamefont
  {Guti{\'{e}}rrez}, \citenamefont {Smith}, \citenamefont {Lulushi},
  \citenamefont {Janardan},\ and\ \citenamefont {Brown}}]{Gutierrez:2016dm}%
  \BibitemOpen
  \bibfield  {author} {\bibinfo {author} {\bibfnamefont {M.}~\bibnamefont
  {Guti{\'{e}}rrez}}, \bibinfo {author} {\bibfnamefont {C.}~\bibnamefont
  {Smith}}, \bibinfo {author} {\bibfnamefont {L.}~\bibnamefont {Lulushi}},
  \bibinfo {author} {\bibfnamefont {S.}~\bibnamefont {Janardan}},\ and\
  \bibinfo {author} {\bibfnamefont {K.~R.}\ \bibnamefont {Brown}},\ }\bibfield
  {title} {\bibinfo {title} {{Errors and pseudothresholds for incoherent and
  coherent noise}},\ }\href {https://doi.org/10.1103/PhysRevA.94.042338}
  {\bibfield  {journal} {\bibinfo  {journal} {Physical Review A}\ }\textbf
  {\bibinfo {volume} {94}},\ \bibinfo {pages} {042338} (\bibinfo {year}
  {2016})}\BibitemShut {NoStop}%
\bibitem [{\citenamefont {Huang}\ \emph {et~al.}(2019)\citenamefont {Huang},
  \citenamefont {Doherty},\ and\ \citenamefont {Flammia}}]{Huang:2019gj}%
  \BibitemOpen
  \bibfield  {author} {\bibinfo {author} {\bibfnamefont {E.}~\bibnamefont
  {Huang}}, \bibinfo {author} {\bibfnamefont {A.~C.}\ \bibnamefont {Doherty}},\
  and\ \bibinfo {author} {\bibfnamefont {S.}~\bibnamefont {Flammia}},\
  }\bibfield  {title} {\bibinfo {title} {{Performance of quantum error
  correction with coherent errors}},\ }\href
  {https://doi.org/10.1103/PhysRevA.99.022313} {\bibfield  {journal} {\bibinfo
  {journal} {Physical Review A}\ }\textbf {\bibinfo {volume} {99}},\ \bibinfo
  {pages} {022313} (\bibinfo {year} {2019})}\BibitemShut {NoStop}%
\bibitem [{\citenamefont {Greenbaum}\ and\ \citenamefont
  {Dutton}(2018)}]{Greenbaum:2018ce}%
  \BibitemOpen
  \bibfield  {author} {\bibinfo {author} {\bibfnamefont {D.}~\bibnamefont
  {Greenbaum}}\ and\ \bibinfo {author} {\bibfnamefont {Z.}~\bibnamefont
  {Dutton}},\ }\bibfield  {title} {\bibinfo {title} {{Modeling coherent errors
  in quantum error correction}},\ }\href
  {https://doi.org/10.1088/2058-9565/aa9a06} {\bibfield  {journal} {\bibinfo
  {journal} {Quantum Science and Technology}\ }\textbf {\bibinfo {volume}
  {3}},\ \bibinfo {pages} {015007} (\bibinfo {year} {2018})}\BibitemShut
  {NoStop}%
\bibitem [{\citenamefont {Iverson}\ and\ \citenamefont
  {Preskill}(2020)}]{Iverson:2020fe}%
  \BibitemOpen
  \bibfield  {author} {\bibinfo {author} {\bibfnamefont {J.~K.}\ \bibnamefont
  {Iverson}}\ and\ \bibinfo {author} {\bibfnamefont {J.}~\bibnamefont
  {Preskill}},\ }\bibfield  {title} {\bibinfo {title} {{Coherence in logical
  quantum channels}},\ }\href {https://doi.org/10.1088/1367-2630/ab8e5c}
  {\bibfield  {journal} {\bibinfo  {journal} {New Journal of Physics}\ }\textbf
  {\bibinfo {volume} {22}},\ \bibinfo {pages} {073066} (\bibinfo {year}
  {2020})}\BibitemShut {NoStop}%
\bibitem [{\citenamefont {Bluvstein}\ \emph {et~al.}(2024)\citenamefont
  {Bluvstein} \emph {et~al.}}]{Bluvstein:2024ht}%
  \BibitemOpen
  \bibfield  {author} {\bibinfo {author} {\bibfnamefont {D.}~\bibnamefont
  {Bluvstein}} \emph {et~al.},\ }\bibfield  {title} {\bibinfo {title} {{Logical
  quantum processor based on reconfigurable atom arrays}},\ }\href
  {https://doi.org/10.1038/s41586-023-06927-3} {\bibfield  {journal} {\bibinfo
  {journal} {Nature}\ }\textbf {\bibinfo {volume} {626}},\ \bibinfo {pages}
  {58} (\bibinfo {year} {2024})}\BibitemShut {NoStop}%
\bibitem [{\citenamefont {Gross}\ \emph {et~al.}(2024)\citenamefont {Gross},
  \citenamefont {Genois}, \citenamefont {Debroy}, \citenamefont {Zhang},
  \citenamefont {Mruczkiewicz}, \citenamefont {Cian},\ and\ \citenamefont
  {Jiang}}]{gross2024characterizing}%
  \BibitemOpen
  \bibfield  {author} {\bibinfo {author} {\bibfnamefont {J.~A.}\ \bibnamefont
  {Gross}}, \bibinfo {author} {\bibfnamefont {{\'{E}}.}~\bibnamefont {Genois}},
  \bibinfo {author} {\bibfnamefont {D.~M.}\ \bibnamefont {Debroy}}, \bibinfo
  {author} {\bibfnamefont {Y.}~\bibnamefont {Zhang}}, \bibinfo {author}
  {\bibfnamefont {W.}~\bibnamefont {Mruczkiewicz}}, \bibinfo {author}
  {\bibfnamefont {Z.-P.}\ \bibnamefont {Cian}},\ and\ \bibinfo {author}
  {\bibfnamefont {Z.}~\bibnamefont {Jiang}},\ }\bibfield  {title} {\bibinfo
  {title} {{Characterizing coherent errors using matrix-element
  amplification}},\ }\href {https://doi.org/10.1038/s41534-024-00917-7}
  {\bibfield  {journal} {\bibinfo  {journal} {npj Quantum Information}\
  }\textbf {\bibinfo {volume} {10}},\ \bibinfo {pages} {123} (\bibinfo {year}
  {2024})}\BibitemShut {NoStop}%
\bibitem [{\citenamefont {Kurilovich}\ \emph {et~al.}()\citenamefont
  {Kurilovich} \emph {et~al.}}]{Kurilovich_corr}%
  \BibitemOpen
  \bibfield  {author} {\bibinfo {author} {\bibfnamefont {V.~D.}\ \bibnamefont
  {Kurilovich}} \emph {et~al.},\ }\bibfield  {title} {\bibinfo {title}
  {{Correlated Error Bursts in a Gap-Engineered Superconducting Qubit Array}},\
  }\Eprint {https://arxiv.org/abs/2506.18228} {arXiv:2506.18228} \BibitemShut
  {NoStop}%
\bibitem [{\citenamefont {Nigmatullin}\ \emph {et~al.}(2025)\citenamefont
  {Nigmatullin}, \citenamefont {H{\'{e}}mery}, \citenamefont {Ghanem},
  \citenamefont {Moses}, \citenamefont {Gresh}, \citenamefont {Siegfried},
  \citenamefont {Mills}, \citenamefont {Gatterman}, \citenamefont {Hewitt},
  \citenamefont {Granet},\ and\ \citenamefont
  {Dreyer}}]{nigmatullin2025experimental}%
  \BibitemOpen
  \bibfield  {author} {\bibinfo {author} {\bibfnamefont {R.}~\bibnamefont
  {Nigmatullin}}, \bibinfo {author} {\bibfnamefont {K.}~\bibnamefont
  {H{\'{e}}mery}}, \bibinfo {author} {\bibfnamefont {K.}~\bibnamefont
  {Ghanem}}, \bibinfo {author} {\bibfnamefont {S.}~\bibnamefont {Moses}},
  \bibinfo {author} {\bibfnamefont {D.}~\bibnamefont {Gresh}}, \bibinfo
  {author} {\bibfnamefont {P.}~\bibnamefont {Siegfried}}, \bibinfo {author}
  {\bibfnamefont {M.}~\bibnamefont {Mills}}, \bibinfo {author} {\bibfnamefont
  {T.}~\bibnamefont {Gatterman}}, \bibinfo {author} {\bibfnamefont
  {N.}~\bibnamefont {Hewitt}}, \bibinfo {author} {\bibfnamefont
  {E.}~\bibnamefont {Granet}},\ and\ \bibinfo {author} {\bibfnamefont
  {H.}~\bibnamefont {Dreyer}},\ }\bibfield  {title} {\bibinfo {title}
  {{Experimental demonstration of breakeven for a compact fermionic
  encoding}},\ }\href {https://doi.org/10.1038/s41567-025-02931-8} {\bibfield
  {journal} {\bibinfo  {journal} {Nature Physics}\ }\textbf {\bibinfo {volume}
  {21}},\ \bibinfo {pages} {1319} (\bibinfo {year} {2025})}\BibitemShut
  {NoStop}%
\bibitem [{\citenamefont {Bluvstein}\ \emph {et~al.}()\citenamefont {Bluvstein}
  \emph {et~al.}}]{Bluvstein_arch}%
  \BibitemOpen
  \bibfield  {author} {\bibinfo {author} {\bibfnamefont {D.}~\bibnamefont
  {Bluvstein}} \emph {et~al.},\ }\bibfield  {title} {\bibinfo {title}
  {{Architectural mechanisms of a universal fault-tolerant quantum computer}},\
  }\Eprint {https://arxiv.org/abs/2506.20661} {arXiv:2506.20661} \BibitemShut
  {NoStop}%
\bibitem [{\citenamefont {Cong}\ \emph {et~al.}(2022)\citenamefont {Cong},
  \citenamefont {Levine}, \citenamefont {Keesling}, \citenamefont {Bluvstein},
  \citenamefont {Wang},\ and\ \citenamefont {Lukin}}]{Cong_PhysRevX.12.021049}%
  \BibitemOpen
  \bibfield  {author} {\bibinfo {author} {\bibfnamefont {I.}~\bibnamefont
  {Cong}}, \bibinfo {author} {\bibfnamefont {H.}~\bibnamefont {Levine}},
  \bibinfo {author} {\bibfnamefont {A.}~\bibnamefont {Keesling}}, \bibinfo
  {author} {\bibfnamefont {D.}~\bibnamefont {Bluvstein}}, \bibinfo {author}
  {\bibfnamefont {S.-T.}\ \bibnamefont {Wang}},\ and\ \bibinfo {author}
  {\bibfnamefont {M.~D.}\ \bibnamefont {Lukin}},\ }\bibfield  {title} {\bibinfo
  {title} {{Hardware-Efficient, Fault-Tolerant Quantum Computation with Rydberg
  Atoms}},\ }\href {https://doi.org/10.1103/PhysRevX.12.021049} {\bibfield
  {journal} {\bibinfo  {journal} {Phys. Rev. X}\ }\textbf {\bibinfo {volume}
  {12}},\ \bibinfo {pages} {021049} (\bibinfo {year} {2022})}\BibitemShut
  {NoStop}%
\bibitem [{\citenamefont {Wallman}\ and\ \citenamefont
  {Emerson}(2016)}]{Wallman:2016be}%
  \BibitemOpen
  \bibfield  {author} {\bibinfo {author} {\bibfnamefont {J.~J.}\ \bibnamefont
  {Wallman}}\ and\ \bibinfo {author} {\bibfnamefont {J.}~\bibnamefont
  {Emerson}},\ }\bibfield  {title} {\bibinfo {title} {{Noise tailoring for
  scalable quantum computation via randomized compiling}},\ }\href
  {https://doi.org/10.1103/PhysRevA.94.052325} {\bibfield  {journal} {\bibinfo
  {journal} {Physical Review A}\ }\textbf {\bibinfo {volume} {94}},\ \bibinfo
  {pages} {052325} (\bibinfo {year} {2016})}\BibitemShut {NoStop}%
\bibitem [{\citenamefont {Zhang}\ \emph {et~al.}(2022)\citenamefont {Zhang},
  \citenamefont {Majumder}, \citenamefont {Leung}, \citenamefont {Crain},
  \citenamefont {Wang}, \citenamefont {Fang}, \citenamefont {Debroy},
  \citenamefont {Kim},\ and\ \citenamefont
  {Brown}}]{Zhang_PhysRevApplied.17.034074}%
  \BibitemOpen
  \bibfield  {author} {\bibinfo {author} {\bibfnamefont {B.}~\bibnamefont
  {Zhang}}, \bibinfo {author} {\bibfnamefont {S.}~\bibnamefont {Majumder}},
  \bibinfo {author} {\bibfnamefont {P.~H.}\ \bibnamefont {Leung}}, \bibinfo
  {author} {\bibfnamefont {S.}~\bibnamefont {Crain}}, \bibinfo {author}
  {\bibfnamefont {Y.}~\bibnamefont {Wang}}, \bibinfo {author} {\bibfnamefont
  {C.}~\bibnamefont {Fang}}, \bibinfo {author} {\bibfnamefont {D.~M.}\
  \bibnamefont {Debroy}}, \bibinfo {author} {\bibfnamefont {J.}~\bibnamefont
  {Kim}},\ and\ \bibinfo {author} {\bibfnamefont {K.~R.}\ \bibnamefont
  {Brown}},\ }\bibfield  {title} {\bibinfo {title} {{Hidden Inverses: Coherent
  Error Cancellation at the Circuit Level}},\ }\href
  {https://doi.org/10.1103/PhysRevApplied.17.034074} {\bibfield  {journal}
  {\bibinfo  {journal} {Phys. Rev. Appl.}\ }\textbf {\bibinfo {volume} {17}},\
  \bibinfo {pages} {034074} (\bibinfo {year} {2022})}\BibitemShut {NoStop}%
\bibitem [{\citenamefont {Hashim}\ \emph {et~al.}(2021)\citenamefont {Hashim},
  \citenamefont {Naik}, \citenamefont {Morvan}, \citenamefont {Ville},
  \citenamefont {Mitchell}, \citenamefont {Kreikebaum}, \citenamefont {Davis},
  \citenamefont {Smith}, \citenamefont {Iancu}, \citenamefont {O'Brien},
  \citenamefont {Hincks}, \citenamefont {Wallman}, \citenamefont {Emerson},\
  and\ \citenamefont {Siddiqi}}]{Hashim:2021gr}%
  \BibitemOpen
  \bibfield  {author} {\bibinfo {author} {\bibfnamefont {A.}~\bibnamefont
  {Hashim}}, \bibinfo {author} {\bibfnamefont {R.~K.}\ \bibnamefont {Naik}},
  \bibinfo {author} {\bibfnamefont {A.}~\bibnamefont {Morvan}}, \bibinfo
  {author} {\bibfnamefont {J.-L.}\ \bibnamefont {Ville}}, \bibinfo {author}
  {\bibfnamefont {B.}~\bibnamefont {Mitchell}}, \bibinfo {author}
  {\bibfnamefont {J.~M.}\ \bibnamefont {Kreikebaum}}, \bibinfo {author}
  {\bibfnamefont {M.}~\bibnamefont {Davis}}, \bibinfo {author} {\bibfnamefont
  {E.}~\bibnamefont {Smith}}, \bibinfo {author} {\bibfnamefont
  {C.}~\bibnamefont {Iancu}}, \bibinfo {author} {\bibfnamefont {K.~P.}\
  \bibnamefont {O'Brien}}, \bibinfo {author} {\bibfnamefont {I.}~\bibnamefont
  {Hincks}}, \bibinfo {author} {\bibfnamefont {J.~J.}\ \bibnamefont {Wallman}},
  \bibinfo {author} {\bibfnamefont {J.}~\bibnamefont {Emerson}},\ and\ \bibinfo
  {author} {\bibfnamefont {I.}~\bibnamefont {Siddiqi}},\ }\bibfield  {title}
  {\bibinfo {title} {{Randomized Compiling for Scalable Quantum Computing on a
  Noisy Superconducting Quantum Processor}},\ }\href
  {https://doi.org/10.1103/PhysRevX.11.041039} {\bibfield  {journal} {\bibinfo
  {journal} {Physical Review X}\ }\textbf {\bibinfo {volume} {11}},\ \bibinfo
  {pages} {041039} (\bibinfo {year} {2021})}\BibitemShut {NoStop}%
\bibitem [{\citenamefont {Dennis}\ \emph {et~al.}(2002)\citenamefont {Dennis},
  \citenamefont {Kitaev}, \citenamefont {Landahl},\ and\ \citenamefont
  {Preskill}}]{Dennis:2002ds}%
  \BibitemOpen
  \bibfield  {author} {\bibinfo {author} {\bibfnamefont {E.}~\bibnamefont
  {Dennis}}, \bibinfo {author} {\bibfnamefont {A.}~\bibnamefont {Kitaev}},
  \bibinfo {author} {\bibfnamefont {A.}~\bibnamefont {Landahl}},\ and\ \bibinfo
  {author} {\bibfnamefont {J.}~\bibnamefont {Preskill}},\ }\bibfield  {title}
  {\bibinfo {title} {{Topological quantum memory}},\ }\href
  {https://doi.org/10.1063/1.1499754} {\bibfield  {journal} {\bibinfo
  {journal} {J. Math. Phys.}\ }\textbf {\bibinfo {volume} {43}},\ \bibinfo
  {pages} {4452} (\bibinfo {year} {2002})}\BibitemShut {NoStop}%
\bibitem [{\citenamefont {Bombin}\ \emph {et~al.}(2012)\citenamefont {Bombin},
  \citenamefont {Andrist}, \citenamefont {Ohzeki}, \citenamefont {Katzgraber},\
  and\ \citenamefont {Martin-Delgado}}]{Bombin:2012km}%
  \BibitemOpen
  \bibfield  {author} {\bibinfo {author} {\bibfnamefont {H.}~\bibnamefont
  {Bombin}}, \bibinfo {author} {\bibfnamefont {R.~S.}\ \bibnamefont {Andrist}},
  \bibinfo {author} {\bibfnamefont {M.}~\bibnamefont {Ohzeki}}, \bibinfo
  {author} {\bibfnamefont {H.~G.}\ \bibnamefont {Katzgraber}},\ and\ \bibinfo
  {author} {\bibfnamefont {M.~A.}\ \bibnamefont {Martin-Delgado}},\ }\bibfield
  {title} {\bibinfo {title} {{Strong Resilience of Topological Codes to
  Depolarization}},\ }\href {https://doi.org/10.1103/PhysRevX.2.021004}
  {\bibfield  {journal} {\bibinfo  {journal} {Physical Review X}\ }\textbf
  {\bibinfo {volume} {2}},\ \bibinfo {pages} {021004} (\bibinfo {year}
  {2012})}\BibitemShut {NoStop}%
\bibitem [{\citenamefont {Wootton}\ and\ \citenamefont
  {Loss}(2012)}]{Wootton:2012cb}%
  \BibitemOpen
  \bibfield  {author} {\bibinfo {author} {\bibfnamefont {J.~R.}\ \bibnamefont
  {Wootton}}\ and\ \bibinfo {author} {\bibfnamefont {D.}~\bibnamefont {Loss}},\
  }\bibfield  {title} {\bibinfo {title} {{High Threshold Error Correction for
  the Surface Code}},\ }\href {https://doi.org/10.1103/PhysRevLett.109.160503}
  {\bibfield  {journal} {\bibinfo  {journal} {Physical Review Letters}\
  }\textbf {\bibinfo {volume} {109}},\ \bibinfo {pages} {160503} (\bibinfo
  {year} {2012})}\BibitemShut {NoStop}%
\bibitem [{\citenamefont {Chubb}\ and\ \citenamefont
  {Flammia}(2021)}]{Chubb:2021cn}%
  \BibitemOpen
  \bibfield  {author} {\bibinfo {author} {\bibfnamefont {C.~T.}\ \bibnamefont
  {Chubb}}\ and\ \bibinfo {author} {\bibfnamefont {S.~T.}\ \bibnamefont
  {Flammia}},\ }\bibfield  {title} {\bibinfo {title} {{Statistical mechanical
  models for quantum codes with correlated noise}},\ }\href
  {https://doi.org/10.4171/AIHPD/105} {\bibfield  {journal} {\bibinfo
  {journal} {Annales de l'Institut Henri Poincar{\'{e}} D}\ }\textbf {\bibinfo
  {volume} {8}},\ \bibinfo {pages} {269} (\bibinfo {year} {2021})}\BibitemShut
  {NoStop}%
\bibitem [{\citenamefont {Venn}\ and\ \citenamefont
  {B{\'{e}}ri}(2020)}]{Venn:2020ge}%
  \BibitemOpen
  \bibfield  {author} {\bibinfo {author} {\bibfnamefont {F.}~\bibnamefont
  {Venn}}\ and\ \bibinfo {author} {\bibfnamefont {B.}~\bibnamefont
  {B{\'{e}}ri}},\ }\bibfield  {title} {\bibinfo {title} {{Error-correction and
  noise-decoherence thresholds for coherent errors in planar-graph surface
  codes}},\ }\href {https://doi.org/10.1103/PhysRevResearch.2.043412}
  {\bibfield  {journal} {\bibinfo  {journal} {Phys. Rev. Res.}\ }\textbf
  {\bibinfo {volume} {2}},\ \bibinfo {pages} {043412} (\bibinfo {year}
  {2020})}\BibitemShut {NoStop}%
\bibitem [{\citenamefont {Venn}\ \emph {et~al.}(2023)\citenamefont {Venn},
  \citenamefont {Behrends},\ and\ \citenamefont {B{\'{e}}ri}}]{Venn:2023fp}%
  \BibitemOpen
  \bibfield  {author} {\bibinfo {author} {\bibfnamefont {F.}~\bibnamefont
  {Venn}}, \bibinfo {author} {\bibfnamefont {J.}~\bibnamefont {Behrends}},\
  and\ \bibinfo {author} {\bibfnamefont {B.}~\bibnamefont {B{\'{e}}ri}},\
  }\bibfield  {title} {\bibinfo {title} {{Coherent-Error Threshold for Surface
  Codes from Majorana Delocalization}},\ }\href
  {https://doi.org/10.1103/PhysRevLett.131.060603} {\bibfield  {journal}
  {\bibinfo  {journal} {Physical Review Letters}\ }\textbf {\bibinfo {volume}
  {131}},\ \bibinfo {pages} {060603} (\bibinfo {year} {2023})}\BibitemShut
  {NoStop}%
\bibitem [{\citenamefont {Behrends}\ \emph {et~al.}(2024)\citenamefont
  {Behrends}, \citenamefont {Venn},\ and\ \citenamefont
  {B{\'{e}}ri}}]{Behrends:2024bs}%
  \BibitemOpen
  \bibfield  {author} {\bibinfo {author} {\bibfnamefont {J.}~\bibnamefont
  {Behrends}}, \bibinfo {author} {\bibfnamefont {F.}~\bibnamefont {Venn}},\
  and\ \bibinfo {author} {\bibfnamefont {B.}~\bibnamefont {B{\'{e}}ri}},\
  }\bibfield  {title} {\bibinfo {title} {{Surface codes, quantum circuits, and
  entanglement phases}},\ }\href
  {https://doi.org/10.1103/PhysRevResearch.6.013137} {\bibfield  {journal}
  {\bibinfo  {journal} {Physical Review Research}\ }\textbf {\bibinfo {volume}
  {6}},\ \bibinfo {pages} {013137} (\bibinfo {year} {2024})}\BibitemShut
  {NoStop}%
\bibitem [{\citenamefont {Behrends}\ and\ \citenamefont
  {B{\'{e}}ri}()}]{behrends2024statistical}%
  \BibitemOpen
  \bibfield  {author} {\bibinfo {author} {\bibfnamefont {J.}~\bibnamefont
  {Behrends}}\ and\ \bibinfo {author} {\bibfnamefont {B.}~\bibnamefont
  {B{\'{e}}ri}},\ }\bibfield  {title} {\bibinfo {title} {Statistical mechanical
  mapping and maximum-likelihood thresholds for the surface code under generic
  single-qubit coherent errors},\ }\Eprint {https://arxiv.org/abs/2410.22436}
  {arXiv:2410.22436} \BibitemShut {NoStop}%
\bibitem [{\citenamefont {Chen}\ and\ \citenamefont
  {Grover}(2024{\natexlab{a}})}]{Chen2024}%
  \BibitemOpen
  \bibfield  {author} {\bibinfo {author} {\bibfnamefont {Y.-H.}\ \bibnamefont
  {Chen}}\ and\ \bibinfo {author} {\bibfnamefont {T.}~\bibnamefont {Grover}},\
  }\bibfield  {title} {\bibinfo {title} {Unconventional topological mixed-state
  transition and critical phase induced by self-dual coherent errors},\ }\href
  {https://doi.org/10.1103/PhysRevB.110.125152} {\bibfield  {journal} {\bibinfo
   {journal} {Phys. Rev. B}\ }\textbf {\bibinfo {volume} {110}},\ \bibinfo
  {pages} {125152} (\bibinfo {year} {2024}{\natexlab{a}})}\BibitemShut
  {NoStop}%
\bibitem [{\citenamefont {Lee}\ and\ \citenamefont {Moon}()}]{Lee2024}%
  \BibitemOpen
  \bibfield  {author} {\bibinfo {author} {\bibfnamefont {S.}~\bibnamefont
  {Lee}}\ and\ \bibinfo {author} {\bibfnamefont {E.-G.}\ \bibnamefont {Moon}},\
  }\bibfield  {title} {\bibinfo {title} {{Mixed-State Topological Order under
  Coherent Noises}},\ }\Eprint {https://arxiv.org/abs/2411.03441}
  {arXiv:2411.03441} \BibitemShut {NoStop}%
\bibitem [{\citenamefont {Bao}\ and\ \citenamefont {Anand}()}]{Bao2024}%
  \BibitemOpen
  \bibfield  {author} {\bibinfo {author} {\bibfnamefont {Y.}~\bibnamefont
  {Bao}}\ and\ \bibinfo {author} {\bibfnamefont {S.}~\bibnamefont {Anand}},\
  }\bibfield  {title} {\bibinfo {title} {{Phases of decodability in the surface
  code with unitary errors}},\ }\Eprint {https://arxiv.org/abs/2411.05785}
  {arXiv:2411.05785} \BibitemShut {NoStop}%
\bibitem [{\citenamefont {Fan}\ \emph {et~al.}(2024)\citenamefont {Fan},
  \citenamefont {Bao}, \citenamefont {Altman},\ and\ \citenamefont
  {Vishwanath}}]{Fan:2024ku}%
  \BibitemOpen
  \bibfield  {author} {\bibinfo {author} {\bibfnamefont {R.}~\bibnamefont
  {Fan}}, \bibinfo {author} {\bibfnamefont {Y.}~\bibnamefont {Bao}}, \bibinfo
  {author} {\bibfnamefont {E.}~\bibnamefont {Altman}},\ and\ \bibinfo {author}
  {\bibfnamefont {A.}~\bibnamefont {Vishwanath}},\ }\bibfield  {title}
  {\bibinfo {title} {{Diagnostics of Mixed-State Topological Order and
  Breakdown of Quantum Memory}},\ }\href
  {https://doi.org/10.1103/PRXQuantum.5.020343} {\bibfield  {journal} {\bibinfo
   {journal} {PRX Quantum}\ }\textbf {\bibinfo {volume} {5}},\ \bibinfo {pages}
  {020343} (\bibinfo {year} {2024})}\BibitemShut {NoStop}%
\bibitem [{\citenamefont {Lee}\ \emph {et~al.}(2023)\citenamefont {Lee},
  \citenamefont {Jian},\ and\ \citenamefont {Xu}}]{Lee:2023fe}%
  \BibitemOpen
  \bibfield  {author} {\bibinfo {author} {\bibfnamefont {J.~Y.}\ \bibnamefont
  {Lee}}, \bibinfo {author} {\bibfnamefont {C.-M.}\ \bibnamefont {Jian}},\ and\
  \bibinfo {author} {\bibfnamefont {C.}~\bibnamefont {Xu}},\ }\bibfield
  {title} {\bibinfo {title} {{Quantum Criticality Under Decoherence or Weak
  Measurement}},\ }\href {https://doi.org/10.1103/PRXQuantum.4.030317}
  {\bibfield  {journal} {\bibinfo  {journal} {PRX Quantum}\ }\textbf {\bibinfo
  {volume} {4}},\ \bibinfo {pages} {030317} (\bibinfo {year}
  {2023})}\BibitemShut {NoStop}%
\bibitem [{\citenamefont {Chen}\ and\ \citenamefont
  {Grover}(2024{\natexlab{b}})}]{Chen:2024jh}%
  \BibitemOpen
  \bibfield  {author} {\bibinfo {author} {\bibfnamefont {Y.-H.}\ \bibnamefont
  {Chen}}\ and\ \bibinfo {author} {\bibfnamefont {T.}~\bibnamefont {Grover}},\
  }\bibfield  {title} {\bibinfo {title} {{Separability Transitions in
  Topological States Induced by Local Decoherence}},\ }\href
  {https://doi.org/10.1103/PhysRevLett.132.170602} {\bibfield  {journal}
  {\bibinfo  {journal} {Physical Review Letters}\ }\textbf {\bibinfo {volume}
  {132}},\ \bibinfo {pages} {170602} (\bibinfo {year}
  {2024}{\natexlab{b}})}\BibitemShut {NoStop}%
\bibitem [{\citenamefont {Sang}\ \emph {et~al.}(2024)\citenamefont {Sang},
  \citenamefont {Zou},\ and\ \citenamefont {Hsieh}}]{Sang2023}%
  \BibitemOpen
  \bibfield  {author} {\bibinfo {author} {\bibfnamefont {S.}~\bibnamefont
  {Sang}}, \bibinfo {author} {\bibfnamefont {Y.}~\bibnamefont {Zou}},\ and\
  \bibinfo {author} {\bibfnamefont {T.~H.}\ \bibnamefont {Hsieh}},\ }\bibfield
  {title} {\bibinfo {title} {Mixed-state quantum phases: Renormalization and
  quantum error correction},\ }\href
  {https://doi.org/10.1103/PhysRevX.14.031044} {\bibfield  {journal} {\bibinfo
  {journal} {Phys. Rev. X}\ }\textbf {\bibinfo {volume} {14}},\ \bibinfo
  {pages} {031044} (\bibinfo {year} {2024})}\BibitemShut {NoStop}%
\bibitem [{\citenamefont {Myerson-Jain}\ \emph {et~al.}(2025)\citenamefont
  {Myerson-Jain}, \citenamefont {Hughes},\ and\ \citenamefont
  {Xu}}]{Myerson:2025bx}%
  \BibitemOpen
  \bibfield  {author} {\bibinfo {author} {\bibfnamefont {N.}~\bibnamefont
  {Myerson-Jain}}, \bibinfo {author} {\bibfnamefont {T.~L.}\ \bibnamefont
  {Hughes}},\ and\ \bibinfo {author} {\bibfnamefont {C.}~\bibnamefont {Xu}},\
  }\bibfield  {title} {\bibinfo {title} {{Decoherence through Ancilla Anyon
  Reservoirs}},\ }\href {https://doi.org/10.1103/PhysRevLett.134.096503}
  {\bibfield  {journal} {\bibinfo  {journal} {Physical Review Letters}\
  }\textbf {\bibinfo {volume} {134}},\ \bibinfo {pages} {096503} (\bibinfo
  {year} {2025})}\BibitemShut {NoStop}%
\bibitem [{\citenamefont {Su}\ \emph {et~al.}(2024)\citenamefont {Su},
  \citenamefont {Yang},\ and\ \citenamefont {Jian}}]{Su:2024je}%
  \BibitemOpen
  \bibfield  {author} {\bibinfo {author} {\bibfnamefont {K.}~\bibnamefont
  {Su}}, \bibinfo {author} {\bibfnamefont {Z.}~\bibnamefont {Yang}},\ and\
  \bibinfo {author} {\bibfnamefont {C.-M.}\ \bibnamefont {Jian}},\ }\bibfield
  {title} {\bibinfo {title} {{Tapestry of dualities in decohered quantum error
  correction codes}},\ }\href {https://doi.org/10.1103/PhysRevB.110.085158}
  {\bibfield  {journal} {\bibinfo  {journal} {Physical Review B}\ }\textbf
  {\bibinfo {volume} {110}},\ \bibinfo {pages} {085158} (\bibinfo {year}
  {2024})}\BibitemShut {NoStop}%
\bibitem [{\citenamefont {Li}\ and\ \citenamefont {Mong}(2025)}]{Li2024}%
  \BibitemOpen
  \bibfield  {author} {\bibinfo {author} {\bibfnamefont {Z.}~\bibnamefont
  {Li}}\ and\ \bibinfo {author} {\bibfnamefont {R.~S.~K.}\ \bibnamefont
  {Mong}},\ }\bibfield  {title} {\bibinfo {title} {Replica topological order in
  quantum mixed states and quantum error correction},\ }\href
  {https://doi.org/10.1103/PhysRevB.111.125106} {\bibfield  {journal} {\bibinfo
   {journal} {Phys. Rev. B}\ }\textbf {\bibinfo {volume} {111}},\ \bibinfo
  {pages} {125106} (\bibinfo {year} {2025})}\BibitemShut {NoStop}%
\bibitem [{\citenamefont {Lavasani}\ and\ \citenamefont
  {Vijay}(2025)}]{Lavasani2024}%
  \BibitemOpen
  \bibfield  {author} {\bibinfo {author} {\bibfnamefont {A.}~\bibnamefont
  {Lavasani}}\ and\ \bibinfo {author} {\bibfnamefont {S.}~\bibnamefont
  {Vijay}},\ }\bibfield  {title} {\bibinfo {title} {Stability of gapped quantum
  matter and error-correction with adiabatic noise},\ }\href
  {https://doi.org/10.1103/PhysRevResearch.7.023166} {\bibfield  {journal}
  {\bibinfo  {journal} {Phys. Rev. Res.}\ }\textbf {\bibinfo {volume} {7}},\
  \bibinfo {pages} {023166} (\bibinfo {year} {2025})}\BibitemShut {NoStop}%
\bibitem [{\citenamefont {Lyons}()}]{Lyons2024}%
  \BibitemOpen
  \bibfield  {author} {\bibinfo {author} {\bibfnamefont {A.}~\bibnamefont
  {Lyons}},\ }\bibfield  {title} {\bibinfo {title} {{Understanding Stabilizer
  Codes Under Local Decoherence Through a General Statistical Mechanics
  Mapping}},\ }\Eprint {https://arxiv.org/abs/2403.03955} {arXiv:2403.03955}
  \BibitemShut {NoStop}%
\bibitem [{\citenamefont {Sohal}\ and\ \citenamefont
  {Prem}(2025)}]{Sohal:2025ec}%
  \BibitemOpen
  \bibfield  {author} {\bibinfo {author} {\bibfnamefont {R.}~\bibnamefont
  {Sohal}}\ and\ \bibinfo {author} {\bibfnamefont {A.}~\bibnamefont {Prem}},\
  }\bibfield  {title} {\bibinfo {title} {{Noisy Approach to Intrinsically
  Mixed-State Topological Order}},\ }\href
  {https://doi.org/10.1103/PRXQuantum.6.010313} {\bibfield  {journal} {\bibinfo
   {journal} {PRX Quantum}\ }\textbf {\bibinfo {volume} {6}},\ \bibinfo {pages}
  {010313} (\bibinfo {year} {2025})}\BibitemShut {NoStop}%
\bibitem [{\citenamefont {Lu}(2024)}]{Lu2024}%
  \BibitemOpen
  \bibfield  {author} {\bibinfo {author} {\bibfnamefont {T.-C.}\ \bibnamefont
  {Lu}},\ }\bibfield  {title} {\bibinfo {title} {Disentangling transitions in
  topological order induced by boundary decoherence},\ }\href
  {https://doi.org/10.1103/PhysRevB.110.125145} {\bibfield  {journal} {\bibinfo
   {journal} {Phys. Rev. B}\ }\textbf {\bibinfo {volume} {110}},\ \bibinfo
  {pages} {125145} (\bibinfo {year} {2024})}\BibitemShut {NoStop}%
\bibitem [{\citenamefont {Sang}\ and\ \citenamefont {Hsieh}(2025)}]{Sang2024}%
  \BibitemOpen
  \bibfield  {author} {\bibinfo {author} {\bibfnamefont {S.}~\bibnamefont
  {Sang}}\ and\ \bibinfo {author} {\bibfnamefont {T.~H.}\ \bibnamefont
  {Hsieh}},\ }\bibfield  {title} {\bibinfo {title} {{Stability of Mixed-State
  Quantum Phases via Finite Markov Length}},\ }\href
  {https://doi.org/10.1103/PhysRevLett.134.070403} {\bibfield  {journal}
  {\bibinfo  {journal} {Phys. Rev. Lett.}\ }\textbf {\bibinfo {volume} {134}},\
  \bibinfo {pages} {070403} (\bibinfo {year} {2025})}\BibitemShut {NoStop}%
\bibitem [{\citenamefont {Ellison}\ and\ \citenamefont
  {Cheng}(2025)}]{Ellison:2025kp}%
  \BibitemOpen
  \bibfield  {author} {\bibinfo {author} {\bibfnamefont {T.~D.}\ \bibnamefont
  {Ellison}}\ and\ \bibinfo {author} {\bibfnamefont {M.}~\bibnamefont
  {Cheng}},\ }\bibfield  {title} {\bibinfo {title} {{Toward a Classification of
  Mixed-State Topological Orders in Two Dimensions}},\ }\href
  {https://doi.org/10.1103/PRXQuantum.6.010315} {\bibfield  {journal} {\bibinfo
   {journal} {PRX Quantum}\ }\textbf {\bibinfo {volume} {6}},\ \bibinfo {pages}
  {010315} (\bibinfo {year} {2025})}\BibitemShut {NoStop}%
\bibitem [{\citenamefont {Zhang}\ \emph {et~al.}(2025)\citenamefont {Zhang},
  \citenamefont {Agrawal},\ and\ \citenamefont {Vijay}}]{Zhang2024}%
  \BibitemOpen
  \bibfield  {author} {\bibinfo {author} {\bibfnamefont {Z.}~\bibnamefont
  {Zhang}}, \bibinfo {author} {\bibfnamefont {U.}~\bibnamefont {Agrawal}},\
  and\ \bibinfo {author} {\bibfnamefont {S.}~\bibnamefont {Vijay}},\ }\bibfield
   {title} {\bibinfo {title} {Quantum communication and mixed-state order in
  decohered symmetry-protected topological states},\ }\href
  {https://doi.org/10.1103/PhysRevB.111.115141} {\bibfield  {journal} {\bibinfo
   {journal} {Phys. Rev. B}\ }\textbf {\bibinfo {volume} {111}},\ \bibinfo
  {pages} {115141} (\bibinfo {year} {2025})}\BibitemShut {NoStop}%
\bibitem [{\citenamefont {Oshima}\ \emph {et~al.}(2025)\citenamefont {Oshima},
  \citenamefont {Mochizuki}, \citenamefont {Hamazaki},\ and\ \citenamefont
  {Fuji}}]{Oshima2024}%
  \BibitemOpen
  \bibfield  {author} {\bibinfo {author} {\bibfnamefont {H.}~\bibnamefont
  {Oshima}}, \bibinfo {author} {\bibfnamefont {K.}~\bibnamefont {Mochizuki}},
  \bibinfo {author} {\bibfnamefont {R.}~\bibnamefont {Hamazaki}},\ and\
  \bibinfo {author} {\bibfnamefont {Y.}~\bibnamefont {Fuji}},\ }\bibfield
  {title} {\bibinfo {title} {Topology and spectrum in measurement-induced phase
  transitions},\ }\href {https://doi.org/10.1103/PhysRevLett.134.240401}
  {\bibfield  {journal} {\bibinfo  {journal} {Phys. Rev. Lett.}\ }\textbf
  {\bibinfo {volume} {134}},\ \bibinfo {pages} {240401} (\bibinfo {year}
  {2025})}\BibitemShut {NoStop}%
\bibitem [{\citenamefont {Lessa}\ \emph {et~al.}(2025)\citenamefont {Lessa},
  \citenamefont {Ma}, \citenamefont {Zhang}, \citenamefont {Bi}, \citenamefont
  {Cheng},\ and\ \citenamefont {Wang}}]{Lessa_PRXQuantum.6.010344}%
  \BibitemOpen
  \bibfield  {author} {\bibinfo {author} {\bibfnamefont {L.~A.}\ \bibnamefont
  {Lessa}}, \bibinfo {author} {\bibfnamefont {R.}~\bibnamefont {Ma}}, \bibinfo
  {author} {\bibfnamefont {J.-H.}\ \bibnamefont {Zhang}}, \bibinfo {author}
  {\bibfnamefont {Z.}~\bibnamefont {Bi}}, \bibinfo {author} {\bibfnamefont
  {M.}~\bibnamefont {Cheng}},\ and\ \bibinfo {author} {\bibfnamefont
  {C.}~\bibnamefont {Wang}},\ }\bibfield  {title} {\bibinfo {title}
  {Strong-to-weak spontaneous symmetry breaking in mixed quantum states},\
  }\href {https://doi.org/10.1103/PRXQuantum.6.010344} {\bibfield  {journal}
  {\bibinfo  {journal} {PRX Quantum}\ }\textbf {\bibinfo {volume} {6}},\
  \bibinfo {pages} {010344} (\bibinfo {year} {2025})}\BibitemShut {NoStop}%
\bibitem [{\citenamefont {Rakovszky}\ \emph {et~al.}(2024)\citenamefont
  {Rakovszky}, \citenamefont {Gopalakrishnan},\ and\ \citenamefont {von
  Keyserlingk}}]{RGK_PhysRevX.14.041031}%
  \BibitemOpen
  \bibfield  {author} {\bibinfo {author} {\bibfnamefont {T.}~\bibnamefont
  {Rakovszky}}, \bibinfo {author} {\bibfnamefont {S.}~\bibnamefont
  {Gopalakrishnan}},\ and\ \bibinfo {author} {\bibfnamefont {C.}~\bibnamefont
  {von Keyserlingk}},\ }\bibfield  {title} {\bibinfo {title} {Defining stable
  phases of open quantum systems},\ }\href
  {https://doi.org/10.1103/PhysRevX.14.041031} {\bibfield  {journal} {\bibinfo
  {journal} {Phys. Rev. X}\ }\textbf {\bibinfo {volume} {14}},\ \bibinfo
  {pages} {041031} (\bibinfo {year} {2024})}\BibitemShut {NoStop}%
\bibitem [{\citenamefont {Umegaki}(1962)}]{Umegaki:1962ee}%
  \BibitemOpen
  \bibfield  {author} {\bibinfo {author} {\bibfnamefont {H.}~\bibnamefont
  {Umegaki}},\ }\bibfield  {title} {\bibinfo {title} {{Conditional expectation
  in an operator algebra. IV. Entropy and information}},\ }\href
  {https://doi.org/10.2996/kmj/1138844604} {\bibfield  {journal} {\bibinfo
  {journal} {Kodai Mathematical Journal}\ }\textbf {\bibinfo {volume} {14}},\
  \bibinfo {pages} {59} (\bibinfo {year} {1962})}\BibitemShut {NoStop}%
\bibitem [{\citenamefont {Schumacher}\ and\ \citenamefont
  {Nielsen}(1996)}]{Schumacher:1996gt}%
  \BibitemOpen
  \bibfield  {author} {\bibinfo {author} {\bibfnamefont {B.}~\bibnamefont
  {Schumacher}}\ and\ \bibinfo {author} {\bibfnamefont {M.~A.}\ \bibnamefont
  {Nielsen}},\ }\bibfield  {title} {\bibinfo {title} {{Quantum data processing
  and error correction}},\ }\href {https://doi.org/10.1103/PhysRevA.54.2629}
  {\bibfield  {journal} {\bibinfo  {journal} {Physical Review A}\ }\textbf
  {\bibinfo {volume} {54}},\ \bibinfo {pages} {2629} (\bibinfo {year}
  {1996})}\BibitemShut {NoStop}%
\bibitem [{\citenamefont {Lloyd}(1997)}]{Lloyd:1997if}%
  \BibitemOpen
  \bibfield  {author} {\bibinfo {author} {\bibfnamefont {S.}~\bibnamefont
  {Lloyd}},\ }\bibfield  {title} {\bibinfo {title} {{Capacity of the noisy
  quantum channel}},\ }\href {https://doi.org/10.1103/PhysRevA.55.1613}
  {\bibfield  {journal} {\bibinfo  {journal} {Physical Review A}\ }\textbf
  {\bibinfo {volume} {55}},\ \bibinfo {pages} {1613} (\bibinfo {year}
  {1997})}\BibitemShut {NoStop}%
\bibitem [{\citenamefont {{\.{Z}}yczkowski}\ \emph {et~al.}(1998)\citenamefont
  {{\.{Z}}yczkowski}, \citenamefont {Horodecki}, \citenamefont {Sanpera},\ and\
  \citenamefont {Lewenstein}}]{Zyczkowski:1998ks}%
  \BibitemOpen
  \bibfield  {author} {\bibinfo {author} {\bibfnamefont {K.}~\bibnamefont
  {{\.{Z}}yczkowski}}, \bibinfo {author} {\bibfnamefont {P.}~\bibnamefont
  {Horodecki}}, \bibinfo {author} {\bibfnamefont {A.}~\bibnamefont {Sanpera}},\
  and\ \bibinfo {author} {\bibfnamefont {M.}~\bibnamefont {Lewenstein}},\
  }\bibfield  {title} {\bibinfo {title} {{Volume of the set of separable
  states}},\ }\href {https://doi.org/10.1103/PhysRevA.58.883} {\bibfield
  {journal} {\bibinfo  {journal} {Physical Review A}\ }\textbf {\bibinfo
  {volume} {58}},\ \bibinfo {pages} {883} (\bibinfo {year} {1998})}\BibitemShut
  {NoStop}%
\bibitem [{\citenamefont {Lee}\ and\ \citenamefont {Vidal}(2013)}]{Lee:2013ia}%
  \BibitemOpen
  \bibfield  {author} {\bibinfo {author} {\bibfnamefont {Y.~A.}\ \bibnamefont
  {Lee}}\ and\ \bibinfo {author} {\bibfnamefont {G.}~\bibnamefont {Vidal}},\
  }\bibfield  {title} {\bibinfo {title} {{Entanglement negativity and
  topological order}},\ }\href {https://doi.org/10.1103/PhysRevA.88.042318}
  {\bibfield  {journal} {\bibinfo  {journal} {Physical Review A}\ }\textbf
  {\bibinfo {volume} {88}},\ \bibinfo {pages} {042318} (\bibinfo {year}
  {2013})}\BibitemShut {NoStop}%
\bibitem [{\citenamefont {Bao}\ \emph {et~al.}()\citenamefont {Bao},
  \citenamefont {Fan}, \citenamefont {Vishwanath},\ and\ \citenamefont
  {Altman}}]{Bao2023}%
  \BibitemOpen
  \bibfield  {author} {\bibinfo {author} {\bibfnamefont {Y.}~\bibnamefont
  {Bao}}, \bibinfo {author} {\bibfnamefont {R.}~\bibnamefont {Fan}}, \bibinfo
  {author} {\bibfnamefont {A.}~\bibnamefont {Vishwanath}},\ and\ \bibinfo
  {author} {\bibfnamefont {E.}~\bibnamefont {Altman}},\ }\bibfield  {title}
  {\bibinfo {title} {Mixed-state topological order and the errorfield double
  formulation of decoherence-induced transitions},\ }\Eprint
  {https://arxiv.org/abs/2301.05687} {arXiv:2301.05687} \BibitemShut {NoStop}%
\bibitem [{\citenamefont {Huang}\ \emph {et~al.}()\citenamefont {Huang},
  \citenamefont {Colmenarez}, \citenamefont {Müller},\ and\ \citenamefont
  {Diehl}}]{huang2024coh}%
  \BibitemOpen
  \bibfield  {author} {\bibinfo {author} {\bibfnamefont {Z.-M.}\ \bibnamefont
  {Huang}}, \bibinfo {author} {\bibfnamefont {L.}~\bibnamefont {Colmenarez}},
  \bibinfo {author} {\bibfnamefont {M.}~\bibnamefont {Müller}},\ and\ \bibinfo
  {author} {\bibfnamefont {S.}~\bibnamefont {Diehl}},\ }\bibfield  {title}
  {\bibinfo {title} {Coherent information as a mixed-state topological order
  parameter of fermions},\ }\Eprint {https://arxiv.org/abs/2412.12279}
  {arXiv:2412.12279} \BibitemShut {NoStop}%
\bibitem [{\citenamefont {Colmenarez}\ \emph {et~al.}(2024)\citenamefont
  {Colmenarez}, \citenamefont {Huang}, \citenamefont {Diehl},\ and\
  \citenamefont {M{\"{u}}ller}}]{Colmenarez:2024iz}%
  \BibitemOpen
  \bibfield  {author} {\bibinfo {author} {\bibfnamefont {L.}~\bibnamefont
  {Colmenarez}}, \bibinfo {author} {\bibfnamefont {Z.-M.}\ \bibnamefont
  {Huang}}, \bibinfo {author} {\bibfnamefont {S.}~\bibnamefont {Diehl}},\ and\
  \bibinfo {author} {\bibfnamefont {M.}~\bibnamefont {M{\"{u}}ller}},\
  }\bibfield  {title} {\bibinfo {title} {{Accurate optimal quantum error
  correction thresholds from coherent information}},\ }\href
  {https://doi.org/10.1103/PhysRevResearch.6.L042014} {\bibfield  {journal}
  {\bibinfo  {journal} {Physical Review Research}\ }\textbf {\bibinfo {volume}
  {6}},\ \bibinfo {pages} {L042014} (\bibinfo {year} {2024})}\BibitemShut
  {NoStop}%
\bibitem [{\citenamefont {Colmenarez}\ \emph {et~al.}()\citenamefont
  {Colmenarez}, \citenamefont {Kim},\ and\ \citenamefont
  {Müller}}]{Colmenarez2024er}%
  \BibitemOpen
  \bibfield  {author} {\bibinfo {author} {\bibfnamefont {L.}~\bibnamefont
  {Colmenarez}}, \bibinfo {author} {\bibfnamefont {S.}~\bibnamefont {Kim}},\
  and\ \bibinfo {author} {\bibfnamefont {M.}~\bibnamefont {Müller}},\
  }\bibfield  {title} {\bibinfo {title} {Fundamental thresholds for
  computational and erasure errors via the coherent information},\ }\Eprint
  {https://arxiv.org/abs/2412.16727} {arXiv:2412.16727} \BibitemShut {NoStop}%
\bibitem [{\citenamefont {Kitaev}(1997)}]{Kitaev:1997kq}%
  \BibitemOpen
  \bibfield  {author} {\bibinfo {author} {\bibfnamefont {A.~Y.}\ \bibnamefont
  {Kitaev}},\ }\bibfield  {title} {\bibinfo {title} {{Quantum computations:
  algorithms and error correction}},\ }\href
  {https://doi.org/10.1070/RM1997v052n06ABEH002155} {\bibfield  {journal}
  {\bibinfo  {journal} {Russian Mathematical Surveys}\ }\textbf {\bibinfo
  {volume} {52}},\ \bibinfo {pages} {1191} (\bibinfo {year}
  {1997})}\BibitemShut {NoStop}%
\bibitem [{\citenamefont {Bravyi}\ and\ \citenamefont
  {Kitaev}(1998)}]{Bravyi1998quantum}%
  \BibitemOpen
  \bibfield  {author} {\bibinfo {author} {\bibfnamefont {S.~B.}\ \bibnamefont
  {Bravyi}}\ and\ \bibinfo {author} {\bibfnamefont {A.~Y.}\ \bibnamefont
  {Kitaev}},\ }\bibfield  {title} {\bibinfo {title} {Quantum codes on a lattice
  with boundary},\ }\Eprint {https://arxiv.org/abs/quant-ph/9811052}
  {arXiv:quant-ph/9811052 [quant-ph]}  (\bibinfo {year} {1998})\BibitemShut
  {NoStop}%
\bibitem [{\citenamefont {Freedman}\ and\ \citenamefont
  {Meyer}(1998)}]{Freedman1998projective}%
  \BibitemOpen
  \bibfield  {author} {\bibinfo {author} {\bibfnamefont {M.~H.}\ \bibnamefont
  {Freedman}}\ and\ \bibinfo {author} {\bibfnamefont {D.~A.}\ \bibnamefont
  {Meyer}},\ }\bibfield  {title} {\bibinfo {title} {Projective plane and planar
  quantum codes},\ }\Eprint {https://arxiv.org/abs/quant-ph/9810055}
  {arXiv:quant-ph/9810055 [quant-ph]}  (\bibinfo {year} {1998})\BibitemShut
  {NoStop}%
\bibitem [{\citenamefont {{Google Quantum AI}}(2023)}]{Acharya:2023fl}%
  \BibitemOpen
  \bibfield  {author} {\bibinfo {author} {\bibnamefont {{Google Quantum AI}}},\
  }\bibfield  {title} {\bibinfo {title} {{Suppressing quantum errors by scaling
  a surface code logical qubit}},\ }\href
  {https://doi.org/10.1038/s41586-022-05434-1} {\bibfield  {journal} {\bibinfo
  {journal} {Nature}\ }\textbf {\bibinfo {volume} {614}},\ \bibinfo {pages}
  {676} (\bibinfo {year} {2023})}\BibitemShut {NoStop}%
\bibitem [{\citenamefont {{Google Quantum AI}}(2024)}]{Acharya2024}%
  \BibitemOpen
  \bibfield  {author} {\bibinfo {author} {\bibnamefont {{Google Quantum AI}}},\
  }\bibfield  {title} {\bibinfo {title} {{Quantum error correction below the
  surface code threshold}},\ }\bibfield  {journal} {\bibinfo  {journal}
  {Nature}\ }\href {https://doi.org/10.1038/s41586-024-08449-y}
  {10.1038/s41586-024-08449-y} (\bibinfo {year} {2024})\BibitemShut {NoStop}%
\bibitem [{\citenamefont {{Goole Quantum AI}}()}]{eickbusch2024dynamic}%
  \BibitemOpen
  \bibfield  {author} {\bibinfo {author} {\bibnamefont {{Goole Quantum AI}}},\
  }\bibfield  {title} {\bibinfo {title} {Demonstrating dynamic surface codes},\
  }\Eprint {https://arxiv.org/abs/2412.14360} {arXiv:2412.14360} \BibitemShut
  {NoStop}%
\bibitem [{\citenamefont {Merz}\ and\ \citenamefont
  {Chalker}(2002{\natexlab{a}})}]{Merz:2002gj}%
  \BibitemOpen
  \bibfield  {author} {\bibinfo {author} {\bibfnamefont {F.}~\bibnamefont
  {Merz}}\ and\ \bibinfo {author} {\bibfnamefont {J.~T.}\ \bibnamefont
  {Chalker}},\ }\bibfield  {title} {\bibinfo {title} {{Two-dimensional
  random-bond Ising model, free fermions, and the network model}},\ }\href
  {https://doi.org/10.1103/PhysRevB.65.054425} {\bibfield  {journal} {\bibinfo
  {journal} {Phys. Rev. B}\ }\textbf {\bibinfo {volume} {65}},\ \bibinfo
  {pages} {054425} (\bibinfo {year} {2002}{\natexlab{a}})}\BibitemShut
  {NoStop}%
\bibitem [{\citenamefont {Schultz}\ \emph {et~al.}(1964)\citenamefont
  {Schultz}, \citenamefont {Mattis},\ and\ \citenamefont
  {Lieb}}]{Schultz:1964fv}%
  \BibitemOpen
  \bibfield  {author} {\bibinfo {author} {\bibfnamefont {T.~D.}\ \bibnamefont
  {Schultz}}, \bibinfo {author} {\bibfnamefont {D.~C.}\ \bibnamefont
  {Mattis}},\ and\ \bibinfo {author} {\bibfnamefont {E.~H.}\ \bibnamefont
  {Lieb}},\ }\bibfield  {title} {\bibinfo {title} {{Two-Dimensional Ising Model
  as a Soluble Problem of Many Fermions}},\ }\href
  {https://doi.org/10.1103/RevModPhys.36.856} {\bibfield  {journal} {\bibinfo
  {journal} {Rev. Mod. Phys.}\ }\textbf {\bibinfo {volume} {36}},\ \bibinfo
  {pages} {856} (\bibinfo {year} {1964})}\BibitemShut {NoStop}%
\bibitem [{\citenamefont {Hauschild}\ and\ \citenamefont
  {Pollmann}(2018)}]{Hauschild:2018bp}%
  \BibitemOpen
  \bibfield  {author} {\bibinfo {author} {\bibfnamefont {J.}~\bibnamefont
  {Hauschild}}\ and\ \bibinfo {author} {\bibfnamefont {F.}~\bibnamefont
  {Pollmann}},\ }\bibfield  {title} {\bibinfo {title} {{Efficient numerical
  simulations with Tensor Networks: Tensor Network Python (TeNPy)}},\ }\href
  {https://doi.org/10.21468/SciPostPhysLectNotes.5} {\bibfield  {journal}
  {\bibinfo  {journal} {SciPost Physics Lecture Notes}\ }\textbf {\bibinfo
  {volume} {5}},\ \bibinfo {pages} {5} (\bibinfo {year} {2018})}\BibitemShut
  {NoStop}%
\bibitem [{\citenamefont {Cirac}\ \emph {et~al.}(2021)\citenamefont {Cirac},
  \citenamefont {P{\'{e}}rez-Garc{\'{i}}a}, \citenamefont {Schuch},\ and\
  \citenamefont {Verstraete}}]{Cirac:2021gx}%
  \BibitemOpen
  \bibfield  {author} {\bibinfo {author} {\bibfnamefont {J.~I.}\ \bibnamefont
  {Cirac}}, \bibinfo {author} {\bibfnamefont {D.}~\bibnamefont
  {P{\'{e}}rez-Garc{\'{i}}a}}, \bibinfo {author} {\bibfnamefont
  {N.}~\bibnamefont {Schuch}},\ and\ \bibinfo {author} {\bibfnamefont
  {F.}~\bibnamefont {Verstraete}},\ }\bibfield  {title} {\bibinfo {title}
  {{Matrix product states and projected entangled pair states: Concepts,
  symmetries, theorems}},\ }\href
  {https://doi.org/10.1103/RevModPhys.93.045003} {\bibfield  {journal}
  {\bibinfo  {journal} {Reviews of Modern Physics}\ }\textbf {\bibinfo {volume}
  {93}},\ \bibinfo {pages} {045003} (\bibinfo {year} {2021})}\BibitemShut
  {NoStop}%
\bibitem [{\citenamefont {Ma}\ \emph {et~al.}(2023)\citenamefont {Ma},
  \citenamefont {Hanks},\ and\ \citenamefont {Kim}}]{Ma:2023ia}%
  \BibitemOpen
  \bibfield  {author} {\bibinfo {author} {\bibfnamefont {Y.}~\bibnamefont
  {Ma}}, \bibinfo {author} {\bibfnamefont {M.}~\bibnamefont {Hanks}},\ and\
  \bibinfo {author} {\bibfnamefont {M.~S.}\ \bibnamefont {Kim}},\ }\bibfield
  {title} {\bibinfo {title} {{Non–Pauli Errors Can Be Efficiently Sampled in
  Qudit Surface Codes}},\ }\href
  {https://doi.org/10.1103/PhysRevLett.131.200602} {\bibfield  {journal}
  {\bibinfo  {journal} {Physical Review Letters}\ }\textbf {\bibinfo {volume}
  {131}},\ \bibinfo {pages} {200602} (\bibinfo {year} {2023})}\BibitemShut
  {NoStop}%
\bibitem [{\citenamefont {Fowler}\ \emph {et~al.}(2012)\citenamefont {Fowler},
  \citenamefont {Mariantoni}, \citenamefont {Martinis},\ and\ \citenamefont
  {Cleland}}]{Fowler:2012fi}%
  \BibitemOpen
  \bibfield  {author} {\bibinfo {author} {\bibfnamefont {A.~G.}\ \bibnamefont
  {Fowler}}, \bibinfo {author} {\bibfnamefont {M.}~\bibnamefont {Mariantoni}},
  \bibinfo {author} {\bibfnamefont {J.~M.}\ \bibnamefont {Martinis}},\ and\
  \bibinfo {author} {\bibfnamefont {A.~N.}\ \bibnamefont {Cleland}},\
  }\bibfield  {title} {\bibinfo {title} {{Surface codes: Towards practical
  large-scale quantum computation}},\ }\href
  {https://doi.org/10.1103/PhysRevA.86.032324} {\bibfield  {journal} {\bibinfo
  {journal} {Physical Review A}\ }\textbf {\bibinfo {volume} {86}},\ \bibinfo
  {pages} {032324} (\bibinfo {year} {2012})}\BibitemShut {NoStop}%
\bibitem [{\citenamefont {Bravyi}\ \emph {et~al.}(2014)\citenamefont {Bravyi},
  \citenamefont {Suchara},\ and\ \citenamefont {Vargo}}]{Bravyi:2014ja}%
  \BibitemOpen
  \bibfield  {author} {\bibinfo {author} {\bibfnamefont {S.}~\bibnamefont
  {Bravyi}}, \bibinfo {author} {\bibfnamefont {M.}~\bibnamefont {Suchara}},\
  and\ \bibinfo {author} {\bibfnamefont {A.}~\bibnamefont {Vargo}},\ }\bibfield
   {title} {\bibinfo {title} {{Efficient algorithms for maximum likelihood
  decoding in the surface code}},\ }\href
  {https://doi.org/10.1103/PhysRevA.90.032326} {\bibfield  {journal} {\bibinfo
  {journal} {Physical Review A}\ }\textbf {\bibinfo {volume} {90}},\ \bibinfo
  {pages} {032326} (\bibinfo {year} {2014})}\BibitemShut {NoStop}%
\bibitem [{\citenamefont {Ho}\ \emph {et~al.}(2015)\citenamefont {Ho},
  \citenamefont {Cincio}, \citenamefont {Moradi}, \citenamefont {Gaiotto},\
  and\ \citenamefont {Vidal}}]{WenWei15}%
  \BibitemOpen
  \bibfield  {author} {\bibinfo {author} {\bibfnamefont {W.~W.}\ \bibnamefont
  {Ho}}, \bibinfo {author} {\bibfnamefont {L.}~\bibnamefont {Cincio}}, \bibinfo
  {author} {\bibfnamefont {H.}~\bibnamefont {Moradi}}, \bibinfo {author}
  {\bibfnamefont {D.}~\bibnamefont {Gaiotto}},\ and\ \bibinfo {author}
  {\bibfnamefont {G.}~\bibnamefont {Vidal}},\ }\bibfield  {title} {\bibinfo
  {title} {{Edge-entanglement spectrum correspondence in a nonchiral
  topological phase and Kramers-Wannier duality}},\ }\href
  {https://doi.org/10.1103/PhysRevB.91.125119} {\bibfield  {journal} {\bibinfo
  {journal} {Physical Review B}\ }\textbf {\bibinfo {volume} {91}},\ \bibinfo
  {pages} {125119} (\bibinfo {year} {2015})}\BibitemShut {NoStop}%
\bibitem [{\citenamefont {Aasen}\ \emph {et~al.}(2016)\citenamefont {Aasen},
  \citenamefont {Mong},\ and\ \citenamefont {Fendley}}]{aasen2016topological}%
  \BibitemOpen
  \bibfield  {author} {\bibinfo {author} {\bibfnamefont {D.}~\bibnamefont
  {Aasen}}, \bibinfo {author} {\bibfnamefont {R.~S.~K.}\ \bibnamefont {Mong}},\
  and\ \bibinfo {author} {\bibfnamefont {P.}~\bibnamefont {Fendley}},\
  }\bibfield  {title} {\bibinfo {title} {{Topological defects on the lattice:
  I. The Ising model}},\ }\href
  {https://doi.org/https://doi.org/10.1088/1751-8113/49/35/354001} {\bibfield
  {journal} {\bibinfo  {journal} {J. Phys. A: Math. Theor.}\ }\textbf {\bibinfo
  {volume} {49}},\ \bibinfo {pages} {354001} (\bibinfo {year}
  {2016})}\BibitemShut {NoStop}%
\bibitem [{\citenamefont {Ji}\ and\ \citenamefont
  {Wen}(2020)}]{JiWen2020categorical}%
  \BibitemOpen
  \bibfield  {author} {\bibinfo {author} {\bibfnamefont {W.}~\bibnamefont
  {Ji}}\ and\ \bibinfo {author} {\bibfnamefont {X.-G.}\ \bibnamefont {Wen}},\
  }\bibfield  {title} {\bibinfo {title} {Categorical symmetry and noninvertible
  anomaly in symmetry-breaking and topological phase transitions},\ }\href
  {https://doi.org/10.1103/PhysRevResearch.2.033417} {\bibfield  {journal}
  {\bibinfo  {journal} {Physical Review Research}\ }\textbf {\bibinfo {volume}
  {2}},\ \bibinfo {pages} {033417} (\bibinfo {year} {2020})}\BibitemShut
  {NoStop}%
\bibitem [{\citenamefont {Lichtman}\ \emph {et~al.}(2021)\citenamefont
  {Lichtman}, \citenamefont {Thorngren}, \citenamefont {Lindner}, \citenamefont
  {Stern},\ and\ \citenamefont {Berg}}]{lichtman2020bulk}%
  \BibitemOpen
  \bibfield  {author} {\bibinfo {author} {\bibfnamefont {T.}~\bibnamefont
  {Lichtman}}, \bibinfo {author} {\bibfnamefont {R.}~\bibnamefont {Thorngren}},
  \bibinfo {author} {\bibfnamefont {N.~H.}\ \bibnamefont {Lindner}}, \bibinfo
  {author} {\bibfnamefont {A.}~\bibnamefont {Stern}},\ and\ \bibinfo {author}
  {\bibfnamefont {E.}~\bibnamefont {Berg}},\ }\bibfield  {title} {\bibinfo
  {title} {{Bulk anyons as edge symmetries: Boundary phase diagrams of
  topologically ordered states}},\ }\href
  {https://doi.org/10.1103/PhysRevB.104.075141} {\bibfield  {journal} {\bibinfo
   {journal} {Physical Review B}\ }\textbf {\bibinfo {volume} {104}},\ \bibinfo
  {pages} {075141} (\bibinfo {year} {2021})}\BibitemShut {NoStop}%
\bibitem [{\citenamefont {Chatterjee}\ and\ \citenamefont
  {Wen}(2023)}]{ChatterjeeWen23}%
  \BibitemOpen
  \bibfield  {author} {\bibinfo {author} {\bibfnamefont {A.}~\bibnamefont
  {Chatterjee}}\ and\ \bibinfo {author} {\bibfnamefont {X.-G.}\ \bibnamefont
  {Wen}},\ }\bibfield  {title} {\bibinfo {title} {Symmetry as a shadow of
  topological order and a derivation of topological holographic principle},\
  }\href {https://doi.org/10.1103/PhysRevB.107.155136} {\bibfield  {journal}
  {\bibinfo  {journal} {Physical Review B}\ }\textbf {\bibinfo {volume}
  {107}},\ \bibinfo {pages} {155136} (\bibinfo {year} {2023})}\BibitemShut
  {NoStop}%
\bibitem [{\citenamefont {Freed}\ \emph {et~al.}()\citenamefont {Freed},
  \citenamefont {Moore},\ and\ \citenamefont {Teleman}}]{FreedMooreTeleman22}%
  \BibitemOpen
  \bibfield  {author} {\bibinfo {author} {\bibfnamefont {D.~S.}\ \bibnamefont
  {Freed}}, \bibinfo {author} {\bibfnamefont {G.~W.}\ \bibnamefont {Moore}},\
  and\ \bibinfo {author} {\bibfnamefont {C.}~\bibnamefont {Teleman}},\
  }\bibfield  {title} {\bibinfo {title} {Topological symmetry in quantum field
  theory},\ }\Eprint {https://arxiv.org/abs/2209.07471} {arXiv:2209.07471}
  \BibitemShut {NoStop}%
\bibitem [{\citenamefont {Moradi}\ \emph {et~al.}(2023)\citenamefont {Moradi},
  \citenamefont {Moosavian},\ and\ \citenamefont {Tiwari}}]{TH23}%
  \BibitemOpen
  \bibfield  {author} {\bibinfo {author} {\bibfnamefont {H.}~\bibnamefont
  {Moradi}}, \bibinfo {author} {\bibfnamefont {S.~F.}\ \bibnamefont
  {Moosavian}},\ and\ \bibinfo {author} {\bibfnamefont {A.}~\bibnamefont
  {Tiwari}},\ }\bibfield  {title} {\bibinfo {title} {{Topological holography:
  Towards a unification of Landau and beyond-Landau physics}},\ }\href
  {https://doi.org/10.21468/SciPostPhysCore.6.4.066} {\bibfield  {journal}
  {\bibinfo  {journal} {SciPost Phys. Core}\ }\textbf {\bibinfo {volume} {6}},\
  \bibinfo {pages} {066} (\bibinfo {year} {2023})}\BibitemShut {NoStop}%
\bibitem [{\citenamefont {Bhardwaj}\ and\ \citenamefont
  {Schafer-Nameki}()}]{Bhardwaj_PtII}%
  \BibitemOpen
  \bibfield  {author} {\bibinfo {author} {\bibfnamefont {L.}~\bibnamefont
  {Bhardwaj}}\ and\ \bibinfo {author} {\bibfnamefont {S.}~\bibnamefont
  {Schafer-Nameki}},\ }\bibfield  {title} {\bibinfo {title} {{Generalized
  Charges, Part II: Non-Invertible Symmetries and the Symmetry TFT}},\ }\Eprint
  {https://arxiv.org/abs/2305.17159} {arXiv:2305.17159} \BibitemShut {NoStop}%
\bibitem [{\citenamefont {Lootens}\ \emph {et~al.}(2023)\citenamefont
  {Lootens}, \citenamefont {Delcamp}, \citenamefont {Ortiz},\ and\
  \citenamefont {Verstraete}}]{Lootens_PRXQuantum.4.020357}%
  \BibitemOpen
  \bibfield  {author} {\bibinfo {author} {\bibfnamefont {L.}~\bibnamefont
  {Lootens}}, \bibinfo {author} {\bibfnamefont {C.}~\bibnamefont {Delcamp}},
  \bibinfo {author} {\bibfnamefont {G.}~\bibnamefont {Ortiz}},\ and\ \bibinfo
  {author} {\bibfnamefont {F.}~\bibnamefont {Verstraete}},\ }\bibfield  {title}
  {\bibinfo {title} {{Dualities in One-Dimensional Quantum Lattice Models:
  Symmetric Hamiltonians and Matrix Product Operator Intertwiners}},\ }\href
  {https://doi.org/10.1103/PRXQuantum.4.020357} {\bibfield  {journal} {\bibinfo
   {journal} {PRX Quantum}\ }\textbf {\bibinfo {volume} {4}},\ \bibinfo {pages}
  {020357} (\bibinfo {year} {2023})}\BibitemShut {NoStop}%
\bibitem [{\citenamefont {Lootens}\ \emph {et~al.}(2024)\citenamefont
  {Lootens}, \citenamefont {Delcamp},\ and\ \citenamefont
  {Verstraete}}]{Lootens_PRXQuantum.5.010338}%
  \BibitemOpen
  \bibfield  {author} {\bibinfo {author} {\bibfnamefont {L.}~\bibnamefont
  {Lootens}}, \bibinfo {author} {\bibfnamefont {C.}~\bibnamefont {Delcamp}},\
  and\ \bibinfo {author} {\bibfnamefont {F.}~\bibnamefont {Verstraete}},\
  }\bibfield  {title} {\bibinfo {title} {{Dualities in One-Dimensional Quantum
  Lattice Models: Topological Sectors}},\ }\href
  {https://doi.org/10.1103/PRXQuantum.5.010338} {\bibfield  {journal} {\bibinfo
   {journal} {PRX Quantum}\ }\textbf {\bibinfo {volume} {5}},\ \bibinfo {pages}
  {010338} (\bibinfo {year} {2024})}\BibitemShut {NoStop}%
\bibitem [{\citenamefont {Bhardwaj}\ \emph {et~al.}(2024)\citenamefont
  {Bhardwaj}, \citenamefont {Bottini}, \citenamefont {Pajer},\ and\
  \citenamefont {Sch\"afer-Nameki}}]{Bhardwaj_PhysRevLett.133.161601}%
  \BibitemOpen
  \bibfield  {author} {\bibinfo {author} {\bibfnamefont {L.}~\bibnamefont
  {Bhardwaj}}, \bibinfo {author} {\bibfnamefont {L.~E.}\ \bibnamefont
  {Bottini}}, \bibinfo {author} {\bibfnamefont {D.}~\bibnamefont {Pajer}},\
  and\ \bibinfo {author} {\bibfnamefont {S.}~\bibnamefont {Sch\"afer-Nameki}},\
  }\bibfield  {title} {\bibinfo {title} {{Categorical Landau Paradigm for
  Gapped Phases}},\ }\href {https://doi.org/10.1103/PhysRevLett.133.161601}
  {\bibfield  {journal} {\bibinfo  {journal} {Physical Review Letters}\
  }\textbf {\bibinfo {volume} {133}},\ \bibinfo {pages} {161601} (\bibinfo
  {year} {2024})}\BibitemShut {NoStop}%
\bibitem [{\citenamefont {Fechisin}\ \emph {et~al.}(2025)\citenamefont
  {Fechisin}, \citenamefont {Tantivasadakarn},\ and\ \citenamefont
  {Albert}}]{Fechisin_PhysRevX.15.011058}%
  \BibitemOpen
  \bibfield  {author} {\bibinfo {author} {\bibfnamefont {C.}~\bibnamefont
  {Fechisin}}, \bibinfo {author} {\bibfnamefont {N.}~\bibnamefont
  {Tantivasadakarn}},\ and\ \bibinfo {author} {\bibfnamefont {V.~V.}\
  \bibnamefont {Albert}},\ }\bibfield  {title} {\bibinfo {title}
  {{Noninvertible Symmetry-Protected Topological Order in a Group-Based Cluster
  State}},\ }\href {https://doi.org/10.1103/PhysRevX.15.011058} {\bibfield
  {journal} {\bibinfo  {journal} {Phys. Rev. X}\ }\textbf {\bibinfo {volume}
  {15}},\ \bibinfo {pages} {011058} (\bibinfo {year} {2025})}\BibitemShut
  {NoStop}%
\bibitem [{\citenamefont {Barbar}\ \emph {et~al.}(2025)\citenamefont {Barbar},
  \citenamefont {Dymarsky},\ and\ \citenamefont
  {Shapere}}]{Barbar_PhysRevLett.134.151603}%
  \BibitemOpen
  \bibfield  {author} {\bibinfo {author} {\bibfnamefont {A.}~\bibnamefont
  {Barbar}}, \bibinfo {author} {\bibfnamefont {A.}~\bibnamefont {Dymarsky}},\
  and\ \bibinfo {author} {\bibfnamefont {A.~D.}\ \bibnamefont {Shapere}},\
  }\bibfield  {title} {\bibinfo {title} {{Global Symmetries, Code Ensembles,
  and Sums over Geometries}},\ }\href
  {https://doi.org/10.1103/PhysRevLett.134.151603} {\bibfield  {journal}
  {\bibinfo  {journal} {Phys. Rev. Lett.}\ }\textbf {\bibinfo {volume} {134}},\
  \bibinfo {pages} {151603} (\bibinfo {year} {2025})}\BibitemShut {NoStop}%
\bibitem [{\citenamefont {Sun}\ \emph {et~al.}(2025)\citenamefont {Sun},
  \citenamefont {Zhang}, \citenamefont {Bi},\ and\ \citenamefont
  {You}}]{Sun_PRXQuantum.6.020333}%
  \BibitemOpen
  \bibfield  {author} {\bibinfo {author} {\bibfnamefont {S.}~\bibnamefont
  {Sun}}, \bibinfo {author} {\bibfnamefont {J.-H.}\ \bibnamefont {Zhang}},
  \bibinfo {author} {\bibfnamefont {Z.}~\bibnamefont {Bi}},\ and\ \bibinfo
  {author} {\bibfnamefont {Y.}~\bibnamefont {You}},\ }\bibfield  {title}
  {\bibinfo {title} {{Holographic View of Mixed-State Symmetry-Protected
  Topological Phases in Open Quantum Systems}},\ }\href
  {https://doi.org/10.1103/PRXQuantum.6.020333} {\bibfield  {journal} {\bibinfo
   {journal} {PRX Quantum}\ }\textbf {\bibinfo {volume} {6}},\ \bibinfo {pages}
  {020333} (\bibinfo {year} {2025})}\BibitemShut {NoStop}%
\bibitem [{\citenamefont {Bottini}\ and\ \citenamefont
  {Sch\"afer-Nameki}(2025)}]{Bottini_PhysRevLett.134.191602}%
  \BibitemOpen
  \bibfield  {author} {\bibinfo {author} {\bibfnamefont {L.~E.}\ \bibnamefont
  {Bottini}}\ and\ \bibinfo {author} {\bibfnamefont {S.}~\bibnamefont
  {Sch\"afer-Nameki}},\ }\bibfield  {title} {\bibinfo {title} {{Construction of
  a Gapless Phase with Haagerup Symmetry}},\ }\href
  {https://doi.org/10.1103/PhysRevLett.134.191602} {\bibfield  {journal}
  {\bibinfo  {journal} {Phys. Rev. Lett.}\ }\textbf {\bibinfo {volume} {134}},\
  \bibinfo {pages} {191602} (\bibinfo {year} {2025})}\BibitemShut {NoStop}%
\bibitem [{\citenamefont {Seifnashri}\ and\ \citenamefont
  {Shao}(2024)}]{Seifnashri_PhysRevLett.133.116601}%
  \BibitemOpen
  \bibfield  {author} {\bibinfo {author} {\bibfnamefont {S.}~\bibnamefont
  {Seifnashri}}\ and\ \bibinfo {author} {\bibfnamefont {S.-H.}\ \bibnamefont
  {Shao}},\ }\bibfield  {title} {\bibinfo {title} {{Cluster State as a
  Noninvertible Symmetry-Protected Topological Phase}},\ }\href
  {https://doi.org/10.1103/PhysRevLett.133.116601} {\bibfield  {journal}
  {\bibinfo  {journal} {Phys. Rev. Lett.}\ }\textbf {\bibinfo {volume} {133}},\
  \bibinfo {pages} {116601} (\bibinfo {year} {2024})}\BibitemShut {NoStop}%
\bibitem [{\citenamefont {Okada}\ and\ \citenamefont
  {Tachikawa}(2024)}]{Okada_PhysRevLett.133.191602}%
  \BibitemOpen
  \bibfield  {author} {\bibinfo {author} {\bibfnamefont {M.}~\bibnamefont
  {Okada}}\ and\ \bibinfo {author} {\bibfnamefont {Y.}~\bibnamefont
  {Tachikawa}},\ }\bibfield  {title} {\bibinfo {title} {{Noninvertible
  Symmetries Act Locally by Quantum Operations}},\ }\href
  {https://doi.org/10.1103/PhysRevLett.133.191602} {\bibfield  {journal}
  {\bibinfo  {journal} {Phys. Rev. Lett.}\ }\textbf {\bibinfo {volume} {133}},\
  \bibinfo {pages} {191602} (\bibinfo {year} {2024})}\BibitemShut {NoStop}%
\bibitem [{\citenamefont {Motamarri}\ \emph {et~al.}()\citenamefont
  {Motamarri}, \citenamefont {McLauchlan},\ and\ \citenamefont
  {B{\'{e}}ri}}]{SymTFT_TC}%
  \BibitemOpen
  \bibfield  {author} {\bibinfo {author} {\bibfnamefont {V.}~\bibnamefont
  {Motamarri}}, \bibinfo {author} {\bibfnamefont {C.}~\bibnamefont
  {McLauchlan}},\ and\ \bibinfo {author} {\bibfnamefont {B.}~\bibnamefont
  {B{\'{e}}ri}},\ }\bibfield  {title} {\bibinfo {title} {{SymTFT out of
  equilibrium: from time crystals to braided drives and Floquet codes}},\
  }\Eprint {https://arxiv.org/abs/2312.17176} {arXiv:2312.17176} \BibitemShut
  {NoStop}%
\bibitem [{\citenamefont {Bombin}\ and\ \citenamefont
  {Martin-Delgado}(2007)}]{Bombin:2007ed}%
  \BibitemOpen
  \bibfield  {author} {\bibinfo {author} {\bibfnamefont {H.}~\bibnamefont
  {Bombin}}\ and\ \bibinfo {author} {\bibfnamefont {M.~A.}\ \bibnamefont
  {Martin-Delgado}},\ }\bibfield  {title} {\bibinfo {title} {{Optimal resources
  for topological two-dimensional stabilizer codes: Comparative study}},\
  }\href {https://doi.org/10.1103/PhysRevA.76.012305} {\bibfield  {journal}
  {\bibinfo  {journal} {Physical Review A}\ }\textbf {\bibinfo {volume} {76}},\
  \bibinfo {pages} {012305} (\bibinfo {year} {2007})}\BibitemShut {NoStop}%
\bibitem [{\citenamefont {Horsman}\ \emph {et~al.}(2012)\citenamefont
  {Horsman}, \citenamefont {Fowler}, \citenamefont {Devitt},\ and\
  \citenamefont {Meter}}]{Horsman:2012ga}%
  \BibitemOpen
  \bibfield  {author} {\bibinfo {author} {\bibfnamefont {D.}~\bibnamefont
  {Horsman}}, \bibinfo {author} {\bibfnamefont {A.~G.}\ \bibnamefont {Fowler}},
  \bibinfo {author} {\bibfnamefont {S.}~\bibnamefont {Devitt}},\ and\ \bibinfo
  {author} {\bibfnamefont {R.~V.}\ \bibnamefont {Meter}},\ }\bibfield  {title}
  {\bibinfo {title} {{Surface code quantum computing by lattice surgery}},\
  }\href {https://doi.org/10.1088/1367-2630/14/12/123011} {\bibfield  {journal}
  {\bibinfo  {journal} {New Journal of Physics}\ }\textbf {\bibinfo {volume}
  {14}},\ \bibinfo {pages} {123011} (\bibinfo {year} {2012})}\BibitemShut
  {NoStop}%
\bibitem [{\citenamefont {Eckstein}\ \emph {et~al.}(2024)\citenamefont
  {Eckstein}, \citenamefont {Han}, \citenamefont {Trebst},\ and\ \citenamefont
  {Zhu}}]{Eckstein:2024ev}%
  \BibitemOpen
  \bibfield  {author} {\bibinfo {author} {\bibfnamefont {F.}~\bibnamefont
  {Eckstein}}, \bibinfo {author} {\bibfnamefont {B.}~\bibnamefont {Han}},
  \bibinfo {author} {\bibfnamefont {S.}~\bibnamefont {Trebst}},\ and\ \bibinfo
  {author} {\bibfnamefont {G.-Y.}\ \bibnamefont {Zhu}},\ }\bibfield  {title}
  {\bibinfo {title} {{Robust Teleportation of a Surface Code and Cascade of
  Topological Quantum Phase Transitions}},\ }\href
  {https://doi.org/10.1103/PRXQuantum.5.040313} {\bibfield  {journal} {\bibinfo
   {journal} {PRX Quantum}\ }\textbf {\bibinfo {volume} {5}},\ \bibinfo {pages}
  {040313} (\bibinfo {year} {2024})}\BibitemShut {NoStop}%
\bibitem [{Note1()}]{Note1}%
  \BibitemOpen
  \bibinfo {note} {In the statistical-mechanics language, this equals the
  typical disorder correlator~\cite {Kadanoff:1971fm,Read:2000hu,Merz:2002ia},
  i.e., the typical energy cost of flipping bonds along the logical $X_\protect
  \text {L}$.}\BibitemShut {Stop}%
\bibitem [{\citenamefont {Merz}\ and\ \citenamefont
  {Chalker}(2002{\natexlab{b}})}]{Merz:2002ia}%
  \BibitemOpen
  \bibfield  {author} {\bibinfo {author} {\bibfnamefont {F.}~\bibnamefont
  {Merz}}\ and\ \bibinfo {author} {\bibfnamefont {J.~T.}\ \bibnamefont
  {Chalker}},\ }\bibfield  {title} {\bibinfo {title} {{Negative scaling
  dimensions and conformal invariance at the Nishimori point in the $\pm J$
  random-bond Ising model}},\ }\href
  {https://doi.org/10.1103/PhysRevB.66.054413} {\bibfield  {journal} {\bibinfo
  {journal} {Physical Review B}\ }\textbf {\bibinfo {volume} {66}},\ \bibinfo
  {pages} {054413} (\bibinfo {year} {2002}{\natexlab{b}})}\BibitemShut
  {NoStop}%
\bibitem [{\citenamefont {Darmawan}()}]{Darmawan2024}%
  \BibitemOpen
  \bibfield  {author} {\bibinfo {author} {\bibfnamefont {A.~S.}\ \bibnamefont
  {Darmawan}},\ }\bibfield  {title} {\bibinfo {title} {{Optimal adaptation of
  surface-code decoders to local noise}},\ }\Eprint
  {https://arxiv.org/abs/2403.08706} {arXiv:2403.08706} \BibitemShut {NoStop}%
\bibitem [{\citenamefont {Cheng}\ \emph {et~al.}()\citenamefont {Cheng},
  \citenamefont {Huang}, \citenamefont {Khemani}, \citenamefont {Gullans},\
  and\ \citenamefont {Ippoliti}}]{Cheng2024}%
  \BibitemOpen
  \bibfield  {author} {\bibinfo {author} {\bibfnamefont {Z.}~\bibnamefont
  {Cheng}}, \bibinfo {author} {\bibfnamefont {E.}~\bibnamefont {Huang}},
  \bibinfo {author} {\bibfnamefont {V.}~\bibnamefont {Khemani}}, \bibinfo
  {author} {\bibfnamefont {M.~J.}\ \bibnamefont {Gullans}},\ and\ \bibinfo
  {author} {\bibfnamefont {M.}~\bibnamefont {Ippoliti}},\ }\bibfield  {title}
  {\bibinfo {title} {{Emergent unitary designs for encoded qubits from coherent
  errors and syndrome measurements}},\ }\Eprint
  {https://arxiv.org/abs/2412.04414} {arXiv:2412.04414} \BibitemShut {NoStop}%
\bibitem [{\citenamefont {Gullans}\ and\ \citenamefont
  {Huse}(2020)}]{Gullans:2020eg}%
  \BibitemOpen
  \bibfield  {author} {\bibinfo {author} {\bibfnamefont {M.~J.}\ \bibnamefont
  {Gullans}}\ and\ \bibinfo {author} {\bibfnamefont {D.~A.}\ \bibnamefont
  {Huse}},\ }\bibfield  {title} {\bibinfo {title} {{Dynamical Purification
  Phase Transition Induced by Quantum Measurements}},\ }\href
  {https://doi.org/10.1103/PhysRevX.10.041020} {\bibfield  {journal} {\bibinfo
  {journal} {Physical Review X}\ }\textbf {\bibinfo {volume} {10}},\ \bibinfo
  {pages} {041020} (\bibinfo {year} {2020})}\BibitemShut {NoStop}%
\bibitem [{\citenamefont {Brand{\~{a}}o}\ and\ \citenamefont
  {Horodecki}(2013)}]{Brandao:2013fz}%
  \BibitemOpen
  \bibfield  {author} {\bibinfo {author} {\bibfnamefont {F.~G. S.~L.}\
  \bibnamefont {Brand{\~{a}}o}}\ and\ \bibinfo {author} {\bibfnamefont
  {M.}~\bibnamefont {Horodecki}},\ }\bibfield  {title} {\bibinfo {title} {{An
  area law for entanglement from exponential decay of correlations}},\ }\href
  {https://doi.org/10.1038/nphys2747} {\bibfield  {journal} {\bibinfo
  {journal} {Nature Physics}\ }\textbf {\bibinfo {volume} {9}},\ \bibinfo
  {pages} {721} (\bibinfo {year} {2013})}\BibitemShut {NoStop}%
\bibitem [{\citenamefont {Brand{\~{a}}o}\ and\ \citenamefont
  {Horodecki}(2015)}]{Brandao:2015cn}%
  \BibitemOpen
  \bibfield  {author} {\bibinfo {author} {\bibfnamefont {F.~G. S.~L.}\
  \bibnamefont {Brand{\~{a}}o}}\ and\ \bibinfo {author} {\bibfnamefont
  {M.}~\bibnamefont {Horodecki}},\ }\bibfield  {title} {\bibinfo {title}
  {{Exponential Decay of Correlations Implies Area Law}},\ }\href
  {https://doi.org/10.1007/s00220-014-2213-8} {\bibfield  {journal} {\bibinfo
  {journal} {Communications in Mathematical Physics}\ }\textbf {\bibinfo
  {volume} {333}},\ \bibinfo {pages} {761} (\bibinfo {year}
  {2015})}\BibitemShut {NoStop}%
\bibitem [{\citenamefont {Cho}(2018)}]{Cho_PhysRevX.8.031009}%
  \BibitemOpen
  \bibfield  {author} {\bibinfo {author} {\bibfnamefont {J.}~\bibnamefont
  {Cho}},\ }\bibfield  {title} {\bibinfo {title} {{Realistic Area-Law Bound on
  Entanglement from Exponentially Decaying Correlations}},\ }\href
  {https://doi.org/10.1103/PhysRevX.8.031009} {\bibfield  {journal} {\bibinfo
  {journal} {Phys. Rev. X}\ }\textbf {\bibinfo {volume} {8}},\ \bibinfo {pages}
  {031009} (\bibinfo {year} {2018})}\BibitemShut {NoStop}%
\bibitem [{\citenamefont {Higgott}()}]{pymatchingv1}%
  \BibitemOpen
  \bibfield  {author} {\bibinfo {author} {\bibfnamefont {O.}~\bibnamefont
  {Higgott}},\ }\bibfield  {title} {\bibinfo {title} {Pymatching: A python
  package for decoding quantum codes with minimum-weight perfect matching},\
  }\Eprint {https://arxiv.org/abs/2105.13082} {arXiv:2105.13082} \BibitemShut
  {NoStop}%
\bibitem [{\citenamefont {Higgott}\ and\ \citenamefont
  {Gidney}(2022)}]{pymatchingv2}%
  \BibitemOpen
  \bibfield  {author} {\bibinfo {author} {\bibfnamefont {O.}~\bibnamefont
  {Higgott}}\ and\ \bibinfo {author} {\bibfnamefont {C.}~\bibnamefont
  {Gidney}},\ }\href@noop {} {\bibinfo {title} {Pymatching v2}},\ \bibinfo
  {howpublished} {\url{https://github.com/oscarhiggott/PyMatching}} (\bibinfo
  {year} {2022})\BibitemShut {NoStop}%
\bibitem [{\citenamefont {Baranger}(1958)}]{Baranger_PhysRev.111.494}%
  \BibitemOpen
  \bibfield  {author} {\bibinfo {author} {\bibfnamefont {M.}~\bibnamefont
  {Baranger}},\ }\bibfield  {title} {\bibinfo {title} {{Problem of Overlapping
  Lines in the Theory of Pressure Broadening}},\ }\href
  {https://doi.org/10.1103/PhysRev.111.494} {\bibfield  {journal} {\bibinfo
  {journal} {Phys. Rev.}\ }\textbf {\bibinfo {volume} {111}},\ \bibinfo {pages}
  {494} (\bibinfo {year} {1958})}\BibitemShut {NoStop}%
\bibitem [{\citenamefont {Bengtsson}\ and\ \citenamefont
  {{\.Z}yczkowski}(2006)}]{bengtsson20006geometry}%
  \BibitemOpen
  \bibfield  {author} {\bibinfo {author} {\bibfnamefont {I.}~\bibnamefont
  {Bengtsson}}\ and\ \bibinfo {author} {\bibfnamefont {K.}~\bibnamefont
  {{\.Z}yczkowski}},\ }\href {https://doi.org/10.1017/CBO9780511535048} {\emph
  {\bibinfo {title} {{Geometry of Quantum States: An Introduction to Quantum
  Entanglement}}}}\ (\bibinfo  {publisher} {Cambridge University Press},\
  \bibinfo {address} {Cambridge, U.K.},\ \bibinfo {year} {2006})\BibitemShut
  {NoStop}%
\bibitem [{\citenamefont {Gilchrist}\ \emph {et~al.}()\citenamefont
  {Gilchrist}, \citenamefont {Terno},\ and\ \citenamefont
  {Wood}}]{Gilchrist_vec}%
  \BibitemOpen
  \bibfield  {author} {\bibinfo {author} {\bibfnamefont {A.}~\bibnamefont
  {Gilchrist}}, \bibinfo {author} {\bibfnamefont {D.~R.}\ \bibnamefont
  {Terno}},\ and\ \bibinfo {author} {\bibfnamefont {C.~J.}\ \bibnamefont
  {Wood}},\ }\bibfield  {title} {\bibinfo {title} {{Vectorization of quantum
  operations and its use}},\ }\Eprint {https://arxiv.org/abs/0911.2539}
  {arXiv:0911.2539} \BibitemShut {NoStop}%
\bibitem [{\citenamefont {Bais}\ and\ \citenamefont
  {Slingerland}(2009)}]{baisCondensateinduced2009}%
  \BibitemOpen
  \bibfield  {author} {\bibinfo {author} {\bibfnamefont {F.~A.}\ \bibnamefont
  {Bais}}\ and\ \bibinfo {author} {\bibfnamefont {J.~K.}\ \bibnamefont
  {Slingerland}},\ }\bibfield  {title} {\bibinfo {title} {Condensate-induced
  transitions between topologically ordered phases},\ }\href
  {https://doi.org/10.1103/PhysRevB.79.045316} {\bibfield  {journal} {\bibinfo
  {journal} {Physical Review B}\ }\textbf {\bibinfo {volume} {79}},\ \bibinfo
  {pages} {045316} (\bibinfo {year} {2009})}\BibitemShut {NoStop}%
\bibitem [{\citenamefont {Kong}(2014)}]{kongAnyon2014}%
  \BibitemOpen
  \bibfield  {author} {\bibinfo {author} {\bibfnamefont {L.}~\bibnamefont
  {Kong}},\ }\bibfield  {title} {\bibinfo {title} {Anyon condensation and
  tensor categories},\ }\href {https://doi.org/10.1016/j.nuclphysb.2014.07.003}
  {\bibfield  {journal} {\bibinfo  {journal} {Nuclear Physics B}\ }\textbf
  {\bibinfo {volume} {886}},\ \bibinfo {pages} {436} (\bibinfo {year}
  {2014})}\BibitemShut {NoStop}%
\bibitem [{\citenamefont {Burnell}(2018)}]{burnellAnyon2018}%
  \BibitemOpen
  \bibfield  {author} {\bibinfo {author} {\bibfnamefont {F.}~\bibnamefont
  {Burnell}},\ }\bibfield  {title} {\bibinfo {title} {Anyon {{Condensation}}
  and {{Its Applications}}},\ }\href
  {https://doi.org/10.1146/annurev-conmatphys-033117-054154} {\bibfield
  {journal} {\bibinfo  {journal} {Annual Review of Condensed Matter Physics}\
  }\textbf {\bibinfo {volume} {9}},\ \bibinfo {pages} {307} (\bibinfo {year}
  {2018})}\BibitemShut {NoStop}%
\bibitem [{\citenamefont {Kesselring}\ \emph {et~al.}(2024)\citenamefont
  {Kesselring}, \citenamefont {{Magdalena de la Fuente}}, \citenamefont
  {Thomsen}, \citenamefont {Eisert}, \citenamefont {Bartlett},\ and\
  \citenamefont {Brown}}]{kesselringAnyon2024}%
  \BibitemOpen
  \bibfield  {author} {\bibinfo {author} {\bibfnamefont {M.~S.}\ \bibnamefont
  {Kesselring}}, \bibinfo {author} {\bibfnamefont {J.~C.}\ \bibnamefont
  {{Magdalena de la Fuente}}}, \bibinfo {author} {\bibfnamefont
  {F.}~\bibnamefont {Thomsen}}, \bibinfo {author} {\bibfnamefont
  {J.}~\bibnamefont {Eisert}}, \bibinfo {author} {\bibfnamefont {S.~D.}\
  \bibnamefont {Bartlett}},\ and\ \bibinfo {author} {\bibfnamefont {B.~J.}\
  \bibnamefont {Brown}},\ }\bibfield  {title} {\bibinfo {title} {Anyon
  {{Condensation}} and the {{Color Code}}},\ }\href
  {https://doi.org/10.1103/PRXQuantum.5.010342} {\bibfield  {journal} {\bibinfo
   {journal} {PRX Quantum}\ }\textbf {\bibinfo {volume} {5}},\ \bibinfo {pages}
  {010342} (\bibinfo {year} {2024})}\BibitemShut {NoStop}%
\bibitem [{\citenamefont {Nayak}\ \emph {et~al.}(2008)\citenamefont {Nayak},
  \citenamefont {Simon}, \citenamefont {Stern}, \citenamefont {Freedman},\ and\
  \citenamefont {Das~Sarma}}]{Nayak_RevModPhys.80.1083}%
  \BibitemOpen
  \bibfield  {author} {\bibinfo {author} {\bibfnamefont {C.}~\bibnamefont
  {Nayak}}, \bibinfo {author} {\bibfnamefont {S.~H.}\ \bibnamefont {Simon}},
  \bibinfo {author} {\bibfnamefont {A.}~\bibnamefont {Stern}}, \bibinfo
  {author} {\bibfnamefont {M.}~\bibnamefont {Freedman}},\ and\ \bibinfo
  {author} {\bibfnamefont {S.}~\bibnamefont {Das~Sarma}},\ }\bibfield  {title}
  {\bibinfo {title} {{Non-Abelian anyons and topological quantum
  computation}},\ }\href {https://doi.org/10.1103/RevModPhys.80.1083}
  {\bibfield  {journal} {\bibinfo  {journal} {Reviews of Modern Physics}\
  }\textbf {\bibinfo {volume} {80}},\ \bibinfo {pages} {1083} (\bibinfo {year}
  {2008})}\BibitemShut {NoStop}%
\bibitem [{\citenamefont {Sachdev}(2011)}]{sachdev_2011}%
  \BibitemOpen
  \bibfield  {author} {\bibinfo {author} {\bibfnamefont {S.}~\bibnamefont
  {Sachdev}},\ }\href {https://doi.org/10.1017/CBO9780511973765} {\emph
  {\bibinfo {title} {Quantum Phase Transitions}}},\ \bibinfo {edition} {2nd}\
  ed.\ (\bibinfo  {publisher} {Cambridge University Press},\ \bibinfo {address}
  {Cambridge, U.K.},\ \bibinfo {year} {2011})\BibitemShut {NoStop}%
\bibitem [{\citenamefont {Fradkin}(2013)}]{fradkin2013field}%
  \BibitemOpen
  \bibfield  {author} {\bibinfo {author} {\bibfnamefont {E.}~\bibnamefont
  {Fradkin}},\ }\href {https://doi.org/CBO9781139015509} {\emph {\bibinfo
  {title} {Field theories of condensed matter physics}}}\ (\bibinfo
  {publisher} {Cambridge University Press},\ \bibinfo {address} {Cambridge,
  U.K.},\ \bibinfo {year} {2013})\BibitemShut {NoStop}%
\bibitem [{\citenamefont {Lee}\ and\ \citenamefont
  {Chan}(2014)}]{Lee_PhysRevX.4.041001}%
  \BibitemOpen
  \bibfield  {author} {\bibinfo {author} {\bibfnamefont {T.~E.}\ \bibnamefont
  {Lee}}\ and\ \bibinfo {author} {\bibfnamefont {C.-K.}\ \bibnamefont {Chan}},\
  }\bibfield  {title} {\bibinfo {title} {{Heralded Magnetism in Non-Hermitian
  Atomic Systems}},\ }\href {https://doi.org/10.1103/PhysRevX.4.041001}
  {\bibfield  {journal} {\bibinfo  {journal} {Phys. Rev. X}\ }\textbf {\bibinfo
  {volume} {4}},\ \bibinfo {pages} {041001} (\bibinfo {year}
  {2014})}\BibitemShut {NoStop}%
\bibitem [{\citenamefont {Biella}\ and\ \citenamefont
  {Schir{\'{o}}}(2021)}]{Biella2021manybodyquantumzeno}%
  \BibitemOpen
  \bibfield  {author} {\bibinfo {author} {\bibfnamefont {A.}~\bibnamefont
  {Biella}}\ and\ \bibinfo {author} {\bibfnamefont {M.}~\bibnamefont
  {Schir{\'{o}}}},\ }\bibfield  {title} {\bibinfo {title} {Many-{B}ody
  {Q}uantum {Z}eno {E}ffect and {M}easurement-{I}nduced {S}ubradiance
  {T}ransition},\ }\href {https://doi.org/10.22331/q-2021-08-19-528} {\bibfield
   {journal} {\bibinfo  {journal} {{Quantum}}\ }\textbf {\bibinfo {volume}
  {5}},\ \bibinfo {pages} {528} (\bibinfo {year} {2021})}\BibitemShut {NoStop}%
\bibitem [{\citenamefont {Turkeshi}\ \emph {et~al.}(2021)\citenamefont
  {Turkeshi}, \citenamefont {Biella}, \citenamefont {Fazio}, \citenamefont
  {Dalmonte},\ and\ \citenamefont
  {Schir{\'{o}}}}]{Turkeshi_PhysRevB.103.224210}%
  \BibitemOpen
  \bibfield  {author} {\bibinfo {author} {\bibfnamefont {X.}~\bibnamefont
  {Turkeshi}}, \bibinfo {author} {\bibfnamefont {A.}~\bibnamefont {Biella}},
  \bibinfo {author} {\bibfnamefont {R.}~\bibnamefont {Fazio}}, \bibinfo
  {author} {\bibfnamefont {M.}~\bibnamefont {Dalmonte}},\ and\ \bibinfo
  {author} {\bibfnamefont {M.}~\bibnamefont {Schir{\'{o}}}},\ }\bibfield
  {title} {\bibinfo {title} {{Measurement-induced entanglement transitions in
  the quantum Ising chain: From infinite to zero clicks}},\ }\href
  {https://doi.org/10.1103/PhysRevB.103.224210} {\bibfield  {journal} {\bibinfo
   {journal} {Physical Review B}\ }\textbf {\bibinfo {volume} {103}},\ \bibinfo
  {pages} {224210} (\bibinfo {year} {2021})}\BibitemShut {NoStop}%
\bibitem [{\citenamefont {Yan}\ \emph {et~al.}()\citenamefont {Yan},
  \citenamefont {Barberena}, \citenamefont {Fisher},\ and\ \citenamefont
  {Vijay}}]{yan2024dissipativedynamicalphasetransition}%
  \BibitemOpen
  \bibfield  {author} {\bibinfo {author} {\bibfnamefont {S.~W.}\ \bibnamefont
  {Yan}}, \bibinfo {author} {\bibfnamefont {D.}~\bibnamefont {Barberena}},
  \bibinfo {author} {\bibfnamefont {M.~P.~A.}\ \bibnamefont {Fisher}},\ and\
  \bibinfo {author} {\bibfnamefont {S.}~\bibnamefont {Vijay}},\ }\bibfield
  {title} {\bibinfo {title} {{Dissipative Dynamical Phase Transition as a
  Complex Ising Model}},\ }\Eprint {https://arxiv.org/abs/2412.09591}
  {arXiv:2412.09591} \BibitemShut {NoStop}%
\bibitem [{\citenamefont {Wen}\ and\ \citenamefont {Zee}(1992)}]{WenZee}%
  \BibitemOpen
  \bibfield  {author} {\bibinfo {author} {\bibfnamefont {X.~G.}\ \bibnamefont
  {Wen}}\ and\ \bibinfo {author} {\bibfnamefont {A.}~\bibnamefont {Zee}},\
  }\bibfield  {title} {\bibinfo {title} {{Classification of Abelian quantum
  Hall states and matrix formulation of topological fluids}},\ }\href
  {https://doi.org/10.1103/PhysRevB.46.2290} {\bibfield  {journal} {\bibinfo
  {journal} {Physical Review B}\ }\textbf {\bibinfo {volume} {46}},\ \bibinfo
  {pages} {2290} (\bibinfo {year} {1992})}\BibitemShut {NoStop}%
\bibitem [{\citenamefont {Wen}(1995)}]{wen1995topological}%
  \BibitemOpen
  \bibfield  {author} {\bibinfo {author} {\bibfnamefont {X.-G.}\ \bibnamefont
  {Wen}},\ }\bibfield  {title} {\bibinfo {title} {{Topological orders and edge
  excitations in fractional quantum Hall states}},\ }\href
  {https://doi.org/10.1080/00018739500101566} {\bibfield  {journal} {\bibinfo
  {journal} {Advances in Physics}\ }\textbf {\bibinfo {volume} {44}},\ \bibinfo
  {pages} {405} (\bibinfo {year} {1995})}\BibitemShut {NoStop}%
\bibitem [{\citenamefont {Haldane}(1995)}]{Haldane_PhysRevLett.74.2090}%
  \BibitemOpen
  \bibfield  {author} {\bibinfo {author} {\bibfnamefont {F.~D.~M.}\
  \bibnamefont {Haldane}},\ }\bibfield  {title} {\bibinfo {title} {{Stability
  of Chiral Luttinger Liquids and Abelian Quantum Hall States}},\ }\href
  {https://doi.org/10.1103/PhysRevLett.74.2090} {\bibfield  {journal} {\bibinfo
   {journal} {Physical Review Letters}\ }\textbf {\bibinfo {volume} {74}},\
  \bibinfo {pages} {2090} (\bibinfo {year} {1995})}\BibitemShut {NoStop}%
\bibitem [{\citenamefont {Wen}(2007)}]{wen2004quantum}%
  \BibitemOpen
  \bibfield  {author} {\bibinfo {author} {\bibfnamefont {X.-G.}\ \bibnamefont
  {Wen}},\ }\href {https://doi.org/10.1093/acprof:oso/9780199227259.001.0001}
  {\emph {\bibinfo {title} {{Quantum Field Theory of Many-Body Systems: From
  the Origin of Sound to an Origin of Light and Electrons}}}}\ (\bibinfo
  {publisher} {Oxford University Press},\ \bibinfo {address} {Oxford, U.K.},\
  \bibinfo {year} {2007})\BibitemShut {NoStop}%
\bibitem [{\citenamefont {Giamarchi}(2003)}]{Giamarchi}%
  \BibitemOpen
  \bibfield  {author} {\bibinfo {author} {\bibfnamefont {T.}~\bibnamefont
  {Giamarchi}},\ }\href
  {https://doi.org/10.1093/acprof:oso/9780198525004.001.0001} {\emph {\bibinfo
  {title} {{Quantum Physics in One Dimension}}}}\ (\bibinfo  {publisher}
  {Oxford University Press},\ \bibinfo {address} {Oxford, U.K.},\ \bibinfo
  {year} {2003})\BibitemShut {NoStop}%
\bibitem [{\citenamefont {Lecheminant}\ \emph {et~al.}(2002)\citenamefont
  {Lecheminant}, \citenamefont {Gogolin},\ and\ \citenamefont
  {Nersesyan}}]{lecheminant2002criticality}%
  \BibitemOpen
  \bibfield  {author} {\bibinfo {author} {\bibfnamefont {P.}~\bibnamefont
  {Lecheminant}}, \bibinfo {author} {\bibfnamefont {A.~O.}\ \bibnamefont
  {Gogolin}},\ and\ \bibinfo {author} {\bibfnamefont {A.~A.}\ \bibnamefont
  {Nersesyan}},\ }\bibfield  {title} {\bibinfo {title} {{Criticality in
  self-dual sine-Gordon models}},\ }\href
  {https://doi.org/10.1016/S0550-3213(02)00474-1} {\bibfield  {journal}
  {\bibinfo  {journal} {Nuclear Physics B}\ }\textbf {\bibinfo {volume}
  {639}},\ \bibinfo {pages} {502} (\bibinfo {year} {2002})}\BibitemShut
  {NoStop}%
\bibitem [{\citenamefont {Calabrese}\ and\ \citenamefont
  {Cardy}(2009)}]{Calabrese:2009dx}%
  \BibitemOpen
  \bibfield  {author} {\bibinfo {author} {\bibfnamefont {P.}~\bibnamefont
  {Calabrese}}\ and\ \bibinfo {author} {\bibfnamefont {J.}~\bibnamefont
  {Cardy}},\ }\bibfield  {title} {\bibinfo {title} {{Entanglement entropy and
  conformal field theory}},\ }\href
  {https://doi.org/10.1088/1751-8113/42/50/504005} {\bibfield  {journal}
  {\bibinfo  {journal} {Journal of Physics A: Mathematical and Theoretical}\
  }\textbf {\bibinfo {volume} {42}},\ \bibinfo {pages} {504005} (\bibinfo
  {year} {2009})}\BibitemShut {NoStop}%
\bibitem [{\citenamefont {Calabrese}\ and\ \citenamefont
  {Cardy}(2004)}]{calabrese2004entanglement}%
  \BibitemOpen
  \bibfield  {author} {\bibinfo {author} {\bibfnamefont {P.}~\bibnamefont
  {Calabrese}}\ and\ \bibinfo {author} {\bibfnamefont {J.}~\bibnamefont
  {Cardy}},\ }\bibfield  {title} {\bibinfo {title} {Entanglement entropy and
  quantum field theory},\ }\href
  {https://doi.org/10.1088/1742-5468/2004/06/P06002} {\bibfield  {journal}
  {\bibinfo  {journal} {J. Stat. Mech.: Theory Exp.}\ }\textbf {\bibinfo
  {volume} {2004}}\bibinfo  {number} { (06)},\ \bibinfo {pages}
  {P06002}}\BibitemShut {NoStop}%
\bibitem [{\citenamefont {Luther}\ and\ \citenamefont
  {Emery}(1974)}]{LutherEmery_PhysRevLett.33.589}%
  \BibitemOpen
\bibfield  {number} {  }\bibfield  {author} {\bibinfo {author} {\bibfnamefont
  {A.}~\bibnamefont {Luther}}\ and\ \bibinfo {author} {\bibfnamefont {V.~J.}\
  \bibnamefont {Emery}},\ }\bibfield  {title} {\bibinfo {title} {{Backward
  Scattering in the One-Dimensional Electron Gas}},\ }\href
  {https://doi.org/10.1103/PhysRevLett.33.589} {\bibfield  {journal} {\bibinfo
  {journal} {Physical Review Letters}\ }\textbf {\bibinfo {volume} {33}},\
  \bibinfo {pages} {589} (\bibinfo {year} {1974})}\BibitemShut {NoStop}%
\bibitem [{\citenamefont {Zhu}\ \emph {et~al.}(2023)\citenamefont {Zhu},
  \citenamefont {Tantivasadakarn}, \citenamefont {Vishwanath}, \citenamefont
  {Trebst},\ and\ \citenamefont {Verresen}}]{Zhu:2023be}%
  \BibitemOpen
  \bibfield  {author} {\bibinfo {author} {\bibfnamefont {G.-Y.}\ \bibnamefont
  {Zhu}}, \bibinfo {author} {\bibfnamefont {N.}~\bibnamefont
  {Tantivasadakarn}}, \bibinfo {author} {\bibfnamefont {A.}~\bibnamefont
  {Vishwanath}}, \bibinfo {author} {\bibfnamefont {S.}~\bibnamefont {Trebst}},\
  and\ \bibinfo {author} {\bibfnamefont {R.}~\bibnamefont {Verresen}},\
  }\bibfield  {title} {\bibinfo {title} {{Nishimori's Cat: Stable Long-Range
  Entanglement from Finite-Depth Unitaries and Weak Measurements}},\ }\href
  {https://doi.org/10.1103/PhysRevLett.131.200201} {\bibfield  {journal}
  {\bibinfo  {journal} {Physical Review Letters}\ }\textbf {\bibinfo {volume}
  {131}},\ \bibinfo {pages} {200201} (\bibinfo {year} {2023})}\BibitemShut
  {NoStop}%
\bibitem [{\citenamefont {Turkeshi}\ and\ \citenamefont
  {Sierant}(2024)}]{Turkeshi:2024dd}%
  \BibitemOpen
  \bibfield  {author} {\bibinfo {author} {\bibfnamefont {X.}~\bibnamefont
  {Turkeshi}}\ and\ \bibinfo {author} {\bibfnamefont {P.}~\bibnamefont
  {Sierant}},\ }\bibfield  {title} {\bibinfo {title} {{Error-Resilience Phase
  Transitions in Encoding-Decoding Quantum Circuits}},\ }\href
  {https://doi.org/10.1103/PhysRevLett.132.140401} {\bibfield  {journal}
  {\bibinfo  {journal} {Physical Review Letters}\ }\textbf {\bibinfo {volume}
  {132}},\ \bibinfo {pages} {140401} (\bibinfo {year} {2024})}\BibitemShut
  {NoStop}%
\bibitem [{\citenamefont {Anand}\ \emph {et~al.}(2025)\citenamefont {Anand},
  \citenamefont {Zaletel},\ and\ \citenamefont {Bao}}]{Anand2025}%
  \BibitemOpen
  \bibfield  {author} {\bibinfo {author} {\bibfnamefont {S.}~\bibnamefont
  {Anand}}, \bibinfo {author} {\bibfnamefont {M.}~\bibnamefont {Zaletel}},\
  and\ \bibinfo {author} {\bibfnamefont {Y.}~\bibnamefont {Bao}},\ }\href@noop
  {} {\bibinfo {title} {Efficient algorithms for sampling and
  maximum-likelihood decoding in the surface code with local noise}} (\bibinfo
  {year} {2025}),\ \bibinfo {note} {in preparation}\BibitemShut {NoStop}%
\bibitem [{Note2()}]{Note2}%
  \BibitemOpen
  \bibinfo {note} {The total number of parameters reduces to 12 when taking
  into account that $\protect \mathcal {E}_j^\protect \mathrm {(f)}$ is
  traceless~\cite {Nielsen:2010ga}.}\BibitemShut {Stop}%
\bibitem [{\citenamefont {Kadanoff}\ and\ \citenamefont
  {Ceva}(1971)}]{Kadanoff:1971fm}%
  \BibitemOpen
  \bibfield  {author} {\bibinfo {author} {\bibfnamefont {L.~P.}\ \bibnamefont
  {Kadanoff}}\ and\ \bibinfo {author} {\bibfnamefont {H.}~\bibnamefont
  {Ceva}},\ }\bibfield  {title} {\bibinfo {title} {{Determination of an
  Operator Algebra for the Two-Dimensional Ising Model}},\ }\href
  {https://doi.org/10.1103/PhysRevB.3.3918} {\bibfield  {journal} {\bibinfo
  {journal} {Physical Review B}\ }\textbf {\bibinfo {volume} {3}},\ \bibinfo
  {pages} {3918} (\bibinfo {year} {1971})}\BibitemShut {NoStop}%
\bibitem [{\citenamefont {Read}\ and\ \citenamefont
  {Ludwig}(2000)}]{Read:2000hu}%
  \BibitemOpen
  \bibfield  {author} {\bibinfo {author} {\bibfnamefont {N.}~\bibnamefont
  {Read}}\ and\ \bibinfo {author} {\bibfnamefont {A.~W.~W.}\ \bibnamefont
  {Ludwig}},\ }\bibfield  {title} {\bibinfo {title} {{Absence of a metallic
  phase in random-bond Ising models in two dimensions: Applications to
  disordered superconductors and paired quantum Hall states}},\ }\href
  {https://doi.org/10.1103/PhysRevB.63.024404} {\bibfield  {journal} {\bibinfo
  {journal} {Physical Review B}\ }\textbf {\bibinfo {volume} {63}},\ \bibinfo
  {pages} {024404} (\bibinfo {year} {2000})}\BibitemShut {NoStop}%
\end{thebibliography}%

\end{document}